\documentclass[10pt,preprint]{sigplanconf}

\usepackage[breaklinks]{hyperref}

\usepackage{amsmath}
\usepackage{amsthm}
\usepackage{amsfonts}
\usepackage{amssymb}
\usepackage{mathtools}
\usepackage{bm}
\usepackage{stmaryrd}
\usepackage{galois}
\usepackage{latexsym}
\usepackage{balance}
\usepackage{array}
\usepackage{tikz}
\usepackage{etoolbox}
\usetikzlibrary{matrix}

\newtheorem{theorem}{Theorem}
\newtheorem{lemma}{Lemma}
\newtheorem{proposition}{Proposition}

\newtheorem{definition}{Definition}

\newcommand\litM{M}
\newcommand\litTime{Time}
\newcommand\litStore{Store}
\newcommand\litVal{Val}
\newcommand\litKAddr{KAddr}
\newcommand\litKStore{KStore}
\newcommand\litEnv{Env}
\newcommand\litClo{Clo}
\newcommand\litelse{else}

\newcommand{\itop}[1]{\operatorname{\mathit{#1}}}
\newcommand{\ttop}[1]{\operatorname{\mathtt{#1}}}
\newcommand{\ttbfop}[1]{\ttop{\boldsymbol{\mathbf{#1}}}}
\newcommand{\ttbfbin}[1]{\mathbin{\mathtt{\mathbf{\boldsymbol{#1}}}}}
\newcommand{\keyword}[1]{\underline{\ttbfop{#1}}}
\newcommand{\keywordop}[1]{\ttbfbin{#1}}
\newcommand{\Concrete}[1]{\mathtt{\boldsymbol{\mathbf{#1}}}}
\newcommand{\Abstract}[1]{\widehat{\mathtt{\boldsymbol{\mathbf{#1}}}}}

\DeclareMathSizes{10}{8}{6}{4}

\begin{document}


\setlength{\pdfpageheight}{\paperheight}
\setlength{\pdfpagewidth}{\paperwidth}

\exclusivelicense
\conferenceinfo{OOPSLA '15}{October 25--30, 2015, Pittsburgh, PA, USA} 
\copyrightyear{2015} 
\copyrightdata{978-1-4503-3689-5/15/10} 
\doi{2814270.2814308}

%





\newcommand{\zerodisplayskips}{%
  \setlength{\abovedisplayskip}{0pt}
  \setlength{\belowdisplayskip}{0pt}
  \setlength{\abovedisplayshortskip}{0pt}
  \setlength{\belowdisplayshortskip}{0pt}}
\appto{\small}{\zerodisplayskips}

\setlength{\textfloatsep}{8pt}

\title{Galois Transformers and Modular Abstract Interpreters}
\subtitle{Reusable Metatheory for Program Analysis}
\authorinfo{David Darais}{University of Maryland, USA}{darais@cs.umd.edu}
\authorinfo{Matthew Might}{University of Utah, USA}{might@cs.utah.edu}
\authorinfo{David Van Horn}{University of Maryland, USA}{dvanhorn@cs.umd.edu}
\maketitle

\begin{abstract}

The design and implementation of static analyzers has become increasingly
systematic.  Yet for a given language or analysis feature, it often requires
tedious and error prone work to implement an analyzer and prove it sound. In
short, static analysis features and their proofs of soundness do not compose
well, causing a dearth of reuse in both implementation and metatheory.

\par 

We solve the problem of systematically constructing static analyzers by
introducing \emph{Galois transformers}: monad transformers that transport
Galois connection properties.  In concert with a monadic interpreter, we define
a library of monad transformers that implement building blocks for classic
analysis parameters like context, path, and heap (in)sensitivity. Moreover,
these can be composed together \emph{independent of the language being
analyzed}.

\par 

Significantly, a Galois transformer can be proved sound once and for all,
making it a reusable analysis component.  As new analysis features and
abstractions are developed and mixed in, soundness proofs need not be
reconstructed, as the composition of a monad transformer stack is sound by
virtue of its constituents.  Galois transformers provide a viable foundation
for reusable and composable metatheory for program analysis.

\par 

Finally, these Galois transformers shift the level of abstraction in analysis
design and implementation to a level where non-specialists have the ability to
synthesize sound analyzers over a number of parameters.

\end{abstract}

\category{F.3.2}{Semantics of Programming Languages}{Program analysis}
\keywords abstract interpretation, monads, Galois connections, program analysis

\section{Introduction}\label{introduction}

\par

Traditional practice in program analysis via abstract interpretation is
to fix a language (as a concrete semantics) and an abstraction (as an
abstraction map, concretization map or Galois connection) before
constructing a static analyzer that is sound with respect to both the
abstraction and the concrete semantics. Thus, each pairing of
abstraction and semantics requires a one-off manual derivation of the
static analyzer and construction of its proof of soundness.

\par

Work has focused on endowing abstractions with knobs, levers, and dials
to tune precision and compute efficiently. These parameters come with
overloaded meanings such as object, context, path and heap
sensitivities, or some combination thereof. These efforts develop
families of analyses \emph{for a specific language} and prove the
framework sound.

\par

But this framework approach suffers from many of the same drawbacks as
the one-off analyzers. They are language-specific, preventing reuse of
concepts across languages, and require similar re-implementations and
soundness proofs. This process is still manual, tedious, difficult and
error-prone. And, changes to the structure of the parameter-space
require a completely new proof of soundness. And, it prevents fruitful
insights and results developed in one paradigm from being applied to
others, e.g., functional to object-oriented and \emph{vice versa}.

\par

We propose an automated alternative to structuring and implementing
program analysis. Inspired by \citeauthor*{dvanhorn:Liang1995Monad}'s
\emph{Monad
Transformers and Modular Interpreters} \cite{dvanhorn:Liang1995Monad},
we propose to start with concrete interpreters written in a specific
monadic style. Changing the monad will transform the concrete
interpreter into an abstract interpreter. As we show, classical program
abstractions can be embodied as language-independent monads. Moreover,
these abstractions can be written as monad \emph{transformers}, thereby
allowing their composition to achieve new forms of analysis. We show
that these monad transformers obey the properties of
\emph{Galois connections} \cite{dvanhorn:Cousot1979Systematic} and
introduce the concept of a \emph{Galois transformer}, a monad
transformer which transports Galois connection properties.

\par

Most significantly, Galois transformers are proven sound once and for
all. Abstract interpreters, which take the form of monad transformer
stacks coupled with a monadic interpreter, inherit the soundness
properties of each element in the stack. This approach enables reuse of
abstractions across languages and lays the foundation for a modular
metatheory of program analysis.

\par

\paragraph{Setup}

We describe a simple programming language and a garbage-collecting
allocating semantics as the starting point of analysis design
(Section~\ref{semantics}). We then briefly discuss three types of path
and flow sensitivity and their corresponding variations in analysis
precision (Section~\ref{path-and-flow-sensitivity-in-analysis}).

\par

\paragraph{Monadic Abstract Interpreters}

We develop an abstract interpreter for our example language as a monadic
function with parameters (Sections~\ref{analysis-parameters}
and~\ref{the-interpreter}), one of which is a monadic effect interface
combining state and nondeterminism effects
(Section~\ref{the-analysis-monad}). These monadic effects---state and
nondeterminism---encode arbitrary relational small-step state-machine
semantics and correspond to state-machine components and relational
nondeterminism, respectively.

\par

Interpreters written in this style are reasoned about using various
laws, including monadic effect laws, and are verified correct
independent of any particular choice of parameters. Likewise, choices
for these parameters are proven correct in isolation from their
instantiation. When instantiated, our generic interpreter recovers the
concrete semantics and a family of abstract interpreters with variations
in abstract domain, abstract garbage collection, call-site sensitivity,
object sensitivity, and path and flow sensitivity
(Section~\ref{recovering-analyses}). Furthermore, each derived abstract
interpreter is proven correct by construction through a reusable,
semantics independent proof framework
(Section~\ref{a-compositional-monadic-framework}).

\par

\paragraph{Isolating Path and Flow Sensitivity}

We give specific monads for instantiating the interpreter from
Section~\ref{the-interpreter} to path-sensitive, flow-sensitive and
flow-insensitive analyses
(Section~\ref{varying-path-and-flow-sensitivity}). This leads to an
isolated understanding of path and flow sensitivity as mere variations
in the monad used for execution. Furthermore, these monads are language
independent, allowing one to reuse the same path and flow sensitivity
machinery for any language of interest, and compose seamlessly with
other analysis parameters.

\par

\paragraph{Galois Transformers}

To ease the construction of monads for building abstract interpreters
and their proofs of correctness, we develop a framework of Galois
transformers (Section~\ref{a-compositional-monadic-framework}). Galois
transformers are an extension of monad transformers which transport
Galois connection properties (Section~\ref{galois-transformers-1}). The
Galois transformer framework allows us to both execute and justify the
correctness of an abstract interpreter piecewise for each transformer.
Galois transformers are language independent and they are proven correct
once and for all in isolation from a particular semantics.

\par

\paragraph{Implementation}

We implement our technique as a Haskell library and example client
analysis (Section~\ref{implementation-1}). Developers are able to reuse
our language-independent framework for prototyping the design space of
analysis features for their language of choice. Our implementation is
publicly available on
Hackage\footnote{http://hackage.haskell.org/package/maam}, Haskell's
package manager.

\par

\par

\begin{figure}

\begin{alignat*}{4}
i  \in  &  \mathbb{Z}  &  {}  & x  \in   \ensuremath{\itop{Var}}  \\
a  \in  &  \ensuremath{\itop{Atom}}  &  {}  &  \Coloneqq  i  \;|\;  x  \;|\;   \keyword{\lambda} (x).e \\
 \oplus   \in  &  \ensuremath{\itop{IOp}}  &  {}  &  \Coloneqq   \keywordop{+}   \;|\;   \keywordop{-}  \\
 \odot   \in  &  \ensuremath{\itop{Op}}  &  {}  &  \Coloneqq   \oplus   \;|\;   \keywordop{ \mathbin{\bm{@}} }  \\
e  \in  &  \ensuremath{\itop{Exp}}  &  {}  &  \Coloneqq  a  \;|\;  e  \odot  e  \;|\;   \keyword{if0} (e) \{ e \}  \{ e \}  \\
 \  \\
 \tau   \in  &  \ensuremath{\itop{Time}}  &  {}  &  \coloneqq   \mathbb{Z}  \\
l  \in  &  \ensuremath{\itop{Addr}}  &  {}  &  \coloneqq   \ensuremath{\itop{Var}}   \times   \ensuremath{\itop{Time}}  \\
 \rho   \in  &  \ensuremath{\itop{Env}}  &  {}  &  \coloneqq   \ensuremath{\itop{Var}}   \rightharpoonup   \ensuremath{\itop{Addr}}  \\
 \sigma   \in  &  \ensuremath{\itop{Store}}  &  {}  &  \coloneqq   \ensuremath{\itop{Addr}}   \rightharpoonup   \ensuremath{\itop{Val}}  \\
c  \in  &  \ensuremath{\itop{Clo}}  &  {}  &  \Coloneqq   \langle  \keyword{\lambda} (x).e, \rho  \rangle  \\
v  \in  &  \ensuremath{\itop{Val}}  &  {}  &  \Coloneqq  i  \;|\;  c \\
 \kappa l  \in  &  \ensuremath{\itop{KAddr}}  &  {}  &  \coloneqq   \ensuremath{\itop{Time}}  \\
 \kappa  \sigma   \in  &  \ensuremath{\itop{KStore}}  &  {}  &  \coloneqq   \ensuremath{\itop{KAddr}}   \rightharpoonup   \ensuremath{\itop{Frame}}   \times   \ensuremath{\itop{KAddr}}  \\
fr  \in  &  \ensuremath{\itop{Frame}}  &  {}  &  \Coloneqq   \langle  \square   \odot  e, \rho  \rangle   \;|\;   \langle v  \odot   \square  \rangle   \;|\;   \langle  \keyword{if0} ( \square ) \{ e \}  \{ e \} , \rho  \rangle  \\
 \varsigma   \in  &  \Sigma  &  {}  &  \Coloneqq   \langle e, \rho , \sigma , \kappa l, \kappa  \sigma , \tau  \rangle 
\end{alignat*}

\caption{ $ \ensuremath{\ttbfop{\lambda{}IF}} $ Syntax and Concrete
State Space } \label{SS} \end{figure}

\par

\par

\paragraph{Contributions}

We make the following contributions:

\par

\begin{itemize}
\itemsep1pt\parskip0pt\parsep0pt
\item
  A methodology for constructing monadic abstract interpreters based on
  \emph{monadic effects}.
\item
  A compositional, language-independent framework for constructing
  monads with varying analysis properties based on \emph{monad
  transformers}.
\item
  A compositional, language-independent proof framework for constructing
  Galois connections and end-to-end correctness proofs based on
  \emph{Galois transformers}, an extension of monad transformers which
  transports Galois connection properties.
\item
  Two new general purpose monad transformers for nondeterminism which
  are not present in any previous work on monad transformers (even
  outside static analysis literature). Although applicable to settings
  other than static analysis, these two transformers give rise naturally
  to variations in path and flow sensitivity when applied to abstract
  interpreters.
\item
  An isolated understanding of path and flow sensitivity in analysis as
  properties of the interpreter monad, which we develop independently of
  other analysis features.
\end{itemize}

\par

Collectively, these contributions make progress toward a reusable
metatheory for program analysis.

\par

\section{Semantics}\label{semantics}

\par

To demonstrate our framework we design an abstract interpreter for
$ \ensuremath{\ttbfop{\lambda{}IF}} $, a simple applied lambda calculus
shown in Figure~\ref{SS}. $ \ensuremath{\ttbfop{\lambda{}IF}} $ extends
traditional lambda calculus with integers, addition, subtraction and
conditionals. We write $ \keywordop{ \mathbin{\bm{@}} } $ as explicit
abstract syntax for function application. The state-space $ \Sigma $ for
$ \ensuremath{\ttbfop{\lambda{}IF}} $ makes allocation explicit using
two separate stores for values ($ \ensuremath{\itop{Store}} $) and for
the stack ($ \ensuremath{\itop{KStore}} $).

\par

\begin{figure}

\begin{align*}
&\hspace{0em} A \llbracket  \underline{\hspace{0.5em}}  \rrbracket  :  \ensuremath{\itop{Atom}}   \rightarrow   (\ensuremath{\itop{Env}}   \times   \ensuremath{\itop{Store}}   \rightharpoonup   \ensuremath{\itop{Val}})  \\
&\hspace{0em} A \llbracket i \rrbracket ( \rho , \sigma )  \coloneqq  i \\
&\hspace{0em} A \llbracket x \rrbracket ( \rho , \sigma )  \coloneqq   \sigma ( \rho (x)) \\
&\hspace{0em} A \llbracket  \keyword{\lambda} (x).e \rrbracket ( \rho , \sigma )  \coloneqq   \langle  \keyword{\lambda} (x).e, \rho  \rangle   \\
&\hspace{0em}  \delta  \llbracket  \underline{\hspace{0.5em}}  \rrbracket  :  \ensuremath{\itop{IOp}}   \rightarrow  ( \mathbb{Z}   \times   \mathbb{Z}   \rightarrow   \mathbb{Z} ) \\
&\hspace{0em}  \delta  \llbracket  \keywordop{+}  \rrbracket (i _1 ,i _2 )  \coloneqq  i _1  + i _2  \\
&\hspace{0em}  \delta  \llbracket  \keywordop{-}  \rrbracket (i _1 ,i _2 )  \coloneqq  i _1  - i _2  \\
&\hspace{0em}  \  \\
&\hspace{0em}  \underline{\hspace{0.5em}}{  \rightsquigarrow  }\underline{\hspace{0.5em}}  :  \mathcal{P} ( \Sigma   \times   \Sigma ) \\
&\hspace{0em}  \langle e _1   \odot  e _2 , \rho , \sigma , \kappa l, \kappa  \sigma , \tau  \rangle   \rightsquigarrow   \langle e _1 , \rho , \sigma , \tau , \kappa  \sigma ', \tau +1 \rangle   \;\;\text{\footnotesize{}where}\;\;   \\
&\hspace{2em}  \kappa  \sigma '  \coloneqq   \kappa  \sigma  \lbrack  \tau   \mapsto   \langle  \langle  \square   \odot  e _2 , \rho  \rangle , \kappa l \rangle  \rbrack  \\
&\hspace{0em}  \langle  \keyword{if0} (e _1 ) \{ e _2  \}  \{ e _3  \} , \rho , \sigma , \kappa l, \kappa  \sigma , \tau  \rangle   \rightsquigarrow   \langle e _1 , \rho , \sigma , \tau , \kappa  \sigma ', \tau +1 \rangle   \;\;\text{\footnotesize{}where}\;\;  \\
&\hspace{2em}  \kappa  \sigma '  \coloneqq   \kappa  \sigma  \lbrack  \tau   \mapsto   \langle  \langle  \keyword{if0} ( \square ) \{ e _2  \}  \{ e _3  \} , \rho  \rangle , \kappa l \rangle  \rbrack  \\
&\hspace{0em}  \langle a, \rho , \sigma , \kappa l, \kappa  \sigma , \tau  \rangle   \rightsquigarrow   \langle e, \rho ', \sigma , \tau , \kappa  \sigma ', \tau +1 \rangle   \;\;\text{\footnotesize{}where}\;\;  \\
&\hspace{2em}  \begin{aligned}[l]  &  \langle  \langle  \square   \odot  e, \rho ' \rangle , \kappa l' \rangle   \coloneqq   \kappa  \sigma ( \kappa l)  \\  &  \kappa  \sigma '  \coloneqq   \kappa  \sigma  \lbrack  \tau   \mapsto   \langle  \langle A \llbracket a \rrbracket ( \rho , \sigma )  \odot   \square  \rangle , \kappa l' \rangle  \rbrack   \end{aligned}  \\
&\hspace{0em}  \langle a, \rho , \sigma , \kappa l, \kappa  \sigma , \tau  \rangle   \rightsquigarrow   \langle e, \rho '', \sigma ', \kappa l', \kappa  \sigma , \tau +1 \rangle   \;\;\text{\footnotesize{}where}\;\;   \\
&\hspace{2em}  \begin{aligned}[l]  &  \langle  \langle  \langle  \keyword{\lambda} (x).e, \rho ' \rangle   \keywordop{ \mathbin{\bm{@}} }   \square  \rangle , \kappa l' \rangle   \coloneqq   \kappa  \sigma ( \kappa l)  \\  &  \rho ''  \coloneqq   \rho ' \lbrack x  \mapsto   \langle x, \tau  \rangle  \rbrack   \\  &  \sigma '  \coloneqq   \sigma  \lbrack  \langle x, \tau  \rangle   \mapsto  A \llbracket a \rrbracket ( \rho , \sigma ) \rbrack   \end{aligned}  \\
&\hspace{0em}  \langle i _2 , \rho , \sigma , \kappa l, \kappa  \sigma , \tau  \rangle   \rightsquigarrow   \langle i, \rho , \sigma , \kappa l', \kappa  \sigma , \tau +1 \rangle   \;\;\text{\footnotesize{}where}\;\;   \\
&\hspace{2em}  \begin{aligned}[l]  &  \langle  \langle i _1   \oplus   \square  \rangle , \kappa l' \rangle   \coloneqq   \kappa  \sigma ( \kappa l)  \\  & i  \coloneqq   \delta  \llbracket  \oplus  \rrbracket (i _1 ,i _2 )  \end{aligned}  \\
&\hspace{0em}  \langle i, \rho , \sigma , \kappa l, \kappa  \sigma , \tau  \rangle   \rightsquigarrow   \langle e, \rho ', \sigma , \kappa l', \kappa  \sigma , \tau +1 \rangle   \;\;\text{\footnotesize{}where}\;\;   \\
&\hspace{2em}  \begin{aligned}[l]  &  \langle  \langle  \keyword{if0} ( \square ) \{ e _1  \}  \{ e _2  \} , \rho ' \rangle , \kappa l' \rangle   \coloneqq   \kappa  \sigma ( \kappa l)  \\  & e  \coloneqq  e _1   \;\;\text{\footnotesize{}when}\;\;  i = 0  \;;\;  e _2   \;\;\text{\footnotesize{}when}\;\;  i  \neq  0  \end{aligned} 
\end{align*}

\caption{Concrete Semantics} \label{ConcreteSemantics} \end{figure}

\par

\begin{figure}

\begin{align*}
&\hspace{0em}  \underline{\hspace{0.5em}}{  \rightsquigarrow   ^{ \ensuremath{\itop{gc}} }   }\underline{\hspace{0.5em}}  :  \mathcal{P} ( \Sigma   \times   \Sigma ) \\
&\hspace{0em}  \varsigma   \rightsquigarrow   ^{ \ensuremath{\itop{gc}} }    \varsigma '  \;\;\text{\footnotesize{}where}\;\;   \varsigma   \rightsquigarrow   \varsigma ' \\
&\hspace{0em}  \langle e, \rho , \sigma , \kappa l, \kappa  \sigma , \tau  \rangle   \rightsquigarrow   ^{ \ensuremath{\itop{gc}} }    \langle e, \rho , \sigma ', \kappa l, \kappa  \sigma ', \tau  \rangle   \;\;\text{\footnotesize{}where}\;\;   \\
&\hspace{2em}  \begin{aligned}[l]   \kappa  \sigma ' &  \coloneqq   \{  \kappa l  \mapsto   \kappa  \sigma ( \kappa l)  \;|\;   \kappa l  \in   \ensuremath{\itop{KR}}(  \kappa l, \kappa  \sigma )  \}   \\   \sigma ' &  \coloneqq   \{ l  \mapsto   \sigma (l)  \;|\;  l  \in   \ensuremath{\itop{R}}( e, \rho , \sigma , \kappa l, \kappa  \sigma ) \}   \end{aligned}  \\
&\hspace{0em}  \  \\
&\hspace{0em}  \ensuremath{\itop{KR}}  :  \ensuremath{\itop{KAddr}}   \times   \ensuremath{\itop{KStore}}   \rightarrow   \mathcal{P}  (\ensuremath{\itop{KAddr}})  \\
&\hspace{0em}  \ensuremath{\itop{KR}}(  \kappa l, \kappa  \sigma )  \coloneqq   \mu (X).  \\
&\hspace{2em} X  \cup   \{  \kappa l \}   \cup   \{   \pi  _2 ( \kappa  \sigma ( \kappa l))  \;|\;   \kappa l  \in  X  \}  \\
&\hspace{0em}  \ensuremath{\itop{R}}  :  \ensuremath{\itop{Exp}}   \times   \ensuremath{\itop{Env}}   \times   \ensuremath{\itop{Store}}   \times   \ensuremath{\itop{KAddr}}   \times   \ensuremath{\itop{KStore}}   \rightarrow   \mathcal{P}  (\ensuremath{\itop{Addr}})  \\
&\hspace{0em}  \ensuremath{\itop{R}}( e, \rho , \sigma , \kappa l, \kappa  \sigma )  \coloneqq   \mu (X). \\
&\hspace{2em} X  \cup   \{  \rho (x)  \;|\;  x  \in   \ensuremath{\itop{FV}}( e) \}  \\
&\hspace{2em}  \cup   \{ l  \;|\;  l  \in   \ensuremath{\itop{R-Frm}}(  \pi  _1 ( \kappa  \sigma ( \kappa l)))  \;;\;   \kappa l  \in   \ensuremath{\itop{KR}}(  \kappa l, \kappa  \sigma ) \}  \\
&\hspace{2em}  \cup   \{ l'  \;|\;  l'  \in   \ensuremath{\itop{R-Val}}(  \sigma (l))  \;;\;  l  \in  X \}  \\
&\hspace{0em}  \  \\
&\hspace{0em}  \ensuremath{\itop{R-Frm}}  :  \ensuremath{\itop{Frame}}   \rightarrow   \mathcal{P}  (\ensuremath{\itop{Addr}})  \\
&\hspace{0em}  \ensuremath{\itop{R-Frm}}(  \langle  \square   \odot  e, \rho  \rangle )  \coloneqq   \{  \rho (x)  \;|\;  x  \in   \ensuremath{\itop{FV}}( e) \}  \\
&\hspace{0em}  \ensuremath{\itop{R-Frm}}(  \langle v  \odot   \square  \rangle )  \coloneqq   \ensuremath{\itop{R-Val}}( v) \\
&\hspace{0em}  \ensuremath{\itop{R-Frm}}(  \langle  \keyword{if0} ( \square ) \{ e _2  \}  \{ e _3  \} , \rho  \rangle )  \coloneqq   \{  \rho (x)  \;|\;  x  \in   \ensuremath{\itop{FV}}( e _1 )  \cup   \ensuremath{\itop{FV}}( e _2 ) \}  \\
&\hspace{0em}  \ensuremath{\itop{R-Val}}   \in   \ensuremath{\itop{Val}}   \rightarrow   \mathcal{P}  (\ensuremath{\itop{Addr}})  \\
&\hspace{0em}  \ensuremath{\itop{R-Val}}( i)  \coloneqq   \{  \}  \\
&\hspace{0em}  \ensuremath{\itop{R-Val}}(  \langle  \keyword{\lambda} (x).e, \rho  \rangle )  \coloneqq   \{  \rho (y)  \;|\;  y  \in   \ensuremath{\itop{FV}}(  \keyword{\lambda} (x).e) \}  \\
&\hspace{0em}  \  \\
&\hspace{0em} collect :  \mathcal{P} ( \Sigma ) \\
&\hspace{0em} collect  \coloneqq   \mu (X).X  \cup   \{  \varsigma  _0  \}   \cup   \{   \varsigma '  \;|\;   \varsigma   \rightsquigarrow   ^{ \ensuremath{\itop{gc}} }    \varsigma '  \;;\;   \varsigma   \in  X  \}   \;\;\text{\footnotesize{}where}\;\;  \\
&\hspace{2em}  \varsigma  _0   \coloneqq   \langle e _0 , \bot , \bot ,0, \bot ,1 \rangle 
\end{align*}

\caption{Garbage Collected Collecting Semantics}
\label{GCCollectingSemantics} \end{figure}

\par

Guided by the syntax and semantics of
$ \ensuremath{\ttbfop{\lambda{}IF}} $ we develop interpretation
parameters in Section~\ref{analysis-parameters}, a monadic interpreter
in Section~\ref{the-interpreter}, and both concrete and abstract
instantiations for the interpretation parameters in
Section~\ref{recovering-analyses}. The variations in path and flow
sensitivity developed in sections
\ref{varying-path-and-flow-sensitivity} and
\ref{a-compositional-monadic-framework} are independent of this (or any
other) semantics.

\par

We define semantics for atomic expressions and primitive operators
denotationally with
$A \llbracket  \underline{\hspace{0.5em}}  \rrbracket $ and
$ \delta  \llbracket  \underline{\hspace{0.5em}}  \rrbracket $, and to
compound expressions relationally with
$ \underline{\hspace{0.5em}}{  \rightsquigarrow  }\underline{\hspace{0.5em}} $,
shown in Figure~\ref{ConcreteSemantics}.

\par

Our abstract interpreter supports abstract garbage
collection~\cite{dvanhorn:Might:2006:GammaCFA}, the concrete analogue of
which is just standard garbage collection. We include abstract garbage
collection for two reasons. First, it is one of the few techniques that
results in both performance \emph{and} precision improvements for
abstract interpreters. Second, we will systematically recover concrete
and abstract garbage collectors with varying path and flow sensitivities
through a single monadic garbage collector, an axis of generality novel
in this work.

\par

We show the garbage collected semantics in
Figure~\ref{GCCollectingSemantics}, as well as a final collecting
semantics $collect$, which will serve as the starting point for
abstraction. The concrete, garbage-collected collecting semantics
$collect$ and a sound static analyzer will both be recovered from
instantiations of a generic monadic interpreter in
Section~\ref{recovering-analyses}.

\par

The garbage collected semantics
$ \underline{\hspace{0.5em}}{  \rightsquigarrow   ^{ \ensuremath{\itop{gc}} }   }\underline{\hspace{0.5em}} $
is defined with reachability functions $ \ensuremath{\itop{KR}} $ and
$ \ensuremath{\itop{R}} $ which define transitively reachable addresses.
We write $ \mu (X). f(X)$ as the least-fixed-point of the function $f$.
$ \ensuremath{\itop{R}} $ is defined in terms of
$ \ensuremath{\itop{R-Frm}} $ and $ \ensuremath{\itop{R-Val}} $, which
define the immediately reachable locations from a frame and value
respectively. We omit the definition of $ \ensuremath{\itop{FV}} $,
which is the standard recursive definition for computing free variables
of an expression.

\par

\section{Path and Flow Sensitivity in
Analysis}\label{path-and-flow-sensitivity-in-analysis}

\par

We identify three specific variants of path and flow sensitivity in
analysis: path-sensitive, flow-sensitive and flow-insensitive. Our
framework exposes the essence of path and flow sensitivity through a
monadic effect interface in Section~\ref{analysis-parameters}, and we
recover each of these variations through specific monad instances in
Sections~\ref{varying-path-and-flow-sensitivity}
and~\ref{a-compositional-monadic-framework}.

\par

Consider a combination of if-statements in our example language
$ \ensuremath{\ttbfop{\lambda{}IF}} $ (extended with let-bindings) where
an analysis cannot determine the value of $N$: \begin{alignat*}{3}

& 1:  \keyword{let}\;\;  x  \coloneqq            &&  \hspace{1em}  \hspace{1em}  \keyword{in}                  \\
&  \hspace{1em}  \hspace{1em} 2:  \keyword{if0} (N) \{           &&  \hspace{1em}  \hspace{1em} 5:  \keyword{let}\;\;  y  \coloneqq         \\
&  \hspace{1em}  \hspace{1em}  \hspace{1em}  \hspace{1em} 3:  \keyword{if0} (N) \{ 1 \}  \{ 2 \}    &&  \hspace{1em}  \hspace{1em}  \hspace{1em}  \hspace{1em} 6:  \keyword{if0} (N) \{ 5 \}  \{ 6 \}   \\
&  \hspace{1em}  \hspace{1em}  \}   \;\;\keyword{\litelse}\;\;   \{             &&  \hspace{1em}  \hspace{1em}  \keyword{in}                  \\
&  \hspace{1em}  \hspace{1em}  \hspace{1em}  \hspace{1em} 4:  \keyword{if0} (N) \{ 3 \}  \{ 4 \}   \}  &&  \hspace{1em}  \hspace{1em} 7:  \keyword{exit} (x, y)

\end{alignat*} \paragraph{Path-Sensitive} A path-sensitive analysis
tracks both data and control flow precisely. At program points 3 and 4
the analysis considers separate worlds:

\begin{alignat*}{1}
3:  \{ N=0 \}  \quad 4:  \{ N \neq 0 \} 
\end{alignat*}

At program points 5 and 6 the analysis continues in two separate,
precise worlds:

\begin{alignat*}{1}
5,6:  \{ N=0 ,\;\;  x=1 \}   \{ N \neq 0 ,\;\;  x=4 \} 
\end{alignat*}

At program point 7 the analysis correctly correlates $x$ and $y$:

\begin{alignat*}{1}
7:  \{ N=0 ,\;\;  x=1 ,\;\;  y=5 \}   \{ N \neq 0 ,\;\;  x=4 ,\;\;  y=6 \} 
\end{alignat*}

\par

\paragraph{Flow-Sensitive}

A flow-sensitive analysis collects a \emph{single} set of facts for each
variable \emph{at each program point}. At program points 3 and 4, the
analysis considers separate worlds:

\begin{alignat*}{1}
3:  \{ N=0 \}  \quad 4:  \{ N \neq 0 \} 
\end{alignat*}

Each nested if-statement then evaluates only one side of the branch,
resulting in values $1$ and $4$. At program points 5 and 6 the analysis
is only allowed one set of facts, so it must merge the possible values
that $x$ and $N$ could take:

\begin{alignat*}{1}
5,6:  \{ N \in  \mathbb{Z}  ,\;\;  x \in  \{ 1,4 \}  \} 
\end{alignat*}

The analysis then explores both branches at program point 6 resulting in
no correlation between values for $x$ and $y$:

\begin{alignat*}{1}
7:  \{ N \in  \mathbb{Z}  ,\;\;  x \in  \{ 1,4 \}  ,\;\;  y \in  \{ 5,6 \}  \} 
\end{alignat*}

\par

\paragraph{Flow-Insensitive}

A flow-insensitive analysis collects a \emph{single} set of facts about
each variable which must hold true \emph{for the entire program}.
Because the value of $N$ is unknown at \emph{some} point in the program,
the value of $x$ must consider both branches of the nested if-statement.
This results in the global set of facts giving four values to $x$:

\begin{alignat*}{1}
 \{ N \in  \mathbb{Z}  ,\;\;  x \in  \{ 1,2,3,4 \}  ,\;\;  y \in  \{ 5,6 \}  \} 
\end{alignat*}

\par

\section{Analysis Parameters}\label{analysis-parameters}

\par

Before constructing the abstract interpreter we first design its
parameters. The interpreter, which we develop in
Section~\ref{the-interpreter}, will be designed such that variations in
these parameters will recover both concrete and a family of abstract
interpreters, which we show in Section~\ref{recovering-analyses}. To do
this we extend the ideas developed in
\citet{dvanhorn:VanHorn2010Abstracting} with a new parameter for path
and flow sensitivity: the interpreter monad.

\par

There will be three parameters to our abstract interpreter:

\par

\begin{enumerate}
\def\labelenumi{\arabic{enumi}.}
\itemsep1pt\parskip0pt\parsep0pt
\item
  The monad, novel in this work, which captures control effects and
  gives rise to path and flow sensitivity.
\item
  The abstract domain, which captures the abstraction of values like
  integers or datatypes.
\item
  The abstraction for time, which captures call-site and object
  sensitivities.
\end{enumerate}

\par

\noindent
We place each of these parameters behind an abstract interface and leave
their implementations opaque when defining the monadic interpreter in
Section~\ref{the-interpreter}. Each parameter comes with laws which can
be used to reason about the generic interpreter independent of a
particular instantiation. Likewise, an instantiation of the interpreter
need only justify that each parameter meets its local interface, which
we justify in isolation from the generic interpreter.

\par

\subsection{The Analysis Monad}\label{the-analysis-monad}

\par

The monad for the interpreter captures the \emph{effects} of
interpretation. There are two effects in the interpreter: state and
nondeterminism. The state effect will mediate how the interpreter
interacts with state cells in the state space:
$ \ensuremath{\itop{Env}} $, $ \ensuremath{\itop{Store}} $,
$ \ensuremath{\itop{KAddr}} $, $ \ensuremath{\itop{KStore}} $ and
$ \ensuremath{\itop{Time}} $. The nondeterminism effect will mediate
branching in the execution of the interpreter. Path and flow sensitivity
will be recovered by altering how these effects interact in a particular
choice of monad.

\par

We use monadic state and nondeterminism effects to abstract over
arbitrary relational small-step state-machine semantics. State effects
correspond to the components of the state-machine and nondeterminism
effects correspond to potential nondeterminism in the relation's
definition.

\par

We briefly review monad, state and nondeterminism operators and their
laws. For a more detailed presentation see
\citet{dvanhorn:Liang1995Monad},
\citet{davdar:gibbons:2011:monadic-equational-reasoning} and
\citet{davdar:Moggi:1989:Monads}.

\par

\paragraph{Monad Operators}

A type operator $m$ is a monad if it supports
$ \ensuremath{\itop{bind}} $, a sequencing operator, and its unit
$ \ensuremath{\itop{return}} $:

\begin{alignat*}{2}
m & :  \ensuremath{\itop{Type}}   \rightarrow   \ensuremath{\itop{Type}}  \\
 \ensuremath{\itop{return}}  & :  \forall  A, A  \rightarrow  m(A) \\
 \ensuremath{\itop{bind}}  & :  \forall  A B, m(A)  \rightarrow  (A  \rightarrow  m(B))  \rightarrow  m(B)
\end{alignat*}

and obeys left unit, right unit and associativity laws.

\par

We use semicolon notation for $ \ensuremath{\itop{bind}} $--\emph{e.g.}
$x  \leftarrow  X  \;;\;  k(x)$ is sugar for
$ \ensuremath{\itop{bind}}( X)(k)$--and we replace semicolons with line
breaks headed by $ \ensuremath{\ttbfop{do}} $ for multiline monadic
definitions.

\par

\paragraph{State Effect}

A type operator $m$ supports the monadic state effect for a type $s$ if
it supports $ \ensuremath{\itop{get}} $ and $ \ensuremath{\itop{put}} $
actions over $s$: \begin{alignat*}{4}

  s & :  \ensuremath{\itop{Type}}         &  \hspace{1em}   \hspace{1em} \ensuremath{\itop{get}}  & : m(s)       \\
  m & :  \ensuremath{\itop{Type}}   \rightarrow   \ensuremath{\itop{Type}}  &  \hspace{1em}   \hspace{1em} \ensuremath{\itop{put}}  & : s  \rightarrow  m(1)

\end{alignat*} and obeys get-get, get-put, put-get and put-put
laws~\cite{davdar:gibbons:2011:monadic-equational-reasoning}.

\par

\paragraph{Nondeterminism Effect}

A type operator $m$ supports the monadic nondeterminism effect if it
supports an alternation operator $ \mathbin{\langle + \rangle} $ and its
unit $ \ensuremath{\itop{mzero}} $:

\begin{alignat*}{2}
m & :  \ensuremath{\itop{Type}}   \rightarrow   \ensuremath{\itop{Type}}  \\
 \ensuremath{\itop{mzero}}  & :  \forall  A, m(A) \\
 \underline{\hspace{0.5em}}{  \mathbin{\langle + \rangle}  }\underline{\hspace{0.5em}}  & :  \forall  A, m(A)  \times  m(A)  \rightarrow  m(A)
\end{alignat*}

$m(A)$ must have a join-semilattice structure,
$ \ensuremath{\itop{mzero}} $ must be a zero for
$ \ensuremath{\itop{bind}} $, $ \ensuremath{\itop{bind}} $ must
distributes through $ \mathbin{\langle + \rangle} $.

\par

The interpreter in Section~\ref{the-interpreter} will be defined generic
to a monad which supports monad operators, state effects and
nondeterminism effects. As a consequence, we do not reference an
explicit configuration $ \varsigma $ or collections of results; instead
we interact with an interface of state and nondeterminism effects. This
level of indirection will be exploited in
Section~\ref{varying-path-and-flow-sensitivity}, where different monads
will meet the same effect interface but yield different analysis
properties.

\par

\subsection{The Abstract Domain}\label{the-abstract-domain}

\par

To expose the abstract domain we parameterize over
$ \ensuremath{\itop{Val}} $, introduction and elimination forms for
$ \ensuremath{\itop{Val}} $, and the denotation for primitive operators
$ \delta  \llbracket  \underline{\hspace{0.5em}}  \rrbracket $.

\par

$ \ensuremath{\itop{Val}} $ must be a join-semilattice with $ \bot $ and
$ \sqcup $:

\begin{alignat*}{2}
 \bot  :  \ensuremath{\itop{Val}}  &  \hspace{1em}  \hspace{1em}  \underline{\hspace{0.5em}}{  \sqcup  }\underline{\hspace{0.5em}}  :  \ensuremath{\itop{Val}}   \times   \ensuremath{\itop{Val}}   \rightarrow   \ensuremath{\itop{Val}} 
\end{alignat*}

and respect the usual join-semilattice laws. $ \ensuremath{\itop{Val}} $
must be a join-semilattice so it can be merged in updates to
$ \ensuremath{\itop{Store}} $ to preserve soundness.

\par

$ \ensuremath{\itop{Val}} $ must also support introduction and
elimination between finite sets of concrete values $ \mathbb{Z} $ and
$ \ensuremath{\itop{Clo}} $:

\begin{alignat*}{4}
 \ensuremath{\itop{int-I}}  & :  \mathbb{Z}   \rightarrow   \ensuremath{\itop{Val}}  &  \hspace{1em}   \hspace{1em} \ensuremath{\itop{if0-E}}  & :  \ensuremath{\itop{Val}}   \rightarrow   \mathcal{P} (Bool) \\
 \ensuremath{\itop{clo-I}}  & :  \ensuremath{\itop{Clo}}   \rightarrow   \ensuremath{\itop{Val}}  &  \hspace{1em}   \hspace{1em} \ensuremath{\itop{clo-E}}  & :  \ensuremath{\itop{Val}}   \rightarrow   \mathcal{P}  (\ensuremath{\itop{Clo}}) 
\end{alignat*}

Introduction functions inject concrete values into abstract values.
Elimination functions project abstract values into a \emph{finite} set
of concrete observations. For example, we do not require that abstract
values support elimination to integers, only to finite observation of
comparison with zero. The laws for the introduction and elimination
functions induce a Galois connection between
$ \mathcal{P} ( \mathbb{Z} )$ and $ \ensuremath{\itop{Val}} $:

\begin{alignat*}{2}
 \{  \ensuremath{\ttbfop{true}}  \}  &  \subseteq   \ensuremath{\itop{if0-E}}  (\ensuremath{\itop{int-I}}( i))  \ensuremath{\ttbfop{if}}  i = 0 \\
 \{  \ensuremath{\ttbfop{false}}  \}  &  \subseteq   \ensuremath{\itop{if0-E}}  (\ensuremath{\itop{int-I}}( i))  \ensuremath{\ttbfop{if}}  i  \neq  0 \\
 \bigsqcup  _{\mathclap{\substack{ b  \in   \ensuremath{\itop{if0-E}}( v)  \\  i  \in   \theta (b) }}}   \ensuremath{\itop{int-I}}( i) &  \sqsubseteq  v \\
 \;\;\text{\footnotesize{}where}\;\;   \theta  (\ensuremath{\ttbfop{true}})  &  \coloneqq   \{ 0 \}  \\
 \theta  (\ensuremath{\ttbfop{false}})  &  \coloneqq   \{ i  \;|\;  i  \in   \mathbb{Z}   \;;\;  i  \neq  0 \} 
\end{alignat*}

Closures must follow similar laws, inducing a Galois connection between
$ \mathcal{P}  (\ensuremath{\itop{Clo}}) $ and
$ \ensuremath{\itop{Val}} $:

\begin{alignat*}{2}
 \{ c \}  &  \subseteq   \ensuremath{\itop{clo-E}}( cloI(c)) \\
 \bigsqcup  _{\mathclap{\substack{ c  \in   \ensuremath{\itop{clo-E}}( v) }}}   \ensuremath{\itop{clo-I}}( c) &  \sqsubseteq  v
\end{alignat*}

Finally, $ \delta  \llbracket  \underline{\hspace{0.5em}}  \rrbracket $
must be sound w.r.t. the Galois connection between concrete values and
$ \ensuremath{\itop{Val}} $:

\begin{alignat*}{2}
 \ensuremath{\itop{int-I}}( i _1  + i _2 ) &  \sqsubseteq   \delta  \llbracket  \keywordop{+}  \rrbracket  (\ensuremath{\itop{int-I}}( i _1 ) ,\ensuremath{\itop{int-I}}( i _2 )) \\
 \ensuremath{\itop{int-I}}( i _1  - i _2 ) &  \sqsubseteq   \delta  \llbracket  \keywordop{-}  \rrbracket  (\ensuremath{\itop{int-I}}( i _1 ) ,\ensuremath{\itop{int-I}}( i _2 ))
\end{alignat*}

\par

Supporting additional primitive types like booleans, lists, or arbitrary
inductive datatypes is analogous. Introduction functions inject the type
into $ \ensuremath{\itop{Val}} $ and elimination functions project a
finite set of discrete observations. Introduction, elimination and
$ \delta $ operators must all be sound and complete following a Galois
connection discipline.

\par

\subsection{Abstract Time}\label{abstract-time}

\par

The interface we use for abstract time is familiar from
\citet{dvanhorn:VanHorn2010Abstracting}, which introduces abstract time
as a single parameter to control various forms of context sensitivity,
and \citet{dvanhorn:Smaragdakis2011Pick}, which instantiates the
parameter to achieve various forms of object sensitivity. We only
demonstrate call-site sensitivity in this presentation; our
semantics-independent Haskell library supports object sensitivity
following the same methodology.

\par

Abstract time need only support a single operation:
$ \ensuremath{\itop{tick}} $:

\begin{alignat*}{1}
 \ensuremath{\itop{Time}}  :  \ensuremath{\itop{Type}}   \hspace{1em}  \hspace{1em}   \ensuremath{\itop{tick}}  :  \ensuremath{\itop{Exp}}   \times   \ensuremath{\itop{KAddr}}   \times   \ensuremath{\itop{Time}}   \rightarrow   \ensuremath{\itop{Time}} 
\end{alignat*}

Remarkably, we need not state laws for $ \ensuremath{\itop{tick}} $. The
interpreter will merge values which reside at the same address to
preserve soundness. Therefore, any supplied implementations of
$ \ensuremath{\itop{tick}} $ is valid from a soundness perspective.
However, different choices in $ \ensuremath{\itop{tick}} $ will yield
different trade-offs in precision and performance of the abstract
interpreter.

\par

\section{The Interpreter}\label{the-interpreter}

\par

\begin{figure}

\begin{align*}
&\hspace{0em} A ^m  \llbracket  \underline{\hspace{0.5em}}  \rrbracket  :  \ensuremath{\itop{Atom}}   \rightarrow  m (\ensuremath{\itop{Val}})  \\
&\hspace{0em} A ^m  \llbracket i \rrbracket   \coloneqq   \ensuremath{\itop{return}}  (\ensuremath{\itop{int-I}}( i)) \\
&\hspace{0em} A ^m  \llbracket x \rrbracket   \coloneqq   \ensuremath{\ttbfop{do}}  \\
&\hspace{1em}  \rho   \leftarrow   \ensuremath{\itop{get-Env}}   \;;\;   \sigma   \leftarrow   \ensuremath{\itop{get-Store}}  \\
&\hspace{1em}  \ensuremath{\ttbfop{if}}  x  \in   \rho   \;\ttbfop{then}\;   \ensuremath{\itop{return}}(  \sigma ( \rho (x)))  \;\ttbfop{\litelse}\;   \ensuremath{\itop{return}}(  \bot ) \\
&\hspace{0em} A ^m  \llbracket  \keyword{\lambda} (x).e \rrbracket   \coloneqq   \rho   \leftarrow   \ensuremath{\itop{get-Env}}   \;;\;   \ensuremath{\itop{return}}  (\ensuremath{\itop{clo-I}}(  \langle  \keyword{\lambda} (x).e, \rho  \rangle )) \\
&\hspace{0em}  \  \\
&\hspace{0em}  \ensuremath{\itop{step}} ^m   :  \ensuremath{\itop{Exp}}   \rightarrow  m (\ensuremath{\itop{Exp}})  \\
&\hspace{0em}  \ensuremath{\itop{step}} ^m  (e)  \coloneqq   \ensuremath{\ttbfop{do}}  \\
&\hspace{1em}  \ensuremath{\itop{tick}} ^m  (e)  \;;\;   \rho   \leftarrow   \ensuremath{\itop{get-Env}}  \\
&\hspace{1em} e'  \leftarrow   \ensuremath{\ttbfop{case}}  e  \ensuremath{\ttbfop{of}}  \\
&\hspace{2em} e _1   \odot  e _2   \rightarrow   \ensuremath{\itop{push}}(  \langle  \square   \odot  e _2 , \rho  \rangle )  \;;\;   \ensuremath{\itop{return}}( e _1 ) \\
&\hspace{2em}  \keyword{if0} (e _1 ) \{ e _2  \}  \{ e _3  \}   \rightarrow   \ensuremath{\itop{push}}(  \langle  \keyword{if0} ( \square ) \{ e _2  \}  \{ e _3  \} , \rho  \rangle )  \;;\;   \ensuremath{\itop{return}}( e _1 ) \\
&\hspace{2em} a  \rightarrow   \ensuremath{\ttbfop{do}}  \\
&\hspace{3em} v  \leftarrow  A ^m  \llbracket a \rrbracket   \;;\;  fr  \leftarrow   \ensuremath{\itop{pop}}   \\
&\hspace{3em}  \ensuremath{\ttbfop{case}}  fr  \ensuremath{\ttbfop{of}}  \\
&\hspace{4em}  \langle  \square   \odot  e, \rho ' \rangle   \rightarrow   \ensuremath{\itop{put-Env}}(  \rho ')  \;;\;   \ensuremath{\itop{push}}(  \langle v  \odot   \square  \rangle )  \;;\;   \ensuremath{\itop{return}}( e) \\
&\hspace{4em}  \langle v'  \keywordop{ \mathbin{\bm{@}} }   \square  \rangle   \rightarrow   \ensuremath{\ttbfop{do}}  \\
&\hspace{5em}  \tau   \leftarrow   \ensuremath{\itop{get-Time}}   \;;\;   \sigma   \leftarrow   \ensuremath{\itop{get-Store}}   \\
&\hspace{5em}  \langle  \keyword{\lambda} (x).e, \rho ' \rangle   \leftarrow   \ensuremath{\itop{\uparrow_p}}  (\ensuremath{\itop{clo-E}}( v')) \\
&\hspace{5em}  \ensuremath{\itop{put-Env}}(  \rho ' \lbrack x  \mapsto  (x, \tau ) \rbrack )  \\
&\hspace{5em}  \ensuremath{\itop{put-Store}}(  \sigma   \sqcup   \lbrack (x, \tau )  \mapsto  v \rbrack )  \;;\;   \ensuremath{\itop{return}}( e) \\
&\hspace{4em}  \langle v'  \oplus   \square  \rangle   \rightarrow   \ensuremath{\itop{return}}(  \delta  \llbracket  \oplus  \rrbracket (v',v)) \\
&\hspace{4em}  \langle  \keyword{if0} ( \square ) \{ e _1  \}  \{ e _2  \} , \rho ' \rangle   \rightarrow   \ensuremath{\ttbfop{do}}  \\
&\hspace{5em}  \ensuremath{\itop{put-Env}}(  \rho ')  \;;\;  b  \leftarrow   \ensuremath{\itop{\uparrow_p}}  (\ensuremath{\itop{if0-E}}( v))  \;;\;  refine(a,b) \\
&\hspace{5em}  \ensuremath{\ttbfop{if}}( b)  \;\ttbfop{then}\;   \ensuremath{\itop{return}}( e _1 )  \;\ttbfop{\litelse}\;   \ensuremath{\itop{return}}( e _2 ) \\
&\hspace{1em}  \ensuremath{\itop{gc}}( e')  \;;\;   \ensuremath{\itop{return}}( e')
\end{align*}

\caption{Monadic Semantics} \label{InterpreterStep} \end{figure}

\par

\begin{figure}

\begin{align*}
&\hspace{0em}  \ensuremath{\itop{push}}  :  \ensuremath{\itop{Frame}}   \rightarrow  m(1) \\
&\hspace{0em}  \ensuremath{\itop{push}}( fr)  \coloneqq   \ensuremath{\ttbfop{do}}  \\
&\hspace{1em}  \kappa l  \leftarrow   \ensuremath{\itop{get-KAddr}}   \;;\;   \kappa  \sigma   \leftarrow   \ensuremath{\itop{get-KStore}}   \;;\;   \kappa l'  \leftarrow   \ensuremath{\itop{get-Time}}  \\
&\hspace{1em}  \ensuremath{\itop{put-KStore}}(  \kappa  \sigma   \sqcup   \lbrack  \kappa l'  \mapsto   \{ fr \mathbin{::}  \kappa l \}  \rbrack )  \;;\;   \ensuremath{\itop{put-KAddr}}(  \kappa l') \\
&\hspace{0em}  \ensuremath{\itop{pop}}  : m (\ensuremath{\itop{Frame}})  \\
&\hspace{0em}  \ensuremath{\itop{pop}}   \coloneqq   \ensuremath{\ttbfop{do}}  \\
&\hspace{1em}  \kappa l  \leftarrow   \ensuremath{\itop{get-KAddr}}   \;;\;   \kappa  \sigma   \leftarrow   \ensuremath{\itop{get-KStore}}   \;;\;  fr \mathbin{::}  \kappa l'  \leftarrow   \ensuremath{\itop{\uparrow_p}} ( \kappa  \sigma ( \kappa l)) \\
&\hspace{1em}  \ensuremath{\itop{put-KAddr}}(  \kappa l')  \;;\;   \ensuremath{\itop{return}}( fr) \\
&\hspace{0em}  \ensuremath{\itop{\uparrow_p}}  :  \forall  A,  \mathcal{P} (A)  \rightarrow  m(A) \\
&\hspace{0em}  \ensuremath{\itop{\uparrow_p}} ( \{ a _1  .. a _n  \} )  \coloneqq   \ensuremath{\itop{return}}( a _1 )  \mathbin{\langle + \rangle}  ..  \mathbin{\langle + \rangle}   \ensuremath{\itop{return}}( a _n ) \\
&\hspace{0em} refine :  \ensuremath{\itop{Atom}}   \times  Bool  \rightarrow  m(1) \\
&\hspace{0em} refine(i,b)  \coloneqq   \ensuremath{\itop{return}}( 1) \\
&\hspace{0em} refine(x,b)  \coloneqq   \ensuremath{\ttbfop{do}}  \\
&\hspace{2em}  \rho   \leftarrow   \ensuremath{\itop{get-Env}}   \;;\;   \sigma   \leftarrow   \ensuremath{\itop{get-Store}}  \\
&\hspace{2em}  \ensuremath{\itop{put-Store}}(  \sigma  \lbrack  \rho (x)  \mapsto  b \rbrack ) \\
&\hspace{0em}  \ensuremath{\itop{tick}} ^m   :  \ensuremath{\itop{Exp}}   \rightarrow  m(1) \\
&\hspace{0em}  \ensuremath{\itop{tick}} ^m  (e)  \coloneqq   \ensuremath{\ttbfop{do}}  \\
&\hspace{1em}  \tau   \leftarrow   \ensuremath{\itop{get-Time}}   \;;\;   \kappa l  \leftarrow   \ensuremath{\itop{get-KAddr}}  \\
&\hspace{1em}  \ensuremath{\itop{put-Time}}(  \ensuremath{\itop{tick}}( e, \kappa l, \tau )) \\
&\hspace{0em}  \ensuremath{\itop{gc}}  :  \ensuremath{\itop{Exp}}   \rightarrow  m(1) \\
&\hspace{0em}  \ensuremath{\itop{gc}}( e)  \coloneqq   \ensuremath{\ttbfop{do}}  \\
&\hspace{1em}  \rho   \leftarrow   \ensuremath{\itop{get-Env}}   \;;\;   \sigma   \leftarrow   \ensuremath{\itop{get-Store}}   \\
&\hspace{1em}  \kappa l  \leftarrow   \ensuremath{\itop{get-KAddr}}   \;;\;   \kappa  \sigma   \leftarrow   \ensuremath{\itop{get-KStore}}  \\
&\hspace{1em}  \ensuremath{\itop{put-KStore}}(  \{  \kappa l  \mapsto   \kappa  \sigma ( \kappa l)  \;|\;   \kappa l  \in   \ensuremath{\itop{KR}}(  \kappa l, \kappa  \sigma ) \} ) \\
&\hspace{1em}  \ensuremath{\itop{put-Store}}(  \{ l  \mapsto   \sigma (l)  \;|\;  l  \in   \ensuremath{\itop{R}}( e, \rho , \sigma , \kappa l, \kappa  \sigma ))
\end{align*}

\caption{Monadic helper functions} \label{InterpreterHelpers}
\end{figure}

\par

We now present a monadic interpreter for
$ \ensuremath{\ttbfop{\lambda{}IF}} $ parameterized over $m$,
$ \ensuremath{\itop{Val}} $ and $ \ensuremath{\itop{Time}} $ from
Section~\ref{analysis-parameters}. We instantiate these parameters to
obtain an analysis in Section~\ref{recovering-analyses}.

\par

We translate $A \llbracket  \underline{\hspace{0.5em}}  \rrbracket $, a
partial denotation function, to
$A ^m  \llbracket  \underline{\hspace{0.5em}}  \rrbracket $, a total
monadic denotation function, shown in Figure~\ref{InterpreterStep}.

\par

Next we implement $ \ensuremath{\itop{step}} ^m  $, a \emph{monadic}
small-step \emph{function} for compound expressions, also shown in
Figure~\ref{InterpreterStep}. $ \ensuremath{\itop{step}} ^m  $ is a
translation of
$ \underline{\hspace{0.5em}}{  \rightsquigarrow  }\underline{\hspace{0.5em}} $
from a relation to a monadic function with state and nondeterminism
effects.

\par

$ \ensuremath{\itop{step}} ^m  $ uses $ \ensuremath{\itop{push}} $ and
$ \ensuremath{\itop{pop}} $ for manipulating stack frames,
$ \ensuremath{\itop{\uparrow_p}} $ for lifting values from
$ \mathcal{P} $ into $m$, $refine$ for value refinement after branching,
and a monadic version of $ \ensuremath{\itop{tick}} $ called
$ \ensuremath{\itop{tick}} ^m  $, each shown in
Figure~\ref{InterpreterHelpers}. Frames are pushed when the control
expression $e$ is compound and popped when $e$ is atomic. The
interpreter looks deterministic, however the nondeterminism is hidden
behind $ \ensuremath{\itop{\uparrow_p}} $ and monadic bind operations
$x  \leftarrow  e _1   \;;\;  e _2 $. The use of $refine$ enforces a
limited form of path-condition, and will yield each variation of path
and flow sensitivity given the appropriate monad.

\par

We implement abstract garbage collection $ \ensuremath{\itop{gc}} $ in a
general way using the monadic effect interface, also shown in
Figure~\ref{InterpreterHelpers}. $ \ensuremath{\itop{R}} $ and
$ \ensuremath{\itop{KR}} $ are as defined in Section~\ref{semantics}.
Remarkably, this single implementation supports instantiation to
analyses with varying path and flow sensitivities.

\par

\paragraph{Preserving Soundness}

In the monadic interpreter, updates to both the data-store and
stack-store must merge rather than overwrite values. To support
$ \sqcup $ for the stack store we redefine the domain to map to a
powerset of frames:

\begin{align*}
&\hspace{0em}  \kappa  \sigma   \in   \ensuremath{\itop{KStore}}  :  \ensuremath{\itop{KAddr}}   \rightarrow   \mathcal{P}  (\ensuremath{\itop{Frame}}   \times   \ensuremath{\itop{KAddr}}) 
\end{align*}

\par

\paragraph{Execution}

In the concrete semantics, execution takes the form of a
least-fixed-point computation over the collecting semantics $collect$.
This in general requires a join-semilattice structure for some
$ \Sigma $ and a transition system $ \Sigma   \rightarrow   \Sigma $.
However, we no longer have a transition system
$ \Sigma   \rightarrow   \Sigma $; we have a monadic function
$ \ensuremath{\itop{Exp}}   \rightarrow  m (\ensuremath{\itop{Exp}}) $
which cannot be iterated to least-fixed-point to execute the analysis.

\par

To solve this we require the existence of a Galois connection between
monadic actions and some transition system:
$ \Sigma   \rightarrow   \Sigma   \galois{\alpha^{\Sigma\leftrightarrow m}}{\gamma^{\Sigma\leftrightarrow m}}   \ensuremath{\itop{Exp}}   \rightarrow  m (\ensuremath{\itop{Exp}}) $.
This Galois connection allows us to implement the analysis by
transporting our interpreter to the transition system
$ \Sigma   \rightarrow   \Sigma $ through
$ \gamma  ^{  \Sigma  \leftrightarrow m } $, and then iterating to
fixed-point in $ \Sigma $. Furthermore, it serves to \emph{transport
other Galois connections} as part of our correctness framework. This
will allow us to construct Galois connections between monads
$m _1   \galois{\alpha^{m}}{\gamma^{m}}  m _2 $ and derive Galois
connections between transition systems
$ \Sigma  _1   \galois{\alpha^{\Sigma}}{\gamma^{\Sigma}}   \Sigma  _2 $.

\par

An execution of our interpreter is then the least-fixed-point iteration
of $ \ensuremath{\itop{step}} ^m  $ transported through
$ \gamma  ^{  \Sigma  \leftrightarrow m } $:

\begin{align*}
&\hspace{0em} analysis  \coloneqq   \mu (X). X  \sqcup   \varsigma  _0   \sqcup   \gamma  ^{  \Sigma  \leftrightarrow m }  (\ensuremath{\itop{step}} ^m  )(X)
\end{align*}

where $ \varsigma  _0 $ is the injection of the initial program $e _0 $
into $ \Sigma $ and $ \gamma  ^{  \Sigma  \leftrightarrow m } $ has type
$ (\ensuremath{\itop{Exp}}   \rightarrow  m (\ensuremath{\itop{Exp}}) )  \rightarrow  ( \Sigma   \rightarrow   \Sigma )$.

\par

\section{Recovering Analyses}\label{recovering-analyses}

\par

\begin{figure}

\begin{alignat*}{2}
v  \in   \Concrete{\litVal}  &  \coloneqq   \mathcal{P}  (\Concrete{\litClo}   \cup   \mathbb{Z} ) \\
 \tau   \in   \Concrete{\litTime}  &  \coloneqq   (\ensuremath{\itop{Exp}}   \times   \ensuremath{\itop{KAddr}})  ^{ * } 
\end{alignat*}

\hrule

\begin{align*}
&\hspace{0em}  \Concrete {\ensuremath{\itop{int-I}}}   :  \mathbb{Z}   \rightarrow   \Concrete{\litVal}   \\
&\hspace{0em}  \Concrete {\ensuremath{\itop{int-I}}} ( i)  \coloneqq   \{ i \}   \\
&\hspace{0em}  \Concrete {\ensuremath{\itop{if0-E}}}   :  \Concrete{\litVal}   \rightarrow   \mathcal{P} (Bool)  \\
&\hspace{0em}  \Concrete {\ensuremath{\itop{if0-E}}} ( v)  \coloneqq   \{   \ensuremath{\ttbfop{true}}   \;|\;  0  \in  v  \}    \cup   \{   \ensuremath{\ttbfop{false}}   \;|\;   \exists  i  \in  v  \;;\;  i  \neq  0  \}   \\
&\hspace{0em}  \Concrete {\ensuremath{\itop{clo-I}}}   :  \ensuremath{\itop{Clo}}   \rightarrow   \Concrete{\litVal}   \\
&\hspace{0em}  \Concrete {\ensuremath{\itop{clo-I}}} ( c)  \coloneqq   \{ c \}   \\
&\hspace{0em}  \Concrete {\ensuremath{\itop{clo-E}}}   :  \Concrete{\litVal}   \rightarrow   \mathcal{P}  (\ensuremath{\itop{Clo}})   \\
&\hspace{0em}  \Concrete {\ensuremath{\itop{clo-E}}} ( v)  \coloneqq   \{  c  \;|\;  c  \in  v  \}  \\
&\hspace{0em}  \Concrete{\delta}  :  \Concrete{\litVal}   \times   \Concrete{\litVal}   \rightarrow   \Concrete{\litVal}  \\
&\hspace{0em}  \Concrete{\delta}  \llbracket  \keywordop{+}  \rrbracket (v _1 ,v _2 )  \coloneqq   \{  i _1  + i _2   \;|\;  i _1   \in  v _1   \;;\;  i _2   \in  v _2   \}  \\
&\hspace{0em}  \Concrete{\delta}  \llbracket  \keywordop{-}  \rrbracket (v _1 ,v _2 )  \coloneqq   \{  i _1  - i _2   \;|\;  i _1   \in  v _1   \;;\;  i _2   \in  v _2   \}  \\
&\hspace{0em}  \Concrete {\ensuremath{\itop{tick}}}   :  \ensuremath{\itop{Exp}}   \times   \Concrete{\litTime}   \rightarrow   \Concrete{\litTime}  \\
&\hspace{0em}  \Concrete {\ensuremath{\itop{tick}}}   (e, \kappa l, \tau )  \coloneqq  (e, \kappa l) \mathbin{::}  \tau 
\end{align*}

\caption{Concrete Interpreter Values and Time}
\label{concrete-parameters} \end{figure}

\par

In Section~\ref{the-interpreter}, we defined a monadic interpreter with
the uninstantiated parameters from Section~\ref{analysis-parameters}:
$m$, $ \ensuremath{\itop{Val}} $ and $ \ensuremath{\itop{Time}} $. To
recover a concrete interpreter, we instantiate these parameters to
concrete components $ \Concrete{\litM} $, $ \Concrete{\litVal} $ and
$ \Concrete{\litTime} $, and to recover an abstract interpreter we
instantiate them to abstract components $ \Abstract{\litM} $,
$ \Abstract{\litVal} $ and $ \Abstract{\litTime} $. Furthermore, the
concrete transition system $ \Concrete{\Sigma} $ induced by
$ \Concrete{\litM} $ will recover the collecting semantics, which is our
final target of abstraction, and the resulting analysis will take the
form of an abstract transition system $ \Abstract{\Sigma} $ induced by
$ \Abstract{\litM} $.

\par

\subsection{Recovering a Concrete
Interpreter}\label{recovering-a-concrete-interpreter}

\par

\begin{figure}

\begin{alignat*}{2}
 \psi   \in   \Concrete{\Psi}  &  \coloneqq   \Concrete{\litEnv}   \times   \Concrete{\litKAddr}   \times   \Concrete{\litKStore}   \times   \Concrete{\litTime}  \\
 \Concrete{\litM}( A) &  \coloneqq   \Concrete{\Psi}   \times   \Concrete{\litStore}   \rightarrow   \mathcal{P} (A  \times   \Concrete{\Psi}   \times   \Concrete{\litStore})  \\
 \varsigma   \in   \Concrete{\Sigma}  &  \coloneqq   \mathcal{P}  (\ensuremath{\itop{Exp}}   \times   \Concrete{\Psi}   \times   \Concrete{\litStore}) 
\end{alignat*}

\hrule

\begin{align*}
&\hspace{0em}  \Concrete {\ensuremath{\itop{return}}}   :  \forall  A, A  \rightarrow   \Concrete{\litM}( A) \\
&\hspace{0em}  \Concrete {\ensuremath{\itop{return}}} ( x)( \psi ,s)  \coloneqq   \{ (x, \psi ,s) \}  \\
&\hspace{0em}  \Concrete {\ensuremath{\itop{bind}}}   :  \forall  A B,  \Concrete{\litM}( A)  \rightarrow  (A  \rightarrow   \Concrete{\litM}( B))  \rightarrow   \Concrete{\litM}( B) \\
&\hspace{0em}  \Concrete {\ensuremath{\itop{bind}}} ( X)(f)( \psi , \sigma )  \coloneqq   \bigcup  _{ (x, \psi ', \sigma ')  \in  X( \psi , \sigma ) }  f(x)( \psi ', \sigma ') \\
&\hspace{0em}  \Concrete {\ensuremath{\itop{get-Env}}}   :  \Concrete{\litM}(  \Concrete{\litEnv})  \\
&\hspace{0em}  \Concrete {\ensuremath{\itop{get-Env}}} (  \langle  \rho , \kappa l, \kappa  \sigma , \tau  \rangle , \sigma )  \coloneqq   \{ ( \rho , \langle  \rho , \kappa l, \kappa  \sigma , \tau  \rangle , \sigma ) \}  \\
&\hspace{0em}  \Concrete {\ensuremath{\itop{put-Env}}}   :  \Concrete{\litEnv}   \rightarrow   \mathcal{P} (1) \\
&\hspace{0em}  \Concrete {\ensuremath{\itop{put-Env}}} (  \rho ')( \langle  \rho , \kappa l, \kappa  \sigma , \tau  \rangle , \sigma )  \coloneqq   \{ (1, \langle  \rho ', \sigma , \kappa , \tau  \rangle , \sigma ) \}  \\
&\hspace{0em}  \Concrete {\ensuremath{\itop{mzero}}}   :  \forall  A,  \Concrete{\litM}( A) \\
&\hspace{0em}  \Concrete {\ensuremath{\itop{mzero}}} (  \psi , \sigma )  \coloneqq   \{  \}  \\
&\hspace{0em}  \underline{\hspace{0.5em}}{  \Concrete{\mathbin{\langle + \rangle}}  }\underline{\hspace{0.5em}}  :  \forall  A,  \Concrete{\litM}( A)  \times   \Concrete{\litM}( A)  \rightarrow   \Concrete{\litM}( A) \\
&\hspace{0em} (X _1   \Concrete{\mathbin{\langle + \rangle}}  X _2 )( \psi , \sigma )  \coloneqq  X _1 ( \psi , \sigma )  \cup  X _2 ( \psi , \sigma ) \\
&\hspace{0em}  \alpha  ^{  \Concrete{\Sigma}   \leftrightarrow \Concrete{\litM} }   : ( \Concrete{\Sigma}   \rightarrow   \Concrete{\Sigma} )  \rightarrow   (\ensuremath{\itop{Exp}}   \rightarrow   \Concrete{\litM}(  \ensuremath{\itop{Exp}}) ) \\
&\hspace{0em}  \alpha  ^{  \Concrete{\Sigma}   \leftrightarrow \Concrete{\litM} }  (f)(e)( \psi , \sigma )  \coloneqq  f( \{ (e, \psi , \sigma ) \} ) \\
&\hspace{0em}  \gamma  ^{  \Concrete{\Sigma}   \leftrightarrow \Concrete{\litM} }   :  (\ensuremath{\itop{Exp}}   \rightarrow   \Concrete{\litM}(  \ensuremath{\itop{Exp}}) )  \rightarrow  ( \Concrete{\Sigma}   \rightarrow   \Concrete{\Sigma} ) \\
&\hspace{0em}  \gamma  ^{  \Concrete{\Sigma}   \leftrightarrow \Concrete{\litM} }  (f)(e \psi  \sigma  ^{ * } )  \coloneqq   \bigcup  _{ (e, \psi , \sigma )  \in  e \psi  \sigma  ^{ * }  } f(e)( \psi , \sigma )
\end{align*}

\caption{Concrete Interpreter Monad} \label{concrete-monad-parameters}
\end{figure}

\par

To recover a concrete interpreter, we instantiate the generic monadic
interpreter from Section~\ref{the-interpreter} with concrete parameters
$ \Concrete{\litVal} $, $ \Concrete{\delta} $, $ \Concrete{\litTime} $
and $ \Concrete{\litM} $, shown in Figures~\ref{concrete-parameters} and
~\ref{concrete-monad-parameters}.

\par

\paragraph{The Concrete Domain}

\par

We instantiate $ \ensuremath{\itop{Val}} $ to $ \Concrete{\litVal} $, a
powerset of concrete values. $ \Concrete{\litVal} $ has precise
introduction and elimination functions
$ \Concrete {\ensuremath{\itop{int-I}}}  $,
$ \Concrete {\ensuremath{\itop{if0-E}}}  $,
$ \Concrete {\ensuremath{\itop{clo-I}}}  $ and
$ \Concrete {\ensuremath{\itop{clo-E}}}  $, and primitive operator
denotation $ \Concrete{\delta} $.

\par

\paragraph{Concrete Time}

\par

We instantiate $ \ensuremath{\itop{Time}} $ to $ \Concrete{\litTime} $,
which captures the execution context as a sequence of previously visited
expressions. $ \Concrete {\ensuremath{\itop{tick}}}  $ is then a cons
operation.

\par

\paragraph{The Concrete Monad}

\par

We instantiate $m$ to $ \Concrete{\litM} $, a powerset of concrete state
space components. Monadic operators
$ \Concrete {\ensuremath{\itop{bind}}}  $ and
$ \Concrete {\ensuremath{\itop{return}}}  $ encapsulate both
state-passing and set-flattening. State effects return singleton sets
and nondeterminism effects are implemented with set union.

\par

\paragraph{Concrete Execution}

\par

To execute the interpreter we establish the Galois connection
$ \Concrete{\Sigma}   \rightarrow   \Concrete{\Sigma}   \galois{\Concrete{\alpha}^{\Sigma\leftrightarrow  \ensuremath{\itop{M}}} }{\Concrete{\gamma}^{\Sigma\leftrightarrow  \ensuremath{\itop{M}}} }   \ensuremath{\itop{Exp}}   \rightarrow   \Concrete{\litM}(  \ensuremath{\itop{Exp}}) $
and transport the monadic interpreter through
$ \Concrete{\gamma}^{\Sigma\leftrightarrow m} $. The injection for a
program $e _0 $ into $ \Concrete{\Sigma} $ is
$ \varsigma  _0   \coloneqq   \{  \langle e _0 , \bot , \bot , \bullet , \bot , \bullet  \rangle  \} $.

\par

\subsection{Recovering an Abstract
Interpreter}\label{recovering-an-abstract-interpreter}

\par

To recover an abstract interpreter we instantiate the generic monadic
interpreter from Section~\ref{the-interpreter} with abstract parameters
$ \Abstract{\litVal} $, $ \Abstract{\delta} $, $ \Abstract{\litTime} $
and $ \Abstract{\litM} $, shown in Figure~\ref{abstract-parameters}. The
abstract monad operators, effects and transition system are not shown
for $ \Abstract{\litM} $; they are identical to $ \Concrete{\litM} $ but
with abstract components.

\par

\paragraph{The Abstract Domain}

\par

We pick a simple abstraction for integers,
$ \{  \keywordop{-} ,0, \keywordop{+}  \} $, although our technique
scales to other abstract domains. $ \Abstract{\litVal} $ is defined as a
powerset of abstract values. $ \Abstract{\litVal} $ has introduction and
elimination functions $ \Abstract {\ensuremath{\itop{int-I}}}  $,
$ \Abstract {\ensuremath{\itop{if0-E}}}  $,
$ \Abstract {\ensuremath{\itop{clo-I}}}  $ and
$ \Abstract {\ensuremath{\itop{clo-E}}}  $, and primitive operator
denotation $ \Abstract{\delta} $.
$ \Abstract {\ensuremath{\itop{if0-E}}}  $ and $ \Abstract{\delta} $
must be conservative, returning an upper bound of the precise results
returned by their concrete counterparts.

\par

\paragraph{Abstract Time}

\par

Abstract time $ \Abstract{\litTime} $ captures an approximation of the
execution context as a finite sequence of previously visited
expressions. $ \Abstract {\ensuremath{\itop{tick}}}  $ is a cons
operation followed by k-truncation, yielding a kCFA
analysis~\cite{dvanhorn:VanHorn2010Abstracting}.

\par

\paragraph{The Abstract Monad and Execution}

The abstract monad $ \Abstract{\litM} $ is identical to
$ \Concrete{\litM} $ up to the definition of $ \Abstract{\Psi} $. The
induced state space $ \Abstract{\Sigma} $ is finite, and its
least-fixed-point iteration will give a sound and computable analysis.

\par

\begin{figure}

\begin{alignat*}{2}
v  \in   \Abstract{\litVal}  &  \coloneqq   \mathcal{P}  (\Abstract{\litClo}   \cup   \{  \keywordop{-} ,0, \keywordop{+}  \} ) \\
 \tau   \in   \Abstract{\litTime}  &  \coloneqq   (\ensuremath{\itop{Exp}}   \times   \ensuremath{\itop{KAddr}})  ^{ * _k  }  \\
 \psi   \in   \Abstract{\Psi}  &  \coloneqq   \Abstract{\litEnv}   \times   \Abstract{\litKAddr}   \times   \Abstract{\litKStore}   \times   \Abstract{\litTime}  \\
 \Abstract{\litM}( A) &  \coloneqq   \Abstract{\Psi}   \times   \Abstract{\litStore}   \rightarrow   \mathcal{P} (A  \times   \Abstract{\Psi}   \times   \Abstract{\litStore})  \\
 \varsigma   \in   \Abstract{\Sigma}  &  \coloneqq   \mathcal{P}  (\ensuremath{\itop{Exp}}   \times   \Abstract{\Psi}   \times   \Abstract{\litStore}) 
\end{alignat*}

\hrule

\begin{align*}
&\hspace{0em}  \begin{aligned}[l]  &  \Abstract {\ensuremath{\itop{int-I}}}   :  \mathbb{Z}   \rightarrow   \Abstract{\litVal}         &&  \Abstract {\ensuremath{\itop{int-I}}} ( i)  \coloneqq   \begin{aligned}[l]  &  \{  \keywordop{-}  \}   \ensuremath{\ttbfop{if}}  i < 0  \\  &  \{ 0 \}   \ensuremath{\ttbfop{if}}  i = 0  \\  &  \{  \keywordop{+}  \}   \ensuremath{\ttbfop{if}}  i > 0  \end{aligned}   \\  &  \Abstract {\ensuremath{\itop{if0-E}}}   :  \Abstract{\litVal}   \rightarrow   \mathcal{P} (Bool)  &&  \Abstract {\ensuremath{\itop{if0-E}}} ( v)  \coloneqq   \begin{aligned}[l]  &  \{   \ensuremath{\ttbfop{true}}   \;|\;  0  \in  v  \} \ \  \cup   \\  &  \{   \ensuremath{\ttbfop{false}}   \;|\;   \keywordop{-}   \in  v  \vee   \keywordop{+}   \in  v  \}   \end{aligned}   \\  &  \Abstract {\ensuremath{\itop{clo-I}}}   :  \ensuremath{\itop{Clo}}   \rightarrow   \Abstract{\litVal}       &&  \Abstract {\ensuremath{\itop{clo-I}}} ( c)  \coloneqq   \{ c \}   \\  &  \Abstract {\ensuremath{\itop{clo-E}}}   :  \Abstract{\litVal}   \rightarrow   \mathcal{P}  (\ensuremath{\itop{Clo}})    &&  \Abstract {\ensuremath{\itop{clo-E}}} ( v)  \coloneqq   \{  c  \;|\;  c  \in  v  \}   \end{aligned}  \\
&\hspace{0em}  \Abstract{\delta}  :  \Abstract{\litVal}   \times   \Abstract{\litVal}   \rightarrow   \Abstract{\litVal}  \\
&\hspace{0em}  \Abstract{\delta}  \llbracket  \keywordop{+}  \rrbracket (v _1 ,v _2 )  \coloneqq   \\
&\hspace{3em} \  \{  i  \;|\;  0  \in  v _1   \wedge  i  \in  v _2   \}   \cup   \{  i  \;|\;  i  \in  v _1   \wedge  0  \in  v _2   \}  \\
&\hspace{2em}  \cup   \{   \keywordop{+}   \;|\;   \keywordop{+}   \in  v _1   \wedge   \keywordop{+}   \in  v _2   \}   \cup   \{   \keywordop{-}   \;|\;   \keywordop{-}   \in  v _1   \wedge   \keywordop{-}   \in  v _2   \}   \\
&\hspace{2em}  \cup   \{   \keywordop{-} ,0, \keywordop{+}   \;|\;   \keywordop{+}   \in  v _1   \wedge   \keywordop{-}   \in  v _2   \}   \\
&\hspace{2em}  \cup   \{   \keywordop{-} ,0, \keywordop{+}   \;|\;   \keywordop{-}   \in  v _1   \wedge   \keywordop{+}   \in  v _2   \}  \\
&\hspace{0em}  \Abstract{\delta}  \llbracket  \keywordop{-}  \rrbracket (v _1 ,v _2 )  \coloneqq  ...  \;\;\text{\footnotesize{}\emph{analogous}}\;\;  ... \\
&\hspace{0em}  \Abstract {\ensuremath{\itop{tick}}}   :  \ensuremath{\itop{Exp}}   \times   \Abstract{\litTime}   \rightarrow   \Abstract{\litTime}  \\
&\hspace{0em}  \Abstract {\ensuremath{\itop{tick}}} ( e, \kappa l, \tau )  \coloneqq   \lfloor (e, \kappa l) \mathbin{::}  \tau  \rfloor  _k 
\end{align*}

\caption{Abstract Interpreter Parameters} \label{abstract-parameters}
\end{figure}

\par

\subsection{End-to-End Correctness}\label{end-to-end-correctness}

\par

The end-to-end correctness of the abstract instantiation of the
interpreter is factored into three steps: (1) proving the parameterized
monadic interpreter correct for any instantiation of $m$,
$ \ensuremath{\itop{Val}} $ and $ \ensuremath{\itop{Time}} $; (2)
constructing Galois connections
$ \Concrete{\litM}   \galois{\alpha^{m}}{\gamma^{m}}   \Abstract{\litM} $,
$ \Concrete{\litVal}   \galois{\alpha^{v}}{\gamma^{v}}   \Abstract{\litVal} $
and
$ \Concrete{\litTime}   \galois{\alpha^{t}}{\gamma^{t}}   \Abstract{\litTime} $
piecewise; and (3) transporting the combination of (1) and (2) from the
monadic function space $A  \rightarrow  m(B)$ to its induced transition
system $ \Sigma   \rightarrow   \Sigma $. The benefit of our approach is
that the first step is proved once and for all (for a particular
semantics) against \emph{any} instantiation of $m$,
$ \ensuremath{\itop{Val}} $ and $ \ensuremath{\itop{Time}} $ using the
reasoning principles established in Section~\ref{analysis-parameters}.
Furthermore the second step can be proved in isolation of the first, and
the construction of the third step is fully systematic.

\par

We do not give proofs for (1) or the abstractions for
$ \ensuremath{\itop{Val}} $ and $ \ensuremath{\itop{Time}} $ for (2) in
this paper, although the details can be found in prior work
\cite{dvanhorn:Cousot98-5,dvanhorn:VanHorn2010Abstracting}. Rather, we
give definitions and proofs for the monad abstractions for (2) and their
systematic mappings to transition systems for (3) through a
compositional framework in
Section~\ref{a-compositional-monadic-framework}.

\par

The final correctness of the abstract interpreter is established as a
partial order relationship between an abstraction of
$ \gamma  ^{  \Concrete{\Sigma}   \leftrightarrow \Concrete{\litM} }   (\ensuremath{\itop{step}} ^m    \lbrack \Concrete{\litM} \rbrack  )$,
which recovers the collecting semantics, and
$ \gamma  ^{  \Abstract{\Sigma}   \leftrightarrow \Abstract{\litM} }   (\ensuremath{\itop{step}} ^m    \lbrack \Abstract{\litM} \rbrack  )$,
the induced abstract semantics: \begin{proposition}
\label{interpreter-end-to-end}

\begin{alignat*}{1}
 \alpha  ^{  \Concrete{\Sigma}  } ( \gamma  ^{  \Concrete{\Sigma}   \leftrightarrow \Concrete{\litM} }   (\ensuremath{\itop{step}} ^m    \lbrack \Concrete{\litM} \rbrack  ))  \sqsubseteq   \gamma  ^{  \Abstract{\Sigma}   \leftrightarrow \Abstract{\litM} }   (\ensuremath{\itop{step}} ^m    \lbrack \Abstract{\litM} \rbrack  )
\end{alignat*}

\end{proposition} \noindent
The left-hand-side of the relationship is the induced ``best
specification'' of the collecting semantics via Galois connection, and
should be familiar from the literature on abstract interpretation
\cite{dvanhorn:Cousot1979Systematic,dvanhorn:Cousot98-5,dvanhorn:Neilson:1999}.
This end-to-end correctness statement will be justified in a
compositional setting in
Section~\ref{a-compositional-monadic-framework}.

\par

\section{Varying Path and Flow
Sensitivity}\label{varying-path-and-flow-sensitivity}

\par

\par

Sections~\ref{the-interpreter} and~\ref{recovering-analyses} describe
the construction of a path-sensitive analysis using our framework. In
this section, we show an alternate definition for $ \Abstract{\litM} $
which yields a flow-insensitive analysis.
Section~\ref{a-compositional-monadic-framework} will generalize the
definitions from this section into compositional components (monad
transformers) in addition to introducing another definition for
$ \Abstract{\litM} $ which yields a flow-sensitive analysis.

\par

Before going into the details of the flow-insensitive monad, we wish to
build intuition regarding what one would expect from such a development.
Recall the path-sensitive monad $ \Abstract{\litM} $ and its state space
$ \Abstract{\Sigma} $ from section \ref{recovering-analyses}:

\begin{align*}
&\hspace{0em}  \Abstract{\litM}(  \ensuremath{\itop{Exp}})   \coloneqq   \Abstract{\Psi}   \times   \Abstract{\litStore}   \rightarrow   \mathcal{P}  (\ensuremath{\itop{Exp}}   \times   \Abstract{\Psi}   \times   \Abstract{\litStore})  \\
&\hspace{0em}  \Abstract{\Sigma}  (\ensuremath{\itop{Exp}})   \coloneqq   \mathcal{P}  (\ensuremath{\itop{Exp}}   \times   \Abstract{\Psi}   \times   \Abstract{\litStore}) 
\end{align*}

where
$ \Psi   \coloneqq   \Abstract{\litEnv}   \times   \Abstract{\litKAddr}   \times   \Abstract{\litKStore}   \times   \Abstract{\litTime} $.
This is path-sensitive because
$ \Abstract{\Sigma}  (\ensuremath{\itop{Exp}}) $ can represent arbitrary
\emph{relations} between $ (\ensuremath{\itop{Exp}}   \times   \Psi )$
and $ \Abstract{\litStore} $.

\par

As discussed in Section~\ref{path-and-flow-sensitivity-in-analysis}, a
flow-sensitive analysis will give a single set of facts per program
point. This results in the following monad $ \Abstract{\litM} ^{  fs } $
and state space $ \Abstract{\Sigma}  ^{ fs } $ which encode \emph{finite
maps} to $ \Abstract{\litStore} $ rather than relations:

\begin{align*}
&\hspace{0em}  \Abstract{\litM} ^{  fs }  (\ensuremath{\itop{Exp}})   \coloneqq   \Abstract{\Psi}   \times   \Abstract{\litStore}   \rightarrow   \lbrack  (\ensuremath{\itop{Exp}}   \times   \Abstract{\Psi} )  \mapsto   \Abstract{\litStore} \rbrack   \\
&\hspace{0em}  \Abstract{\Sigma}  ^{ fs }  (\ensuremath{\itop{Exp}})   \coloneqq   \lbrack  (\ensuremath{\itop{Exp}}   \times   \Abstract{\Psi} )  \mapsto   \Abstract{\litStore} \rbrack  
\end{align*}

Finally, a flow-insensitive analysis must contain a global set of facts
for each variable, which we achieve by pulling $ \Abstract{\litStore} $
out of the powerset:

\begin{align*}
&\hspace{0em}  \Abstract{\litM} ^{  fi }  (\ensuremath{\itop{Exp}})   \coloneqq   \Abstract{\Psi}   \times   \Abstract{\litStore}   \rightarrow   \mathcal{P}  (\ensuremath{\itop{Exp}}   \times   \Abstract{\Psi} )  \times   \Abstract{\litStore}  \\
&\hspace{0em}  \Abstract{\Sigma}  ^{ fi }  (\ensuremath{\itop{Exp}})   \coloneqq   \mathcal{P}  (\ensuremath{\itop{Exp}}   \times   \Abstract{\Psi} )  \times   \Abstract{\litStore} 
\end{align*}

\par

These three resulting structures, $ \Abstract{\Sigma} $,
$ \Abstract{\Sigma}  ^{ fs } $ and $ \Abstract{\Sigma}  ^{ fi } $,
capture the essence of path-sensitive, flow-sensitive and
flow-insensitive transition systems, and arise naturally from
$ \Abstract{\litM} $, $ \Abstract{\litM} ^{  fs } $ and
$ \Abstract{\litM} ^{  fi } $, which each have monadic structure. We
only describe $ \Abstract{\litM} ^{  fi } $ directly in this section; in
Section~\ref{a-compositional-monadic-framework} we describe a more
compositional set of building blocks, from which $ \Abstract{\litM} $,
$ \Abstract{\litM} ^{  fs } $ and $ \Abstract{\litM} ^{  fi } $ are
recovered.

\par

\subsection{Flow Insensitive Monad}\label{flow-insensitive-monad}

\par

We show the definitions for monad operators, state effects,
nondeterminism effects, and mapping to transition system for the
flow-insensitive monad $ \Abstract{\litM} ^{  fi } $ in
Figure~\ref{flow-insensitive-monad-parameter}.

\par

\begin{figure}

\begin{alignat*}{2}
 \Abstract{\litM} ^{  fi } (A) &  \coloneqq   \Abstract{\Psi}   \times   \Abstract{\litStore}   \rightarrow   \mathcal{P} (A  \times   \Abstract{\Psi} )  \times   \Abstract{\litStore}  \\
 \varsigma   \in   \Abstract{\Sigma}  ^{ fi }  &  \coloneqq   \mathcal{P}  (\ensuremath{\itop{Exp}}   \times   \Abstract{\Psi} )  \times   \Abstract{\litStore} 
\end{alignat*}

\hrule

\begin{align*}
&\hspace{0em}  \Abstract {\ensuremath{\itop{return}}}   ^{ fi }  :  \forall  A, A  \rightarrow   \Abstract{\litM} ^{  fi } (A) \\
&\hspace{0em}  \Abstract {\ensuremath{\itop{return}}}   ^{ fi } (x)( \psi , \sigma )  \coloneqq  ( \{ x, \psi  \} , \sigma ) \\
&\hspace{0em}  \Abstract {\ensuremath{\itop{bind}}}   ^{ fi }  :  \forall  A B,  \Abstract{\litM} ^{  fi } (A)  \rightarrow  (A  \rightarrow   \Abstract{\litM} ^{  fi } (B))  \rightarrow   \Abstract{\litM} ^{  fi } (B) \\
&\hspace{0em}  \Abstract {\ensuremath{\itop{bind}}}   ^{ fi } (X)(f)( \psi , \sigma )  \coloneqq   \\
&\hspace{2em} ( \{ y \psi  _{ 11 }  .. y \psi  _{ 1m _1  }  .. y \psi  _{ n1 }  .. y \psi  _{ nm _n  }  \} , \sigma  _1   \sqcup  ..  \sqcup   \sigma  _n )  \;\;\text{\footnotesize{}where}\;\;  \\
&\hspace{4em} ( \{ (x _1 , \psi  _1 ) .. (x _n , \psi  _n ) \} , \sigma ')  \coloneqq  X( \psi , \sigma ) \\
&\hspace{4em} ( \{ y \psi  _{ i1 }  .. y \psi  _{ im _i  }  \} , \sigma  _i )  \coloneqq  f(x _i )( \psi  _i , \sigma ') \\
&\hspace{0em}  \Abstract {\ensuremath{\itop{get-Env}}}   ^{ fi }  :  \Abstract{\litM} ^{  fi }  (\Abstract{\litEnv})  \\
&\hspace{0em}  \Abstract {\ensuremath{\itop{get-Env}}}   ^{ fi } ( \langle  \rho , \kappa , \tau  \rangle , \sigma )  \coloneqq  ( \{ ( \rho , \langle  \rho , \kappa , \tau  \rangle ) \} , \sigma ) \\
&\hspace{0em}  \Abstract {\ensuremath{\itop{put-Env}}}   ^{ fi }  :  \Abstract{\litEnv}   \rightarrow   \Abstract{\litM} ^{  fi } (1) \\
&\hspace{0em}  \Abstract {\ensuremath{\itop{put-Env}}}   ^{ fi } ( \rho ')( \langle  \rho , \kappa , \tau  \rangle , \sigma )  \coloneqq  ( \{ (1, \langle  \rho ', \kappa , \tau  \rangle ) \} , \sigma ) \\
&\hspace{0em}  \Abstract {\ensuremath{\itop{mzero}}}   ^{ fi }  :  \forall  A,  \Abstract{\litM} ^{  fi } (A) \\
&\hspace{0em}  \Abstract {\ensuremath{\itop{mzero}}}   ^{ fi } ( \psi , \sigma )  \coloneqq  ( \{  \} ,  \bot ) \\
&\hspace{0em}  \underline{\hspace{0.5em}}{  \Abstract{\mathbin{\langle + \rangle}}^{fi}  }\underline{\hspace{0.5em}}  :  \forall  A,  \Abstract{\litM} ^{  fi } (A)  \times   \Abstract{\litM} ^{  fi } (A)  \rightarrow   \Abstract{\litM} ^{  fi }  A  \\
&\hspace{0em} (X _1   \mathbin{\langle + \rangle}  X _2 )( \psi , \sigma )  \coloneqq  (x \psi  ^{ * }  _1   \cup  x \psi  ^{ * }  _2 , \sigma  _1   \sqcup   \sigma  _2 )  \;\;\text{\footnotesize{}where}\;\;   \\
&\hspace{2em} (x \psi  ^{ * }  _i , \sigma  _i )  \coloneqq  X _i ( \psi , \sigma ) \\
&\hspace{0em}  \alpha  ^{  \Abstract{\Sigma}   \leftrightarrow \Abstract{\litM} ^{  fi }  }  : ( \Abstract{\Sigma}  ^{ fi }   \rightarrow   \Abstract{\Sigma}  ^{ fi } )  \rightarrow   (\ensuremath{\itop{Exp}}   \rightarrow   \Abstract{\litM} ^{  fi }  (\ensuremath{\itop{Exp}}) ) \\
&\hspace{0em}  \alpha  ^{  \Abstract{\Sigma}   \leftrightarrow \Abstract{\litM} ^{  fi }  } (f)(e)( \psi , \sigma )  \coloneqq  f( \{ (e, \psi ) \} , \sigma ) \\
&\hspace{0em}  \gamma  ^{  \Abstract{\Sigma}   \leftrightarrow \Abstract{\litM} ^{  fi }  }  :  (\ensuremath{\itop{Exp}}   \rightarrow   \Abstract{\litM} ^{  fi }  (\ensuremath{\itop{Exp}}) )  \rightarrow  ( \Abstract{\Sigma}  ^{ fi }   \rightarrow   \Abstract{\Sigma}  ^{ fi } ) \\
&\hspace{0em}  \gamma  ^{  \Abstract{\Sigma}   \leftrightarrow \Abstract{\litM} ^{  fi }  } (f)(e \psi  ^{ * } , \sigma )  \coloneqq   \\
&\hspace{2em} ( \{ e \psi  _{ 11 }  .. e \psi  _{ n1 }  .. e \psi  _{ nm _n  }  \} ,  \sigma  _1   \sqcup  ..  \sqcup   \sigma  _n )  \;\;\text{\footnotesize{}where}\;\;   \\
&\hspace{4em}  \{ (e _1 , \psi  _1 ) .. (e _n , \psi  _n ) \}   \coloneqq  e \psi  ^{ * }  \\
&\hspace{4em} ( \{ e \psi  _{ i1 }  .. e \psi  _{ im _i  }  \} , \sigma  _i )  \coloneqq  f(e _i )( \psi  _i , \sigma )
\end{align*}

\caption{Flow Insensitive Monad Parameter}
\label{flow-insensitive-monad-parameter} \end{figure}

\par

The $ \Abstract {\ensuremath{\itop{bind}}}   ^{ fi } $ operation
performs the global store merging required to capture a flow-insensitive
analysis. The unit for
$ \Abstract {\ensuremath{\itop{bind}}}   ^{ fi } $ returns one
nondeterminism branch and a single global store. State effects
$ \Abstract {\ensuremath{\itop{get-Env}}}   ^{ fi } $ and
$ \Abstract {\ensuremath{\itop{put-Env}}}   ^{ fi } $ return a single
branch of nondeterminism. Nondeterminism operations union the powerset
and join the store pairwise. Finally, the Galois connection relating
$ \Abstract{\litM} ^{  fi } $ to the state space
$ \Abstract{\Sigma}  ^{ fi } $ also computes powerset unions and store
joins pairwise.

\par

Instantiating the generic monadic interpreter with $ \Concrete{\litM} $,
$ \Abstract{\litM} $ and $ \Abstract{\litM} ^{  fi } $ yields a concrete
interpreter, path-sensitive abstract interpreter, and flow-insensitive
abstract interpreter respectively, purely by changing the underlying
monad. Furthermore, the proofs of abstraction between interpreters and
their induced transition systems is isolated to a proof of abstraction
between monads.

\par

\section{A Compositional Monadic
Framework}\label{a-compositional-monadic-framework}

\par

In our development thus far, any modification to the interpreter
requires redesigning the monad $ \Abstract{\litM} $ and constructing new
proofs relating $ \Abstract{\litM} $ to $ \Concrete{\litM} $. We want to
avoid reconstructing complicated monads for interpreters, especially as
languages and analyses grow and change. Even more, we want to avoid
reconstructing complicated \emph{proofs} that such changes require.
Toward this goal, we introduce a compositional framework for
constructing monads which are correct-by-construction by extending the
well-known structure of monad transformer to that of \emph{Galois
transformer}.

\par

Galois transformers are monad transformers which transport Galois
connections and mappings to an executable transition system. We make
this definition precise and prove our Galois transformers correct in
Section~\ref{galois-transformers-1}. For now we present monad
transformer operations augmented with the computational part of Galois
transformers: the mapping to a transition system, which we called
$ \alpha  ^{  \Concrete{\Sigma}   \leftrightarrow \Concrete{\litM} }  $,
$ \gamma  ^{  \Concrete{\Sigma}   \leftrightarrow \Concrete{\litM} }  $,
$ \alpha  ^{  \Abstract{\Sigma}   \leftrightarrow \Abstract{\litM} ^{  fi }  } $
and
$ \gamma  ^{  \Abstract{\Sigma}   \leftrightarrow \Abstract{\litM} ^{  fi }  } $
in Sections~\ref{recovering-analyses}
and~\ref{varying-path-and-flow-sensitivity}.

\par

There are two monadic effects used in our monadic interpreter: state and
nondeterminism. For state, we review the state monad transformer
$S ^t  \lbrack s \rbrack $, which is
standard\cite{dvanhorn:Liang1995Monad,davdar:Moggi:1989:Monads}, however
we also show how $S ^t  \lbrack s \rbrack $ maps to a transition system
and obeys Galois transformer properties. For nondeterminism we develop
two new monad transformers: $ \mathcal{P}  ^t $ and
$F ^t  \lbrack s \rbrack $. These monad transformers are fully general
purpose, even outside the context of program analysis, and are novel in
this work. Finally we show that $ \mathcal{P}  ^t $ and
$F ^t  \lbrack s \rbrack $ map to transition systems and obey Galois
transformer properties.

\par

To create a monad with various state and nondeterminism effects, one
need only construct some composition of these three monad transformers.
Implementations and proofs for monadic sequencing, state effects,
nondeterminism effects, and mappings to an executable transition system
will come entirely for free. This means that for a language which has a
different state space than the example in this paper, no added effort is
required to construct a monad stack for that language; it will merely
require a different selection and permutation of the same monad
transformer components.

\par

Path and flow sensitivity properties arise from the \emph{order of
composition} of state and nondeterminism monad transformers. Placing
state after nondeterminism
($S ^t  \lbrack s \rbrack   \circ   \mathcal{P}  ^t $ or
$S ^t  \lbrack s \rbrack   \circ  F ^t  \lbrack s' \rbrack $) will
result in $s$ being path-sensitive. Placing state before nondeterminism
($ \mathcal{P}  ^t   \circ  S ^t  \lbrack s \rbrack $ or
$F ^t  \lbrack s' \rbrack   \circ  S ^t  \lbrack s \rbrack $) will
result in $s$ being flow-insensitive. Finally, when
$F ^t  \lbrack s \rbrack $ is used in place of
$S ^t  \lbrack s \rbrack   \circ   \mathcal{P}  ^t $ or
$ \mathcal{P}  ^t   \circ  S ^t  \lbrack s \rbrack $, $s$ will be
flow-sensitive. The combination of all three sensitivities is
$ \ensuremath{\itop{M}}   \coloneqq  S ^t  \lbrack s _1  \rbrack   \circ  F ^t  \lbrack s _2  \rbrack   \circ  S ^t  \lbrack s _3  \rbrack $
which induces the transition system
$ \Sigma  (\ensuremath{\itop{Exp}})   \coloneqq   \lbrack  (\ensuremath{\itop{Exp}}   \times  s _1 )  \mapsto  s _2  \rbrack   \times  s _3 $,
where $s _1 $ is path-sensitive, $s _2 $ is flow-sensitive, and $s _3 $
is flow-insensitive. Using $S ^t  \lbrack s \rbrack $,
$ \mathcal{P}  ^t $ and $F ^t  \lbrack s \rbrack $, one can easily
choose which components of the state space should be path-sensitive,
flow-sensitive or flow-insensitive, purely by the order of monad
composition.

\par

In the following definitions we must refer to
$ \ensuremath{\itop{bind}} $, $ \ensuremath{\itop{return}} $ and other
operations from the underlying monad, which we notate
$ \ensuremath{\itop{bind}} ^m  $, $ \ensuremath{\itop{return}} ^m  $,
$ \leftarrow  ^m $, etc.

\par

\subsection{State Galois Transformer}\label{state-galois-transformer}

\par

\begin{figure}

\begin{alignat*}{2}
S ^t  \lbrack s \rbrack  & :  (\ensuremath{\itop{Type}}   \rightarrow   \ensuremath{\itop{Type}})   \rightarrow   (\ensuremath{\itop{Type}}   \rightarrow   \ensuremath{\itop{Type}})  \\
S ^t  \lbrack s \rbrack (m)(A) &  \coloneqq  s  \rightarrow  m(A  \times  s) \\
 \Pi  ^{ S ^t  }  \lbrack s \rbrack  & :  (\ensuremath{\itop{Type}}   \rightarrow   \ensuremath{\itop{Type}})   \rightarrow   (\ensuremath{\itop{Type}}   \rightarrow   \ensuremath{\itop{Type}})  \\
 \Pi  ^{ S ^t  }  \lbrack s \rbrack ( \Sigma )(A) &  \coloneqq   \Sigma (A  \times  s)
\end{alignat*}

\hrule

\begin{align*}
&\hspace{0em}  \ensuremath{\itop{return}} ^{  S ^t  }  :  \forall  A, A  \rightarrow  S ^t  \lbrack s \rbrack (m)(A) \\
&\hspace{0em}  \ensuremath{\itop{return}} ^{  S ^t  } (x)(s)  \coloneqq   \ensuremath{\itop{return}} ^m  (x,s) \\
&\hspace{0em}  \ensuremath{\itop{bind}} ^{  S ^t  }   \hspace{-3pt}  :  \forall  A B, S ^t  \lbrack s \rbrack (m)(A)  \rightarrow  (A  \rightarrow  S ^t  \lbrack s \rbrack (m)(B))  \rightarrow  S ^t  \lbrack s \rbrack (m)(B) \\
&\hspace{0em}  \ensuremath{\itop{bind}} ^{  S ^t  } (X)(f)(s)  \coloneqq  (x,s')  \leftarrow  ^m  X(s)  \;;\;  f(x)(s') \\
&\hspace{0em}  \ensuremath{\itop{get}} ^{  S ^t  }  : S ^t  \lbrack s \rbrack (m)(s)      \\
&\hspace{0em}  \ensuremath{\itop{get}} ^{  S ^t  } (s)  \coloneqq   \ensuremath{\itop{return}} ^m  (s,s) \\
&\hspace{0em}  \ensuremath{\itop{put}} ^{  S ^t  }  : s  \rightarrow  S ^t  \lbrack s \rbrack (m)(1)  \\
&\hspace{0em}  \ensuremath{\itop{put}} ^{  S ^t  } (s')(s)  \coloneqq   \ensuremath{\itop{return}} ^m  (1,s') \\
&\hspace{0em}  \ensuremath{\itop{mzero}} ^{  S ^t  }  :  \forall  A, S ^t  \lbrack s \rbrack (m)(A) \\
&\hspace{0em}  \ensuremath{\itop{mzero}} ^{  S ^t  } (s)  \coloneqq   \ensuremath{\itop{mzero}} ^m    \\
&\hspace{0em}  \underline{\hspace{0.5em}}{  \mathbin{\langle + \rangle}  ^{ S ^t  }  }\underline{\hspace{0.5em}}  :  \forall  A, S ^t  \lbrack s \rbrack (m)(A)  \times  S ^t  \lbrack s \rbrack (m)(A)  \rightarrow  S ^t  \lbrack s \rbrack (m)(A) \\
&\hspace{0em} (X _1   \mathbin{\langle + \rangle}  ^{ S ^t  }  X _2 )(s)  \coloneqq  X _1 (s)  \mathbin{\langle + \rangle}  ^m  X _2 (s)  \\
&\hspace{0em}  \alpha  ^{ S ^t  }  :  \forall  A B, \\
&\hspace{2em} ( \Pi  ^{ S ^t  }  \lbrack s \rbrack ( \Sigma  ^m )(A)  \rightarrow   \Pi  ^{ S ^t  }  \lbrack s \rbrack ( \Sigma  ^m )(B))  \rightarrow  (A  \rightarrow  S ^t  \lbrack s \rbrack (m)(B)) \\
&\hspace{0em}  \alpha  ^{ S ^t  } (f)(x)(s)  \coloneqq   \alpha  ^m (f)(x,s) \\
&\hspace{0em}  \gamma  ^{ S ^t  }  :  \forall  A B,  \\
&\hspace{2em} (A  \rightarrow  S ^t  \lbrack s \rbrack (m)(B))  \rightarrow  ( \Pi  ^{ S ^t  }  \lbrack s \rbrack ( \Sigma  ^m )(A)  \rightarrow   \Pi  ^{ S ^t  }  \lbrack s \rbrack ( \Sigma  ^m )(B)) \\
&\hspace{0em}  \gamma  ^{ S ^t  } (f)  \coloneqq   \gamma  ^m ( \lambda (x,s). f(x)(s))
\end{align*}

\caption{State Galois Transformer} \label{state-galois-transformer-def}
\end{figure}

\par

The state Galois transformer is shown in
Figure~\ref{state-galois-transformer-def}.
$ \ensuremath{\itop{return}} ^{  S ^t  } $,
$ \ensuremath{\itop{bind}} ^{  S ^t  } $,
$ \ensuremath{\itop{get}} ^{  S ^t  } $ and
$ \ensuremath{\itop{put}} ^{  S ^t  } $ require that $m$ be a monad.
$ \ensuremath{\itop{mzero}} ^{  S ^t  } $ and
$ \underline{\hspace{0.5em}}{  \mathbin{\langle + \rangle}  ^{ S ^t  }  }\underline{\hspace{0.5em}} $
require that $m$ be a monad with nondeterminism effects. And finally,
$ \alpha  ^{ S ^t  } $ and $ \gamma  ^{ S ^t  } $ require that $m$ maps
to $ \Sigma  ^m $ via Galois connection
$ \Sigma (A)  \rightarrow   \Sigma (B)  \galois{\alpha^{m}}{\gamma^{m}}  A  \rightarrow  m(B)$.

\par

\subsection{Nondeterminism Galois
Transformer}\label{nondeterminism-galois-transformer}

\par

\begin{figure}

\begin{alignat*}{2}
 \mathcal{P}  ^t  & :  (\ensuremath{\itop{Type}}   \rightarrow   \ensuremath{\itop{Type}})   \rightarrow   (\ensuremath{\itop{Type}}   \rightarrow   \ensuremath{\itop{Type}})  \\
 \mathcal{P}  ^t (m)(A) &  \coloneqq  m( \mathcal{P} (A)) \\
 \Pi  ^{  \mathcal{P}  ^t  }  & :  (\ensuremath{\itop{Type}}   \rightarrow   \ensuremath{\itop{Type}})   \rightarrow   (\ensuremath{\itop{Type}}   \rightarrow   \ensuremath{\itop{Type}})  \\
 \Pi  ^{  \mathcal{P}  ^t  } ( \Sigma )(A) &  \coloneqq   \Sigma ( \mathcal{P} (A))
\end{alignat*}

\hrule

\begin{align*}
&\hspace{0em}  \ensuremath{\itop{return}} ^{   \mathcal{P}  ^t  }  :  \forall  A, A  \rightarrow   \mathcal{P}  ^t (m)(A) \\
&\hspace{0em}  \ensuremath{\itop{return}} ^{   \mathcal{P}  ^t  } (x)  \coloneqq   \ensuremath{\itop{return}} ^m  ( \{ x \} ) \\
&\hspace{0em}  \ensuremath{\itop{bind}} ^{   \mathcal{P}  ^t  }  :  \forall  A B,  \mathcal{P}  ^t (m)(A)  \rightarrow  (A  \rightarrow   \mathcal{P}  ^t (m)(B))  \rightarrow   \mathcal{P}  ^t (m)(B) \\
&\hspace{0em}  \ensuremath{\itop{bind}} ^{   \mathcal{P}  ^t  } (X)(f)  \coloneqq   \ensuremath{\ttbfop{do}}  \\
&\hspace{2em}  \{ x _1  .. x _n  \}   \leftarrow  ^m  X \\
&\hspace{2em} f(x _1 )  \sqcup  ^m  ..  \sqcup  ^m  f(x _n ) \\
&\hspace{0em}  \ensuremath{\itop{get}} ^{   \mathcal{P}  ^t  }  :  \mathcal{P}  ^t (m)(s) \\
&\hspace{0em}  \ensuremath{\itop{get}} ^{   \mathcal{P}  ^t  }   \coloneqq  s  \leftarrow  ^m   \ensuremath{\itop{get}} ^m    \;;\;   \ensuremath{\itop{return}} ^m  ( \{ s \} ) \\
&\hspace{0em}  \ensuremath{\itop{put}} ^{   \mathcal{P}  ^t  }  : s  \rightarrow   \mathcal{P}  ^t (m)(1) \\
&\hspace{0em}  \ensuremath{\itop{put}} ^{   \mathcal{P}  ^t  } (s)  \coloneqq  u  \leftarrow  ^m   \ensuremath{\itop{put}} ^m  (x)  \;;\;   \ensuremath{\itop{return}} ^m  ( \{ u \} ) \\
&\hspace{0em}  \ensuremath{\itop{mzero}} ^{   \mathcal{P}  ^t  }  :  \forall  A,  \mathcal{P}  ^t (m)(A) \\
&\hspace{0em}  \ensuremath{\itop{mzero}} ^{   \mathcal{P}  ^t  }   \coloneqq   \bot  ^m  \\
&\hspace{0em}  \underline{\hspace{0.5em}}{  \mathbin{\langle + \rangle}  ^{  \mathcal{P}  ^t  }  }\underline{\hspace{0.5em}}  :  \forall  A,  \mathcal{P}  ^t (m)(A) x  \mathcal{P}  ^t (m)(A)  \rightarrow   \mathcal{P}  ^t (m)(A) \\
&\hspace{0em} X _1   \mathbin{\langle + \rangle}  ^{  \mathcal{P}  ^t  }  X _2   \coloneqq  X _1   \sqcup  ^m  X _2  \\
&\hspace{0em}  \alpha  ^{  \mathcal{P}  ^t  }  :  \forall  A B, \\
&\hspace{2em} ( \Pi  ^{  \mathcal{P}  ^t  } ( \Sigma  ^m )(A)  \rightarrow   \Pi  ^{  \mathcal{P}  ^t  } ( \Sigma  ^m )(B))  \rightarrow  (A  \rightarrow   \mathcal{P}  ^t (m)(B)) \\
&\hspace{0em}  \alpha  ^{  \mathcal{P}  ^t  } (f)(x)  \coloneqq   \alpha  ^m (f)( \{ x \} ) \\
&\hspace{0em}  \gamma  ^{  \mathcal{P}  ^t  }  :  \forall  A B,  \\
&\hspace{2em} (A  \rightarrow   \mathcal{P}  ^t (m)(B))  \rightarrow  ( \Pi  ^{  \mathcal{P}  ^t  } ( \Sigma  ^m )(A)  \rightarrow   \Pi  ^{  \mathcal{P}  ^t  } ( \Sigma  ^m )(B)) \\
&\hspace{0em}  \gamma  ^{  \mathcal{P}  ^t  } (f)  \coloneqq   \\
&\hspace{2em}  \gamma  ^m ( \lambda ( \{ x _1  .. x _n  \} ). f(x _1 )  \sqcup  ^m  ..  \sqcup  ^m  f(x _n ))
\end{align*}

\caption{Nondeterminism Galois Transformer}
\label{nondeterminism-galois-transformer-def} \end{figure}

\par

The nondeterminism Galois transformer is shown in
Figure~\ref{nondeterminism-galois-transformer-def}. Crucially,
$ \ensuremath{\itop{return}} ^{   \mathcal{P}  ^t  } $ and
$ \ensuremath{\itop{bind}} ^{   \mathcal{P}  ^t  } $ require that $m$ be
both a monad and a \emph{join-semilattice functor}. We attribute this
requirement (and the difficulty of expressing it in Haskell) as a
possible reason why it has not been discovered thus far. This
functorality of $m$ is instantiated with
$ \mathcal{P} ( \underline{\hspace{0.5em}} )$ using the usual
join-semilattice on powersets: $ \{  \} $ for $ \bot $ and $ \cup $ for
$ \sqcup $. $ \ensuremath{\itop{get}} ^{   \mathcal{P}  ^t  } $ and
$ \ensuremath{\itop{put}} ^{   \mathcal{P}  ^t  } $ require that $m$ be
a monad with state effects. Like the state Galois transformer,
$ \alpha  ^{  \mathcal{P}  ^t  } $ and
$ \gamma  ^{  \mathcal{P}  ^t  } $ require that $m$ maps to
$ \Sigma  ^m $ via Galois connection.

\par

\begin{lemma}
\label{nondeterminism-transformer-laws}{[}$ \mathcal{P}  ^t $ laws{]}
$ \ensuremath{\itop{bind}} ^{   \mathcal{P}  ^t  } $ and
$ \ensuremath{\itop{return}} ^{   \mathcal{P}  ^t  } $ satisfy monad
laws, $ \ensuremath{\itop{get}} ^{   \mathcal{P}  ^t  } $ and
$ \ensuremath{\itop{put}} ^{   \mathcal{P}  ^t  } $ satisfy state monad
laws, and $ \ensuremath{\itop{mzero}} ^{   \mathcal{P}  ^t  } $ and
$ \mathbin{\langle + \rangle}  ^{  \mathcal{P}  ^t  } $ satisfy
nondeterminism monad laws. \end{lemma} \noindent
See our proofs in the extended version of the paper, where the key lemma
in proving monad laws is the join-semilattice functorality of $m$,
namely that:

\begin{alignat*}{1}
 \ensuremath{\itop{return}} ^m  (x  \sqcup  y) &=  \ensuremath{\itop{return}} ^m  (x)  \sqcup  ^m   \ensuremath{\itop{return}} ^m  (y) \\
 \ensuremath{\itop{bind}} ^m  (X  \sqcup  Y)(f) &=  \ensuremath{\itop{bind}} ^m  (X)(f)  \sqcup  ^m   \ensuremath{\itop{bind}} ^m  (Y)(f)
\end{alignat*}

\par

\subsection{Flow Sensitivity Galois
Transformer}\label{flow-sensitivity-galois-transformer}

\par

\begin{figure}

\begin{alignat*}{2}
F ^t  \lbrack s \rbrack  & :  (\ensuremath{\itop{Type}}   \rightarrow   \ensuremath{\itop{Type}})   \rightarrow   (\ensuremath{\itop{Type}}   \rightarrow   \ensuremath{\itop{Type}})  \\
F ^t  \lbrack s \rbrack (m)(A) &  \coloneqq  s  \rightarrow  m( \lbrack A  \mapsto  s \rbrack ) \\
 \Pi  ^{ F ^t  }  \lbrack s \rbrack  & :  (\ensuremath{\itop{Type}}   \rightarrow   \ensuremath{\itop{Type}})   \rightarrow   (\ensuremath{\itop{Type}}   \rightarrow   \ensuremath{\itop{Type}})  \\
 \Pi  ^{ F ^t  }  \lbrack s \rbrack ( \Sigma )(A) &  \coloneqq   \Sigma ( \lbrack A  \mapsto  s \rbrack )
\end{alignat*}

\hrule

\begin{align*}
&\hspace{0em}  \ensuremath{\itop{return}} ^{  F ^t  }  :  \forall  A, A  \rightarrow  F ^t  \lbrack s \rbrack (m)(A) \\
&\hspace{0em}  \ensuremath{\itop{return}} ^{  F ^t  } (x)(s)  \coloneqq   \ensuremath{\itop{return}} ^m  ( \{ x  \mapsto  s \} ) \\
&\hspace{0em}  \ensuremath{\itop{bind}} ^{  F ^t  }   \hspace{-3pt}  :  \forall  A B, F ^t  \lbrack s \rbrack (m)(A)  \hspace{-1pt}  \rightarrow  \hspace{-1pt}  (A  \hspace{-1pt}  \rightarrow  \hspace{-1pt}  F ^t  \lbrack s \rbrack (m)(B))  \hspace{-1pt}  \rightarrow  \hspace{-1pt}  F ^t  \lbrack s \rbrack (m)(B) \\
&\hspace{0em}  \ensuremath{\itop{bind}} ^{  F ^t  } (X)(f)(s)  \coloneqq   \ensuremath{\ttbfop{do}}  \\
&\hspace{2em}  \{ x _1   \mapsto  s _1 ..x _n   \mapsto  s _n  \}   \leftarrow  ^m  X(s) \\
&\hspace{2em} f(x _1 )(s _1 )  \sqcup  ^m  ..  \sqcup  ^m  f(x _n )(s _n ) \\
&\hspace{0em}  \ensuremath{\itop{get}} ^{  F ^t  }  : F ^t  \lbrack s \rbrack (m)(s) \\
&\hspace{0em}  \ensuremath{\itop{get}} ^{  F ^t  } (s)  \coloneqq   \ensuremath{\itop{return}} ^m    \{ s  \mapsto  s \}  \\
&\hspace{0em}  \ensuremath{\itop{put}} ^{  F ^t  }  : s  \rightarrow  F ^t  \lbrack s \rbrack (m)(1) \\
&\hspace{0em}  \ensuremath{\itop{put}} ^{  F ^t  } (s')(s)  \coloneqq   \ensuremath{\itop{return}} ^m    \{ 1  \mapsto  s' \}  \\
&\hspace{0em}  \ensuremath{\itop{mzero}} ^{  F ^t  }  :  \forall  A, F ^t  \lbrack s \rbrack (m)(A) \\
&\hspace{0em}  \ensuremath{\itop{mzero}} ^{  F ^t  } (s)  \coloneqq   \bot  ^m  \\
&\hspace{0em}  \underline{\hspace{0.5em}}{  \mathbin{\langle + \rangle}  ^{ F ^t  }  }\underline{\hspace{0.5em}}  :  \forall  A, F ^t  \lbrack s \rbrack (m)(A) x F ^t  \lbrack s \rbrack (m)(A)  \rightarrow  F ^t  \lbrack s \rbrack (m)(A) \\
&\hspace{0em} (X _1   \mathbin{\langle + \rangle}  ^{ F ^t  }  X _2 )(s)  \coloneqq  X _1 (s)  \sqcup  ^m  X _2 (s) \\
&\hspace{0em}  \alpha  ^{ F ^t  }  :  \forall  A B,  \\
&\hspace{2em} ( \Pi  ^{ F ^t  }  \lbrack s \rbrack ( \Sigma  ^m )(A)  \rightarrow   \Pi  ^{ F ^t  }  \lbrack s \rbrack ( \Sigma  ^m )(B))  \rightarrow  (A  \rightarrow  F ^t  \lbrack s \rbrack (m)(B)) \\
&\hspace{0em}  \alpha  ^{ F ^t  } (f)(x)(s)  \coloneqq   \alpha  ^m (f)( \{ x  \mapsto  s \} ) \\
&\hspace{0em}  \gamma  ^{ F ^t  }  :  \forall  A B, \\
&\hspace{2em} (A  \rightarrow  F ^t  \lbrack s \rbrack (m)(B))  \rightarrow  ( \Pi  ^{ F ^t  }  \lbrack s \rbrack ( \Sigma  ^m )(A)  \rightarrow   \Pi  ^{ F ^t  }  \lbrack s \rbrack ( \Sigma  ^m )(B)) \\
&\hspace{0em}  \gamma  ^{ F ^t  } (f)  \coloneqq   \\
&\hspace{2em}  \gamma  ^m ( \lambda ( \{ x _1   \mapsto  s _1  .. x _n   \mapsto  s _n  \} ). f(x _1 )(s _1 )  \sqcup  ^m  ..  \sqcup  ^m  f(x _n )(s _n ))
\end{align*}

\caption{Flow Sensitivity Galois Transformer}
\label{flow-sensitive-galois-transformer-def} \end{figure}

\par

The flow sensitivity monad transformer, shown in Figure
~\ref{flow-sensitive-galois-transformer-def}, is a unique monad
transformer that combines state and nondeterminism effects, and does not
arise naturally from composing vanilla nondeterminism and state
transformers. The finite map in the definition of
$F ^t  \lbrack s \rbrack $ is what yields flow sensitivity when
instantiated to a monadic interpreter. After instantiation,
$F ^t  \lbrack s \rbrack (m)(A)$ will be
$ \Abstract{\litStore}   \rightarrow    \lbrack \ensuremath{\itop{Exp}}   \times   \Abstract{\Psi}   \rightarrow   \Abstract{\litStore} \rbrack  $,
which maps each possible expression and context to a unique abstract
store.

\par

Like nondeterminism, $ \ensuremath{\itop{return}} ^{  F ^t  } $ and
$ \ensuremath{\itop{bind}} ^{  F ^t  } $ require that $m$ be both a
monad and a \emph{join-semilattice functor}. This functorality of $m$ is
instantiated with
$ \lbrack  \underline{\hspace{0.5em}}   \mapsto  s \rbrack $ using the
usual join-semilattice on finite maps: $ \{  \} $ for $ \bot $ and:

\begin{alignat*}{1}
Y  \sqcup  Z  \coloneqq   \{  x  \mapsto  y  \sqcup  z  \;|\;   \{ x  \mapsto  y \}   \in  X  \wedge   \{ x  \mapsto  z \}   \in  Y \} 
\end{alignat*}

$ \ensuremath{\itop{get}} ^{   \mathcal{P}  ^t  } $ and
$ \ensuremath{\itop{put}} ^{   \mathcal{P}  ^t  } $ require that $m$ be
a monad. Like the nondeterminism Galois transformer,
$ \alpha  ^{  \mathcal{P}  ^t  } $ and
$ \gamma  ^{  \mathcal{P}  ^t  } $ require that $m$ maps to
$ \Sigma  ^m $ via Galois connection.

\par

\begin{lemma} \label{flow-transformer-laws}{[}$F ^t $ laws{]}
$ \ensuremath{\itop{bind}} ^{  F ^t  } $ and
$ \ensuremath{\itop{return}} ^{  F ^t  } $ satisfy monad laws,
$ \ensuremath{\itop{get}} ^{  F ^t  } $ and
$ \ensuremath{\itop{put}} ^{  F ^t  } $ satisfy state monad laws, and
$ \ensuremath{\itop{mzero}} ^{  F ^t  } $ and
$ \mathbin{\langle + \rangle}  ^{ F ^t  } $ satisfy nondeterminism monad
laws. \end{lemma} \noindent
See our proofs in the extended version of the paper. Monad and
nondeterminism laws are are analogous to those for nondeterminism, and
also rely on the join-semilattice functorailty of $m$. State monad laws
are proved by calculation.

\par

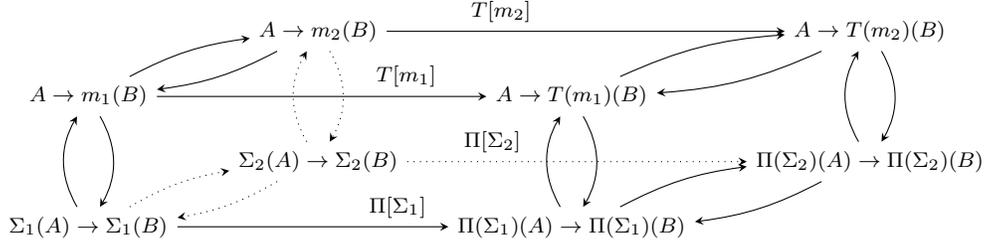
\begin{figure*}
\begin{center}
\begin{tikzpicture}
  \tikzstyle{every node}=[font=\footnotesize]
  \matrix (m) [matrix of math nodes,row sep=1em,column sep=2em]
  {
    & A \rightarrow m_2(B)                & & A \rightarrow T(m_2)(B)                         \\
      A \rightarrow m_1(B)                & & A \rightarrow T(m_1)(B)                       & \\
    & \Sigma_2(A) \rightarrow \Sigma_2(B) & & \Pi(\Sigma_2)(A) \rightarrow \Pi(\Sigma_2)(B)   \\
      \Sigma_1(A) \rightarrow \Sigma_1(B) & & \Pi(\Sigma_1)(A) \rightarrow \Pi(\Sigma_1)(B) & \\
  };
  \path[-stealth]
    (m-2-1) edge [bend left=10]        node [above] {$$}                          (m-1-2)
    (m-1-2) edge [bend left=10]        node [above] {$$}                          (m-2-1)
    (m-4-1) edge [bend left=10,dotted] node [right] {$$}                          (m-3-2)
    (m-3-2) edge [bend left=10,dotted] node [right] {$$}                          (m-4-1)
    (m-4-1) edge [bend left=30]        node [left]  {$$}                          (m-2-1)
    (m-2-1) edge [bend left=30]        node [right] {$$}                          (m-4-1)
    (m-1-2) edge [bend left=30,dotted] node [right] {$$}                          (m-3-2)
    (m-3-2) edge [bend left=30,dotted] node [left]  {$$}                          (m-1-2)
    (m-1-2) edge                       node [above] {$T[m_2]\hspace{8em}$}        (m-1-4)
    (m-2-1) edge                       node [above] {$\hspace{8em}T[m_1]$}        (m-2-3)
    (m-3-2) edge [dotted]              node [above] {$\Pi[\Sigma_2]\hspace{8em}$} (m-3-4)
    (m-4-1) edge                       node [above] {$\hspace{8em}\Pi[\Sigma_1]$} (m-4-3)
    (m-2-3) edge [bend left=10]        node [above] {$$}                          (m-1-4)
    (m-1-4) edge [bend left=10]        node [above] {$$}                          (m-2-3)
    (m-4-3) edge [bend left=10]        node [right] {$$}                          (m-3-4)
    (m-3-4) edge [bend left=10]        node [right] {$$}                          (m-4-3)
    (m-4-3) edge [bend left=30]        node [left]  {$$}                          (m-2-3)
    (m-2-3) edge [bend left=30]        node [right] {$$}                          (m-4-3)
    (m-1-4) edge [bend left=30]        node [right] {$$}                          (m-3-4)
    (m-3-4) edge [bend left=30]        node [left]  {$$}                          (m-1-4)
  ;
\end{tikzpicture}
\end{center}
\caption{Galois Transformer Commuting Cube of Abstractions}
\label{three-d-diagram}
\end{figure*}

\par

\subsection{Galois Transformers}\label{galois-transformers-1}

\par

The capstone of our framework is the fact that monad transformers
$S ^t  \lbrack s \rbrack $, $ \mathcal{P}  ^t $ and
$F ^t  \lbrack s \rbrack $ are also \emph{Galois transformers}.

\par

\begin{definition} A monad transformer $T$ is a Galois transformer with
transition system $ \Pi $ if: \begin{enumerate} \item $T$ transports
Galois connections between monads $m _1 $ and $m _2 $ into Galois
connections between $T(m _1 )$ and $T(m _2 )$:

\begin{center}
\begin{tikzpicture}
  \tikzstyle{every node}=[font=\footnotesize]
  \matrix (m) [matrix of math nodes,row sep=3em,column sep=4em,minimum width=2em]
  {
     A \rightarrow m_2(B) & A \rightarrow T(m_2)(B) \\
     A \rightarrow m_1(B) & A \rightarrow T(m_1)(B) \\
  };
  \path[-stealth]
    (m-1-1) edge [bend left=40] node [right] {$\gamma^m$}     (m-2-1)
            edge                node [below] {$T[m_2]$}       (m-1-2)
    (m-2-1) edge [bend left=40] node [left]  {$\alpha^m$}     (m-1-1)
            edge                node [below] {$T[m_1]$}       (m-2-2)
    (m-1-2) edge [bend left=40] node [right] {$T[\gamma^m]$} (m-2-2)
    (m-2-2) edge [bend left=40] node [left]  {$T[\alpha^m]$} (m-1-2)
  ;
\end{tikzpicture}
\end{center}

$T \lbrack m \rbrack $ must be monotonic, and $T$ must commute with
Galois connections, that is for all $f : A  \rightarrow  m _1 (B)$:

\begin{alignat*}{1}
T \lbrack m _2  \rbrack ( \alpha  ^m (f)) = T \lbrack  \alpha  ^m  \rbrack (T \lbrack m _1  \rbrack (f))
\end{alignat*}

\item 

$ \Pi $ transports Galois connections between induced transition systems
$ \Sigma  _1 $ and $ \Sigma  _2 $ into Galois connections between
$ \Pi ( \Sigma  _1 )$ and $ \Pi ( \Sigma  _2 )$:

\begin{center}
\begin{tikzpicture}
  \tikzstyle{every node}=[font=\footnotesize]
  \matrix (m) [matrix of math nodes,row sep=3em,column sep=4em,minimum width=2em]
  {
     \Sigma_2(A) \rightarrow \Sigma_2(B) & \Pi(\Sigma_2)(A) \rightarrow \Pi(\Sigma_2)(B) \\
     \Sigma_1(A) \rightarrow \Sigma_1(B) & \Pi(\Sigma_1)(A) \rightarrow \Pi(\Sigma_1)(B) \\
  };
  \path[-stealth]
    (m-1-1) edge [bend left=40] node [right] {$\gamma^\Sigma$}       (m-2-1)
            edge                node [below] {$\Pi[\Sigma_2]$}       (m-1-2)
    (m-2-1) edge [bend left=40] node [left]  {$\alpha^\Sigma$}       (m-1-1)
            edge                node [below] {$\Pi[\Sigma_1]$}       (m-2-2)
    (m-1-2) edge [bend left=40] node [right] {$\Pi[\gamma^\Sigma]$}  (m-2-2)
    (m-2-2) edge [bend left=40] node [left]  {$\Pi[\alpha^\Sigma]$}  (m-1-2)
  ;
\end{tikzpicture}
\end{center}

$ \Pi  \lbrack  \Sigma  \rbrack $ must be monotonic, and $ \Pi $ must
commute with Galois connections, that is for all
$f :  \Sigma  _1 (A)  \rightarrow   \Sigma  _1 (B)$:

\begin{alignat*}{1}
 \Pi  \lbrack  \Sigma  _2  \rbrack ( \alpha  ^{  \Sigma  } (f)) =  \Pi  \lbrack  \alpha  ^{  \Sigma  }  \rbrack ( \Pi  \lbrack  \Sigma  _1  \rbrack (f))
\end{alignat*}

\item 

$T$ and $ \Pi $ transport transition system mappings between $m$ and
$ \Sigma $ into transition system mappings between $T(m)$ and
$ \Pi ( \Sigma )$:

\begin{center}
\begin{tikzpicture}
  \tikzstyle{every node}=[font=\footnotesize]
  \matrix (m) [matrix of math nodes,row sep=3em,column sep=4em,minimum width=2em]
  {
           A \rightarrow m(B)      &              A \rightarrow T(m)(B)        \\
   \Sigma(A) \rightarrow \Sigma(B) & \Pi(\Sigma)(A) \rightarrow \Pi(\Sigma)(B) \\
  };
  \path[-stealth]
    (m-1-1) edge [bend left=40] node [right] {$\gamma^{\Sigma\leftrightarrow m}$}       (m-2-1)
            edge                node [below] {$T[m]$}                                   (m-1-2)
    (m-2-1) edge [bend left=40] node [left]  {$\alpha^{\Sigma\leftrightarrow m}$}       (m-1-1)
            edge                node [below] {$\Pi[\Sigma]$}                            (m-2-2)
    (m-1-2) edge [bend left=40] node [right] {$T[\gamma^{\Sigma\leftrightarrow m}]$}    (m-2-2)
    (m-2-2) edge [bend left=40] node [left]  {$T[\alpha^{\Sigma\leftrightarrow m}]$}    (m-1-2)
  ;
\end{tikzpicture}
\end{center}

$T \lbrack  \gamma  ^{  \Sigma  \leftrightarrow m }  \rbrack $ must
commute asymmetrically (in the partial order) with $T$ and $ \Pi $, that
is for all functions $f : A  \rightarrow  m(B)$:

\begin{alignat*}{1}
 \Pi  \lbrack  \Sigma  \rbrack ( \gamma  ^{  \Sigma  \leftrightarrow m } (f))  \sqsubseteq  T \lbrack  \gamma  ^{  \Sigma  \leftrightarrow m }  \rbrack (T \lbrack m \rbrack (f))
\end{alignat*}

\end{enumerate} \end{definition}

\par

\par

\begin{lemma}[Galois Transformer Properties]
\label{are-galois-transformers} $S ^t  \lbrack s \rbrack $,
$ \mathcal{P}  ^t $ and $F ^t  \lbrack s \rbrack $ are Galois
transformers. \end{lemma} \noindent
Definitions for $ \alpha  ^{  \Sigma  \leftrightarrow  \gamma  } $ and
$ \gamma  ^{  \Sigma  \leftrightarrow  \gamma  } $ from property (3) are
shown in Figures~\ref{state-galois-transformer-def},
\ref{nondeterminism-galois-transformer-def}
and~\ref{flow-sensitive-galois-transformer-def}. Definitions of other
Galois connections and commutativity proofs are given in the appendix.

\par

These three properties of Galois transformers snap together in a
three-dimensional diagram, shown in Figure~\ref{three-d-diagram} which
relates abstractions between monads $m _1 $ and $m _2 $ and their
transition systems $ \Sigma  _1 $ and $ \Sigma  _2 $ to their actions
under $T$ and $ \Pi $. The left-hand side of the cube is a commuting
square of abstractions between $m _1 $, $m _2 $, $ \Sigma  _1 $ and
$ \Sigma  _2 $. The right-hand side of the cube is constructed from the
composition of properties (1) through (3) as the front, top, back, and
bottom faces of the cube, and is a commuting square of abstractions
between $T(m _1 )$, $T(m _2 )$, $ \Pi ( \Sigma  _1 )$ and
$ \Pi ( \Sigma  _2 )$. The whole cube commutes, by combining the
commuting properties of the left face and the commuting properties of
(1) through (3).

\par

\begin{theorem} \label{galois-theorem} If $T$ is a Galois transformer
with transition system $ \Pi $, given a commuting square of abstractions
between monads $m _1 $ and $m _2 $ and their transition systems
$ \Sigma  _1 $ and $ \Sigma  _2 $, $T$ and $ \Pi $ construct a commuting
square of abstractions between monads $T(m _1 )$ and $T(m _2 )$ and
their transition systems $ \Pi ( \Sigma  _1 )$ and
$ \Pi ( \Sigma  _2 )$. \end{theorem} \noindent
The proof is the composition of Galois transformer properties, as shown
in the Figure~\ref{three-d-diagram}.

\par

The consequence of this theorem is that any two compositions of Galois
transformers $T _1   \circ  ..  \circ  T _n $ and
$U _1   \circ  ..  \circ  U _n $ where $U _i $ is an abstraction of
$T _i $ will yield a commuting square of abstractions between monads
$(T _1   \circ  ..  \circ  T _n )(ID)$ and
$(U _1   \circ  ..  \circ  U _n )(ID)$ and their induced transition
systems $( \Pi  ^{ T _1  }   \circ  ...  \circ   \Pi  ^{ T _n  } )(ID)$
and $( \Pi  ^{ U _1  }   \circ  ...  \circ   \Pi  ^{ U _n  } )(ID)$.
This is the first step in proving the resulting abstract interpreter
correct; we need to establish a commuting square of abstractions between
a concrete monad, an abstract monad, and their induced concrete and
abstract transition systems.

\par

\subsection{End-to-End Correctness with Galois
Transformers}\label{end-to-end-correctness-with-galois-transformers}

\par

In the setting of abstract interpretation, we instantiate the Galois
transformer framework described above with two compositions of monad
transformers yielding a commuting square of abstractions between the
concrete monad $ \Concrete{\litM} $, the abstract monad
$ \Abstract{\litM} $, and concrete and abstract transition systems
$ \Concrete{\Sigma} $ and $ \Abstract{\Sigma} $:

\begin{center}
\begin{tikzpicture}
  \tikzstyle{every node}=[font=\footnotesize]
  \matrix (m) [matrix of math nodes,row sep=3em,column sep=4em]
  {
   Exp \rightarrow \Concrete{M}(Exp)                         &  Exp \rightarrow \Abstract{M}(Exp)                        \\
   \Concrete{\Sigma}(Exp) \rightarrow \Concrete{\Sigma}(Exp) & \Abstract{\Sigma}(Exp) \rightarrow \Abstract{\Sigma}(Exp) \\
  };
  \path[-stealth]
    (m-1-1) edge [bend left=10] node [above] {$\alpha^{\Concrete{M}}$}                                  (m-1-2)
    (m-1-2) edge [bend left=10] node [below] {$\gamma^{\Concrete{M}}$}                                  (m-1-1)
    (m-2-1) edge [bend left=10] node [above] {$\alpha^{\Concrete{\Sigma}}$}                             (m-2-2)
    (m-2-2) edge [bend left=10] node [below] {$\gamma^{\Concrete{\Sigma}}$}                             (m-2-1)
    (m-1-1) edge [bend left=40] node [right] {$\gamma^{\Concrete{\Sigma}\leftrightarrow\Concrete{M}}$}  (m-2-1)
    (m-2-1) edge [bend left=40] node [left]  {$\alpha^{\Concrete{\Sigma}\leftrightarrow\Concrete{M}}$}  (m-1-1)
    (m-1-2) edge [bend left=40] node [right] {$\gamma^{\Abstract{\Sigma}\leftrightarrow\Abstract{M}}$}  (m-2-2)
    (m-2-2) edge [bend left=40] node [left]  {$\alpha^{\Abstract{\Sigma}\leftrightarrow\Abstract{M}}$}  (m-1-2)
  ;
\end{tikzpicture}
\end{center}

This diagram shows how to relate monadic interpreters to transition
systems (the vertical axis of the diagram), and concrete semantics to
abstract semantics (the horizontal axis of the diagram). The top half is
where we write the monadic interpreter, and the bottom half is where we
execute the analysis as the least-fixed point of a transition system.

\par

We use this commuting square to systematically relate a recovered
collecting semantics with the induced abstract transition system in the
following theorem:

\par

\begin{theorem} \label{galois-end-to-end-thm} Given a commuting square
of abstraction between $ \Concrete{\litM} $, $ \Abstract{\litM} $,
$ \Concrete{\Sigma} $ and $ \Abstract{\Sigma} $, and a generic monadic
interpreter $ \ensuremath{\itop{step}} ^m  $, if
$collect =  \gamma  ^{  \Concrete{\Sigma}   \leftrightarrow \Concrete{\litM} }   (\ensuremath{\itop{step}} ^m    \lbrack \Concrete{\litM} \rbrack  )$
recovers the collecting semantics, then
$analysis =  \gamma  ^{  \Abstract{\Sigma}   \leftrightarrow \Abstract{\litM} }   (\ensuremath{\itop{step}} ^m    \lbrack \Abstract{\litM} \rbrack  )$
is a sound abstraction of the collecting semantics. \end{theorem}

\par

\begin{proof} Given that $ \ensuremath{\itop{step}} ^m  $ is monotonic
in the monad parameter $m$, instantiating it with $ \Concrete{\litM} $
and $ \Abstract{\litM} $ will result in:

\begin{alignat*}{1}
 \alpha   ^{ \Concrete{\litM} }   (\ensuremath{\itop{step}} ^m    \lbrack \Concrete{\litM} \rbrack  )  \sqsubseteq   \ensuremath{\itop{step}} ^m    \lbrack \Abstract{\litM} \rbrack  
\end{alignat*}

Transporting through
$ \gamma  ^{  \Abstract{\Sigma}   \leftrightarrow \Abstract{\litM} }  $,
which is monotonic by virtue of forming a Galois connection with
$ \alpha  ^{  \Abstract{\Sigma}   \leftrightarrow \Abstract{\litM} }  $,
we have:

\begin{alignat*}{1}
(1) \hspace{1em}   \gamma  ^{  \Abstract{\Sigma}   \leftrightarrow \Abstract{\litM} }  ( \alpha   ^{ \Concrete{\litM} }   (\ensuremath{\itop{step}} ^m    \lbrack \Concrete{\litM} \rbrack  ))  \sqsubseteq   \gamma  ^{  \Abstract{\Sigma}   \leftrightarrow \Abstract{\litM} }   (\ensuremath{\itop{step}} ^m    \lbrack \Abstract{\litM} \rbrack  ) = analysis
\end{alignat*}

\par

Next, we abstract the recovered collecting semantics to form its best
specification for abstraction:

\begin{alignat*}{1}
(2) \hspace{1em}   \alpha  ^{  \Abstract{\Sigma}  } (collect) =  \alpha  ^{  \Abstract{\Sigma}  } ( \gamma  ^{  \Concrete{\Sigma}   \leftrightarrow \Concrete{\litM} }   (\ensuremath{\itop{step}} ^m    \lbrack \Concrete{\litM} \rbrack  ))
\end{alignat*}

\par

Finally, we exploit the commutativity of the square of abstractions
between $ \Concrete{\litM} $, $ \Abstract{\litM} $,
$ \Concrete{\Sigma} $ and $ \Abstract{\Sigma} $ to relate the recovered
collecting semantics with the abstract monadic semantics:

\begin{alignat*}{1}
(3) \hspace{1em}   \alpha  ^{  \Abstract{\Sigma}  } ( \gamma  ^{  \Concrete{\Sigma}   \leftrightarrow \Concrete{\litM} }   (\ensuremath{\itop{step}} ^m    \lbrack \Concrete{\litM} \rbrack  ))  \sqsubseteq   \gamma  ^{  \Abstract{\Sigma}   \leftrightarrow \Abstract{\litM} }  ( \alpha   ^{ \Concrete{\litM} }   (\ensuremath{\itop{step}} ^m    \lbrack \Concrete{\litM} \rbrack  ))
\end{alignat*}

\par

The transitive combination of (1), (2) and (3) establishes the soundness
of the derived abstract execution system w.r.t. the recovered collecting
semantics:
$ \alpha  ^{  \Abstract{\Sigma}  } (collect)  \sqsubseteq  analysis$.

\par

\end{proof}

\par

This theorem proves Proposition~\ref{interpreter-end-to-end} in
Section~\ref{end-to-end-correctness} after instantiating the example to
the Galois transformer framework.

\par

\subsection{Applying the Framework to Our
Semantics}\label{applying-the-framework-to-our-semantics}

\par

Our setting is the ground-truth semantics
$ \underline{\hspace{0.5em}}{  \rightsquigarrow   ^{ \ensuremath{\itop{gc}} }   }\underline{\hspace{0.5em}} $
from Section~\ref{semantics} and the generic interpreter
$ \ensuremath{\itop{step}} ^m  $ from Section~\ref{the-interpreter}.

\par

To recover the concrete collecting semantics, we instantiate
$ \ensuremath{\itop{step}} ^m  $ to the concrete parameters for the
domain and time from Section~\ref{recovering-a-concrete-interpreter},
and synthesize the monad as a combination of state and nondeterminism
Galois transformers:

\begin{alignat*}{1}
 \Concrete{\litM}   \coloneqq  (S ^t  \lbrack  \Concrete{\Psi}  \rbrack   \circ  S ^t   \lbrack \Concrete{\litStore} \rbrack    \circ   \mathcal{P}  ^t )(ID)
\end{alignat*}

\par

To recover a path-sensitive abstract interpreter we instantiate
$ \ensuremath{\itop{step}} ^m  $ to the abstract parameters for the
domain and time from Section~\ref{recovering-an-abstract-interpreter},
and synthesize the monad as a combination of state and nondeterminism
Galois transformers:

\begin{alignat*}{1}
 \Abstract{\litM}   \coloneqq  (S ^t  \lbrack  \Abstract{\Psi}  \rbrack   \circ  S ^t   \lbrack \Abstract{\litStore} \rbrack    \circ   \mathcal{P}  ^t )(ID)
\end{alignat*}

which abstract $ \Concrete{\litM} $ piecewise. Both the implementation
and correctness of the induced abstract transition system are
constructed for free by Theorems~\ref{galois-theorem}
and~\ref{galois-end-to-end-thm}.

\par

To recover a flow-sensitive abstract interpreter we synthesize the monad
as a combination of state and flow-sensitive Galois transformers:

\begin{alignat*}{1}
 \Abstract{\litM} ^{  fs }   \coloneqq  (S ^t  \lbrack  \Abstract{\Psi}  \rbrack   \circ  F ^t   \lbrack \Abstract{\litStore} \rbrack  )(ID)
\end{alignat*}

which abstracts $ \Abstract{\litM} $ piecewise.

\par

Finally, to recover a flow-insensitive abstract interpreter we
synthesize the monad as a permuted combination of state and
nondeterminism Galois transformers:

\begin{alignat*}{1}
 \Abstract{\litM} ^{  ps }   \coloneqq  (S ^t  \lbrack  \Abstract{\Psi}  \rbrack   \circ   \mathcal{P}  ^t   \circ  S ^t   \lbrack \Abstract{\litStore} \rbrack  )(ID)
\end{alignat*}

which abstracts $ \Abstract{\litM} ^{  ps } $ piecewise.

\par

\subsection{Applying the Framework to Another
Semantics}\label{applying-the-framework-to-another-semantics}

\par

Our Galois transformers framework is semantics independent, and the
proofs in Section~\ref{galois-transformers-1} need not be reproved for
another semantic setting. To use our framework and establish an
end-to-end correctness theorem, the user must:

\par

\begin{itemize}
\itemsep1pt\parskip0pt\parsep0pt
\item
  Design a generic monadic interpreter for their semantics using an
  interface of monadic effects
\item
  Prove their interpreter monotonic w.r.t. parameters
\item
  Prove that the induced concrete transition system recovers the
  concrete collecting semantics of interest.
\end{itemize}

\par

\noindent
The user then enjoys the following for free:

\par

\begin{itemize}
\itemsep1pt\parskip0pt\parsep0pt
\item
  A combination of state, nondeterminism and flow-sensitive Galois
  transformers which supports the monadic effect interface unique to the
  semantics.
\item
  The ability to rearrange monad transformers to recover variations in
  path and flow sensitivities.
\item
  An induced, executable abstract interpreter for each stack of monad
  transformers.
\item
  A proof that each induced abstract interpreter is a sound abstraction
  of the collecting semantics, as a result of
  Theorems~\ref{galois-theorem} and~\ref{galois-end-to-end-thm}.
\end{itemize}

\par

\section{Implementation}\label{implementation-1}

\par

We have implemented our framework in Haskell and applied it to compute
analyses for $ \ensuremath{\ttbfop{\lambda{}IF}} $. Our implementation
provides path sensitivity, flow sensitivity, and flow insensitivity as a
semantics-independent monad library. The code shares a striking
resemblance with the math.

\par

Our implementation is suitable for prototyping and exploring the design
space of static analyzers. Our analyzer supports exponentially more
compositions of analysis features than any current analyzer. For
example, our implementation is the first which can combine arbitrary
choices in call-site, object, path and flow sensitivities. Furthermore,
the user can choose different path and flow sensitivities independently
for each component of the state space.

\par

Our implementation \(\texttt{\footnotesize maam}\) supports command-line
flags for garbage collection, mCFA, call-site sensitivity, object
sensitivity, and path and flow sensitivity. \begin{align*} &
\texttt{\footnotesize ./maam prog.lam --gc --mcfa --kcfa=1 --ocfa=2} \\
& \hspace{2em}
\texttt{\footnotesize --data-store=flow-sen --stack-store=path-sen}
\end{align*} Each flag is implemented independently of each other
applied to a single parameterized monadic interpreter. Furthermore,
using Galois transformers allows us to prove each combination correct in
one fell swoop.

\par

A developer wishing to use our library to develop analyzers for their
language of choice inherits as much of the analysis infrastructure as
possible. We provide call-site, object, path and flow sensitivities as
language-independent libraries. To support analysis for a new language a
developer need only implement:

\par

\begin{itemize}
\itemsep1pt\parskip0pt\parsep0pt
\item
  A monadic semantics for their language, using state and nondeterminism
  effects.
\item
  The abstract value domain, and optionally the concrete value domain if
  they wish to recover concrete execution.
\item
  Intentional optimizations for their semantics like garbage collection
  and mcfa.
\end{itemize}

\par

The developer then receives the following for free through our analysis
library:

\par

\begin{itemize}
\itemsep1pt\parskip0pt\parsep0pt
\item
  A family of monads which implement their effect interface and give
  different path and flow sensitivities.
\item
  Mechanisms for call-site and object sensitivities.
\item
  An execution engine for each monad to drive the analysis.
\end{itemize}

\par

Not only is a developer able to reuse our implementation of call-site,
object, path and flow sensitivities, they need not understand the
execution machinery or soundness proofs for them either. They need only
verify that their monadic semantics is monotonic w.r.t. the analysis
parameters, and that their abstract value domain forms a Galois
connection. The execution and correctness of the final analyzer is
constructed automatically given these two properties.

\par

Our implementation is publicly available and can be installed as a cabal
package: {\footnotesize\tt cabal install maam}.

\par

\section{Related Work}\label{related-work}

\par

\paragraph{Overview}

\par

Program analysis comes in many forms such as points-to
\cite{dvanhorn:Andersen1994Program}, flow
\cite{dvanhorn:Jones:1981:LambdaFlow}, or shape analysis
\cite{dvanhorn:Chase1990Analysis}, and the literature is vast. (See
\citet{dvanhorn:hind-paste01,dvanhorn:Midtgaard2012Controlflow} for
surveys.) Much of the research has focused on developing families or
frameworks of analyses that endow the abstraction with a number of
knobs, levers, and dials to tune precision and compute efficiently (some
examples include
\citet{dvanhorn:Shivers:1991:CFA, dvanhorn:nielson-nielson-popl97,
dvanhorn:Milanova2005Parameterized, dvanhorn:VanHorn2010Abstracting};
there are many more). These parameters come in various forms with
overloaded meanings such as object
\cite{dvanhorn:Milanova2005Parameterized,
dvanhorn:Smaragdakis2011Pick}, context
\cite{dvanhorn:Sharir:Interprocedural,
dvanhorn:Shivers:1991:CFA}, path \cite{davdar:das:2002:esp}, and heap
\cite{dvanhorn:VanHorn2010Abstracting} sensitivities, or some
combination thereof \cite{dvanhorn:Kastrinis2013Hybrid}.

\par

These various forms can all be cast in the theory of abstraction
interpretation of \citet{dvanhorn:Cousot:1977:AI,
dvanhorn:Cousot1979Systematic} and understood as computable
approximations of an underlying concrete interpreter. Our work
demonstrates that if this underlying concrete interpreter is written in
monadic style, monad transformers are a useful way to organize and
compose these various kinds of program abstractions in a modular and
language-independent way.

\par

This work is inspired by the trifecta combination of
\citeauthor{dvanhorn:Cousot:1977:AI}'s theory of abstract interpretation
based on Galois connections \cite{dvanhorn:Cousot:1977:AI,
dvanhorn:Cousot1979Systematic, dvanhorn:Cousot98-5},
\citeauthor{davdar:Moggi:1989:Monads}'s original monad transformers
\cite{davdar:Moggi:1989:Monads} which were later popularized in
\citeauthor{dvanhorn:Liang1995Monad}'s \emph{Monad Transformers and
Modular Interpreters} \cite{dvanhorn:Liang1995Monad}, and
\citeauthor{dvanhorn:Sergey2013Monadic}'s \emph{Monadic Abstract
Interpreters} \cite{dvanhorn:Sergey2013Monadic}.

\par

\par

\citet{dvanhorn:Liang1995Monad} first demonstrated how monad
transformers could be used to define building blocks for constructing
(concrete) interpreters. Their interpreter monad
\mbox{\(\mathit{InterpM}\)} bears a strong resemblance to ours. We show
this ``building blocks'' approach to interpreter construction also
extends to \emph{abstract} interpreter construction using Galois
transformers. Moreover, we show that these monad transformers can be
proved sound via a Galois connection to their concrete counterparts,
ensuring the soundness of any stack built from sound blocks of Galois
transformers. Soundness proofs of various forms of analysis are
notoriously brittle with respect to language and analysis features. A
reusable framework of Galois transformers offers a potential way forward
for a modular metatheory of program analysis.

\par

\citet{dvanhorn:Cousot98-5} develops a ``calculational approach'' to
analysis design whereby analyses are not designed and then verified
\emph{post facto}, but rather derived by positing an abstraction and
calculating it from the concrete interpreter using Galois connections.
These calculations are done by hand. Our approach offers the ability to
automate the calculation process for a limited set of abstractions for
small-step state machines, where the abstractions are
correct-by-construction through the composition of monad transformers.

\par

We build directly on the work of Abstracting Abstract Machines (AAM) by
\citet{dvanhorn:VanHorn2010Abstracting} and
\citet{dvanhorn:Smaragdakis2011Pick} in our parameterization of abstract
time to achieve call-site and object sensitivity. We follow the AAM
philosophy of instrumenting a concrete semantics \emph{first} and
performing a systematic abstraction \emph{second}. This greatly
simplifies the Galois connection arguments during systematic
abstraction, at the cost of proving the correctness of the instrumented
semantics.

\par

\paragraph{Monadic Abstract Interpreters}

\par

\citeauthor{dvanhorn:Sergey2013Monadic} first introduced the concept of
writing abstract interpreters in monadic style in \emph{Monadic Abstract
Interpreters} (MAI) \cite{dvanhorn:Sergey2013Monadic}, where variations
in analysis are also recovered through monads.

\par

In MAI, the framework's interface is based on \emph{denotation
functions} for every syntactic form of the language. The denotation
functions in MAI are language-specific and specialized to their example
language. MAI uses a single monad stack fixed to the denotation function
interface: state on top of list. New analyses are achieved through
multiple denotation functions into this single monad. Analyses in MAI
are all fixed to be path-sensitive, and the methodology for
incorporating other path or flow properties is to surgically instrument
the execution of the analysis with a custom Galois connection. Lastly,
the framework provides no reasoning principles or proofs of soundness
for the resulting analysis. A user of MAI must inline the definitions of
each analysis and prove each implementation correct from scratch.

\par

Our framework is based on state and nondeterminism \emph{monadic
effects}. This interface comes equipped with laws, allowing one to
verify the correctness of a monadic interpreter independent of a
particular monad. State and nondeterminism monadic effects capture
arbitrary small-step relational semantics, and are language independent.
Because we place the monadic interpreter behind an interface of effects
with laws, we are able to introduce language-independent monads which
capture flow-sensitivity and flow-insensitivity, and we show how to
compose these features with other analysis design choices. The monadic
effect interface also allows us to separate the monad from the abstract
domain. Finally, our framework is compositional through the use of monad
transformers, and constructs execution engines and end-to-end soundness
proofs for free.

\par

\paragraph{Widening for Control-Flow}

\par

\citeauthor{dvanhorn:Hardekopf2014Widening} also introduce a unifying
account of control flow properties in \emph{Widening for Control-Flow}
(WCF) \cite{dvanhorn:Hardekopf2014Widening}, accounting for path, flow
and call-site sensitivities . WCF achieves this through an
instrumentation of the abstract machine's state space which is allowed
to track arbitrary contextual information, up to the path-history of the
entire execution. WCF also develops a modular proof framework, proving
the bulk of soundness proofs for each instantiation of the
instrumentation at once.

\par

Our work achieves similar goals, although isolating path and flow
sensitivity is not our primary objective. While WCF is based on a
language-dependent instrumentation of the semantics, we achieve
variations in path and flow sensitivity by modifying control properties
of the interpreter through language-independent monads.

\par

Particular strengths of WCF are the wide range of choices for
control-flow sensitivity which are shown to be implementable within the
design, and the modular proof framework. For example, WCF is able to
also account for call-site sensitivity through their design; we must
account for call-site sensitivity through a different mechanism.

\par

Particular strengths of our work is the understanding of path and flow
sensitivity not through instrumentation but through
semantics-independent control properties of the interpreter, and also a
modular proof framework, although modular in a different sense from WCF.
We also show how to compose different path and flow sensitivity choices
for independent components of the state space, like a flow-sensitive
data-store and path-sensitive stack-store, for example.

\par

\section{Conclusion}\label{conclusion}

\par

We have shown that \emph{Galois transformers}, monad transformers that
transport Galois connections and mappings to an executable transition
system, are effective, language-independent building blocks for
constructing program analyzers, and form the basis of a modular,
reusable and composable metatheory for program analysis.

\par

In the end, we hope language independent characterizations of analysis
ingredients will both facilitate the systematic construction of program
analyses and bridge the gap between various communities which often work
in isolation.

\par

\acks

\par

This material is partially based on research sponsored by DARPA under
agreements number AFRL FA8750-15-2-0092 and FA8750-12-2-0106 and by NSF
under CAREER grant 1350344. The U.S. Government is authorized to
reproduce and distribute reprints for Governmental purposes
notwithstanding any copyright notation thereon.

\par

\bibliographystyle{abbrvnat}\bibliography{dvanhorn,davdar}

\par

\appendix

\par

\section{Proofs}\label{proofs}

\par

\subsection{ Lemma \ref{are-galois-transformers} [ Galois Transformers ]
(Section~\ref{galois-transformers-1}) }

\par

\paragraph{State}

\emph{$S ^t  \lbrack s \rbrack $ is a Galois transformer.\quad} Recall
the definition of $S ^t  \lbrack s \rbrack $ and
$ \Pi  ^{ S ^t  }  \lbrack s \rbrack $:

\begin{alignat*}{2}
S ^t  \lbrack s \rbrack (m)(A) &  \coloneqq  s  \rightarrow  m(A  \times  s) \\
 \Pi  ^{ S ^t  }  \lbrack s \rbrack ( \Sigma )(A) &  \coloneqq   \Sigma (A  \times  s)  \rightarrow   \Sigma (A  \times  s)
\end{alignat*}

\par

\noindent
\emph{State Property (1):\quad} The action $S ^t  \lbrack s \rbrack $ on
functions:

\begin{align*}
&\hspace{0em} S ^t  \lbrack s \rbrack  : (A  \rightarrow  m(B))  \rightarrow  A  \rightarrow  S ^t  \lbrack s \rbrack (m)(B) \\
&\hspace{0em} S ^t  \lbrack s \rbrack (f)(x)(s)  \coloneqq  y  \leftarrow  ^m  f(x)  \;;\;   \ensuremath{\itop{return}} ^m  (y,s)
\end{align*}

\par

\noindent
To transport Galois connections, we assume a Galois connection
$A  \rightarrow  m _1 (B)  \galois{\alpha^{m}}{\gamma^{m}}  A  \rightarrow  m _2 (B)$
and define $ \alpha $ and $ \gamma $:

\begin{align*}
&\hspace{0em}  \alpha  : (A  \rightarrow  S ^t  \lbrack s \rbrack (m _1 )(B))  \rightarrow  A  \rightarrow  S ^t  \lbrack s \rbrack (m _2 )(B) \\
&\hspace{0em}  \alpha (f)(x)(s)  \coloneqq   \alpha  ^m ( \lambda (x,s). f(x)(s))(x,s) \\
&\hspace{0em}  \gamma  : (A  \rightarrow  S ^t  \lbrack s \rbrack (m _2 )(B))  \rightarrow  A  \rightarrow  S ^t  \lbrack s \rbrack (m _1 )(B) \\
&\hspace{0em}  \gamma (f)(x)(s)  \coloneqq   \gamma  ^m ( \lambda (x,s). f(x)(s))(x,s)
\end{align*}

\par

\noindent
$ \alpha $ and $ \gamma $ are monotonic by inspection, and extensive and
reductive:

\begin{align*}
&\hspace{0em} extensive :  \forall  fxs, f(x)(s)  \sqsubseteq   \gamma ( \alpha (f))(x)(s) \\
&\hspace{0em}  \gamma ( \alpha (f))(x)(s) \\
&\hspace{0em} =  \gamma  ^m ( \lambda (x,s).  \alpha  ^m ( \lambda (x,s). f(x)(s))(x,s))(x,s) \\
&\hspace{2em}  \Lbag\text{\footnotesize{}  definition of  \(  \alpha  \)  and  \(  \gamma  \)   }\Rbag  \\
&\hspace{0em} =  \gamma  ^m ( \alpha  ^m ( \lambda (x,s). f(x)(s)))(x,s)  \hspace{1em}   \Lbag\text{\footnotesize{}   \(  \eta  \) -reduction  }\Rbag  \\
&\hspace{0em}  \sqsupseteq  ( \lambda (x,s). f(x)(s))(x,s)  \hspace{1em}   \Lbag\text{\footnotesize{}   \(  \gamma  ^m   \circ   \alpha  ^m  \)  extensive  }\Rbag  \\
&\hspace{0em} = f(x)(s)  \hspace{1em}   \Lbag\text{\footnotesize{}   \(  \beta  \) -reduction  }\Rbag   \hspace{1em}  \hspace{1em}   \blacksquare  \\
&\hspace{0em} reductive :  \forall  fxs,  \alpha ( \gamma (f))(x)(s)  \sqsubseteq  f(x)(s) \\
&\hspace{0em}  \alpha ( \gamma (f))(x)(s) \\
&\hspace{0em} =  \alpha  ^m ( \lambda (x,s).  \gamma  ^m ( \lambda (x,s). f(x)(s))(x,s))(x,s) \\
&\hspace{2em}  \Lbag\text{\footnotesize{}  definition of  \(  \alpha  \)  and  \(  \gamma  \)   }\Rbag  \\
&\hspace{0em} =  \alpha  ^m ( \gamma  ^m ( \lambda (x,s). f(x)(s))(x,s))(x,s)  \hspace{1em}   \Lbag\text{\footnotesize{}   \(  \eta  \) -reduction  }\Rbag  \\
&\hspace{0em}  \sqsubseteq  ( \lambda (x,s). f(x)(s))(x,s)  \hspace{1em}   \Lbag\text{\footnotesize{}   \(  \alpha  ^m   \circ   \gamma  ^m  \)  reductive  }\Rbag  \\
&\hspace{0em} = f(x)(s)  \hspace{1em}   \Lbag\text{\footnotesize{}   \(  \beta  \) -reduction  }\Rbag   \hspace{1em}  \hspace{1em}   \blacksquare 
\end{align*}

\par

\noindent
Finally, Property (1) commutes, assuming that the Galois connection
$A  \rightarrow  m _1 (B)  \galois{\alpha^{m}}{\gamma^{m}}  A  \rightarrow  m _2 (B)$
is homomorphic:

\begin{align*}
&\hspace{0em} goal: S ^t  \lbrack s \rbrack  \lbrack m _2  \rbrack ( \alpha  ^m (f))(x)(s) =  \alpha (S ^t  \lbrack s \rbrack  \lbrack m _1  \rbrack (f))(x)(s) \\
&\hspace{0em}  \alpha (S ^t  \lbrack s \rbrack  \lbrack m _1  \rbrack (f))(x)(s) \\
&\hspace{0em} =  \alpha  ^m ( \lambda (x,s). y  \leftarrow  ^{ m _1  }  f(x)  \;;\;   \ensuremath{\itop{return}} ^{  m _1  } (y,s))(s,x) \\
&\hspace{2em}  \Lbag\text{\footnotesize{}  definition of  \(  \alpha  \)  and  \( S ^t  \lbrack s \rbrack  \lbrack m _1  \rbrack  \)   }\Rbag  \\
&\hspace{0em} = ( \lambda (x,s). y  \leftarrow  ^{ m _1  }   \alpha  ^m (f)(x)  \;;\;   \ensuremath{\itop{return}} ^{  m _2  } (y,s))(s,x) \\
&\hspace{2em}  \Lbag\text{\footnotesize{}   \(  \alpha  ^m  \)  homomorphic on  \(  \ensuremath{\itop{bind}} ^{  m _1  }  \)  and  \(  \ensuremath{\itop{return}} ^{  m _1  }  \)   }\Rbag  \\
&\hspace{0em} = y  \leftarrow  ^{ m _2  }   \alpha  ^m (f)(x)  \;;\;   \ensuremath{\itop{return}} ^{  m _2  } (y,s)  \hspace{1em}   \Lbag\text{\footnotesize{}   \(  \beta  \) -reduction  }\Rbag  \\
&\hspace{0em} = S ^t  \lbrack s \rbrack  \lbrack m _2  \rbrack ( \alpha  ^m (f))(s)(x)  \hspace{1em}   \Lbag\text{\footnotesize{}  definition of  \( S ^t  \lbrack s \rbrack  \)   }\Rbag   \hspace{1em}  \hspace{1em}   \blacksquare 
\end{align*}

\par

\noindent
\emph{State Property (2):\quad} The action
$ \Pi  ^{ S ^t  }  \lbrack s \rbrack $ on functions uses the mapping to
monadic functions defined in Property (3):

\begin{align*}
&\hspace{0em}  \Pi  ^{ S ^t  }  \lbrack s \rbrack  : ( \Sigma (A)  \rightarrow   \Sigma (B))  \rightarrow   \Pi  ^{ S ^t  }  \lbrack s \rbrack ( \Sigma )(A)  \rightarrow   \Pi  ^{ S ^t  }  \lbrack s \rbrack ( \Sigma )(B) \\
&\hspace{0em}  \Pi  ^{ S ^t  }  \lbrack s \rbrack (f)( \varsigma )  \coloneqq   \gamma  ^{  \Sigma  \leftrightarrow m } (S ^t  \lbrack s \rbrack ( \alpha  ^{  \Sigma  \leftrightarrow m } (f)))( \varsigma )
\end{align*}

\par

\noindent
To transport Galois connections, we assume a Galois connection
$ \Sigma  _1 (A)  \rightarrow   \Sigma  _1 (B)  \galois{\alpha^{\Sigma}}{\gamma^{\Sigma}}   \Sigma  _2 (A)  \rightarrow   \Sigma  _2 (B)$
and define $ \alpha $ and $ \gamma $ as instantiations of
$ \alpha  ^{  \Sigma  } $ and $ \gamma  ^{  \Sigma  } $:

\begin{align*}
&\hspace{0em}  \alpha  : ( \Pi  ^{ S ^t  }  \lbrack s \rbrack ( \Sigma  _1 )(A)  \rightarrow   \Pi  ^{ S ^t  }  \lbrack s \rbrack ( \Sigma  _1 )(B))  \\
&\hspace{1em}  \rightarrow   \Pi  ^{ S ^t  }  \lbrack s \rbrack ( \Sigma  _2 )(A)  \rightarrow   \Pi  ^{ S ^t  }  \lbrack s \rbrack ( \Sigma  _2 )(B) \\
&\hspace{0em}  \gamma  : ( \Pi  ^{ S ^t  }  \lbrack s \rbrack ( \Sigma  _2 )(A)  \rightarrow   \Pi  ^{ S ^t  }  \lbrack s \rbrack ( \Sigma  _2 )(B))  \\
&\hspace{1em}  \rightarrow   \Pi  ^{ S ^t  }  \lbrack s \rbrack ( \Sigma  _1 )(A)  \rightarrow   \Pi  ^{ S ^t  }  \lbrack s \rbrack ( \Sigma  _1 )(B) \\
&\hspace{0em}  \gamma (f)( \varsigma )  \coloneqq   \gamma  ^{  \Sigma  } (f)( \varsigma )  \hspace{1em}  \hspace{1em}   \alpha (f)( \varsigma )  \coloneqq   \alpha  ^{  \Sigma  } (f)( \varsigma )
\end{align*}

\par

\noindent
Monotonicity, reductive and extensive properties carry over by
definition. Finally, Property (2) commutes, assuming that
$ \alpha  ^{  \Sigma  } $ and $ \alpha  ^m $ commute with both
$ \gamma  ^{  \Sigma  \leftrightarrow m } $ and
$ \alpha  ^{  \Sigma  \leftrightarrow m } $:

\begin{align*}
&\hspace{0em} goal:  \Pi  ^{ S ^t  }  \lbrack s \rbrack  \lbrack  \Sigma  _2  \rbrack ( \alpha  ^{  \Sigma  } (f))( \varsigma ) =  \alpha  ^{  \Sigma  } ( \Pi  ^{ S ^t  }  \lbrack s \rbrack  \lbrack  \Sigma  _1  \rbrack (f))( \varsigma ) \\
&\hspace{0em}  \alpha  ^{  \Sigma  } ( \Pi  ^{ S ^t  }  \lbrack s \rbrack  \lbrack  \Sigma  _1  \rbrack (f)( \varsigma ) \\
&\hspace{0em} =  \alpha  ^{  \Sigma  } ( \gamma  ^{  \Sigma  \leftrightarrow m } (S ^t  \lbrack s \rbrack ( \alpha  ^{  \Sigma  \leftrightarrow m } (f))))( \varsigma )  \hspace{1em}   \Lbag\text{\footnotesize{}  definition of  \(  \Pi  ^{ S ^t  }  \lbrack s \rbrack  \lbrack  \Sigma  _1  \rbrack  \)   }\Rbag  \\
&\hspace{0em} =  \alpha  ^{  \Sigma  } ( \gamma  ^{  \Sigma  \leftrightarrow m } ( \lambda (x)(s). y  \leftarrow  ^{ m _1  }   \alpha  ^{  \Sigma  \leftrightarrow m } (f)(x)  \;;\;   \\
&\hspace{1em}  \ensuremath{\itop{return}} ^{  m _1  } (y,s)))( \varsigma )  \hspace{1em}   \Lbag\text{\footnotesize{}  definition of  \( S ^t  \lbrack s \rbrack  \)   }\Rbag  \\
&\hspace{0em} =  \gamma  ^{  \Sigma  \leftrightarrow m } ( \alpha  ^m ( \lambda (x)(s). y  \leftarrow  ^{ m _1  }   \alpha  ^{  \Sigma  \leftrightarrow m } (f)(x)  \;;\;   \\
&\hspace{1em}  \ensuremath{\itop{return}} ^{  m _1  } (y,s)))( \varsigma )  \hspace{1em}   \Lbag\text{\footnotesize{}   \(  \alpha  ^{  \Sigma  }  \)  and  \(  \gamma  ^{  \Sigma  \leftrightarrow m }  \)  commute  }\Rbag  \\
&\hspace{0em} =  \gamma  ^{  \Sigma  \leftrightarrow m } ( \lambda (x)(s). y  \leftarrow  ^{ m _2  }   \alpha  ^m ( \alpha  ^{  \Sigma  \leftrightarrow m } (f))(x)  \;;\;   \\
&\hspace{1em}  \ensuremath{\itop{return}} ^{  m _2  } (y,s))( \varsigma )  \hspace{1em}   \Lbag\text{\footnotesize{}   \(  \alpha  ^m  \)  homomorphic  }\Rbag  \\
&\hspace{0em} =  \gamma  ^{  \Sigma  \leftrightarrow m } ( \lambda (x)(s). y  \leftarrow  ^{ m _2  }   \alpha  ^{  \Sigma  \leftrightarrow m } ( \alpha  ^{  \Sigma  } (f))(x)  \;;\;  \\
&\hspace{1em}  \ensuremath{\itop{return}} ^{  m _2  } (y,s))( \varsigma )  \hspace{1em}   \Lbag\text{\footnotesize{}   \(  \alpha  ^m  \)  and  \(  \alpha  ^{  \Sigma  \leftrightarrow m }  \)  commute  }\Rbag  \\
&\hspace{0em} =  \gamma  ^{  \Sigma  \leftrightarrow m } (S ^t  \lbrack s \rbrack ( \alpha  ^{  \Sigma  \leftrightarrow m } ( \alpha  ^{  \Sigma  } (f))))( \varsigma )  \hspace{1em}   \Lbag\text{\footnotesize{}  definition of  \( S ^t  \lbrack s \rbrack  \)   }\Rbag  \\
&\hspace{0em} =  \Pi  ^{ S ^t  }  \lbrack s \rbrack  \lbrack  \Sigma  _2  \rbrack ( \alpha  ^{  \Sigma  } (f))( \varsigma )  \hspace{1em}   \Lbag\text{\footnotesize{}  definition of  \(  \Pi  ^{ S ^t  }  \lbrack s \rbrack  \lbrack  \Sigma  _2  \rbrack  \)   }\Rbag   \hspace{1em}  \hspace{1em}   \blacksquare 
\end{align*}

\par

\noindent
\emph{State Property (3):\quad} Assume a Galois connection
$ \Sigma (A)  \rightarrow   \Sigma (B)  \galois{\alpha^{\Sigma\leftrightarrow m}}{\gamma^{\Sigma\leftrightarrow m}}  A  \rightarrow  m(B)$.
The Galois connection between $S ^t  \lbrack s \rbrack (m)$ and
$ \Pi  ^{ S ^t  }  \lbrack s \rbrack ( \Sigma )$ is defined:

\begin{align*}
&\hspace{0em}  \alpha  : ( \Pi  ^{ S ^t  }  \lbrack s \rbrack ( \Sigma )(A)  \rightarrow   \Pi  ^{ S ^t  }  \lbrack s \rbrack ( \Sigma )(B))  \rightarrow  A  \rightarrow  S ^t  \lbrack s \rbrack (m)(B) \\
&\hspace{0em}  \alpha (f)(x)(s)  \coloneqq   \alpha  ^{  \Sigma  \leftrightarrow m } (f)(x,s) \\
&\hspace{0em}  \gamma  : (A  \rightarrow  S ^t  \lbrack s \rbrack (m)(B))  \rightarrow   \Pi  ^{ S ^t  }  \lbrack s \rbrack ( \Sigma )(A)  \rightarrow   \Pi  ^{ S ^t  }  \lbrack s \rbrack ( \Sigma )(B) \\
&\hspace{0em}  \gamma (f)( \varsigma )  \coloneqq   \gamma  ^{  \Sigma  \leftrightarrow m } ( \lambda (x,s)  \rightarrow  f(x)(s))( \varsigma )
\end{align*}

\par

\noindent
$ \alpha $ and $ \gamma $ are monotonic by inspection, and extensive and
reductive:

\begin{align*}
&\hspace{0em} extensive:  \forall  f  \varsigma , f( \varsigma )  \sqsubseteq   \gamma ( \alpha (f))( \varsigma ) \\
&\hspace{0em}  \gamma ( \alpha (f))( \varsigma ) \\
&\hspace{0em} =  \gamma  ^{  \Sigma  \leftrightarrow m } ( \lambda (x,s)  \rightarrow   \alpha  ^{  \Sigma  \leftrightarrow m } (f)(x,s))( \varsigma )  \hspace{1em}   \Lbag\text{\footnotesize{}  definition of  \(  \alpha  \)  and  \(  \gamma  \)   }\Rbag  \\
&\hspace{0em} =  \gamma  ^{  \Sigma  \leftrightarrow m } ( \alpha  ^{  \Sigma  \leftrightarrow m } (f))( \varsigma )  \hspace{1em}   \Lbag\text{\footnotesize{}   \(  \eta  \) -reduction  }\Rbag  \\
&\hspace{0em}  \sqsupseteq  f( \varsigma )  \hspace{1em}   \Lbag\text{\footnotesize{}   \(  \gamma  ^{  \Sigma  \leftrightarrow m }   \circ   \alpha  ^{  \Sigma  \leftrightarrow m }  \)  extensive  }\Rbag   \hspace{1em}  \hspace{1em}   \blacksquare  \\
&\hspace{0em} reductive:  \forall  f x s,  \alpha ( \gamma (f))(x)(s)  \sqsubseteq  f(x)(s) \\
&\hspace{0em}  \alpha ( \gamma (f))(x)(s) \\
&\hspace{0em} =  \alpha  ^{  \Sigma  \leftrightarrow m } ( \gamma  ^{  \Sigma  \leftrightarrow m } ( \lambda (x,s)  \rightarrow  f(x)(s)))(x,s)  \\
&\hspace{2em}  \Lbag\text{\footnotesize{}  definition of  \(  \alpha  \)  and  \(  \gamma  \)   }\Rbag  \\
&\hspace{0em}  \sqsubseteq  ( \lambda (x,s)  \rightarrow  f(x)(s))(x,s)  \hspace{1em}   \Lbag\text{\footnotesize{}   \(  \alpha  ^{  \Sigma  \leftrightarrow m }   \circ   \gamma  ^{  \Sigma  \leftrightarrow m }  \)  reductive  }\Rbag  \\
&\hspace{0em} = f(x)(s)  \hspace{1em}   \Lbag\text{\footnotesize{}   \(  \beta  \) -reduction  }\Rbag   \hspace{1em}  \hspace{1em}   \blacksquare 
\end{align*}

\par

\noindent
Finally, Property (3) commutes:

\begin{align*}
&\hspace{0em} goal:  \Pi  ^{ S ^t  }  \lbrack s \rbrack  \lbrack  \Sigma  \rbrack ( \gamma  ^{  \Sigma  \leftrightarrow m } (f))( \varsigma )  \sqsubseteq   \gamma (S ^t  \lbrack s \rbrack (f))( \varsigma ) \\
&\hspace{0em}  \Pi  ^{ S ^t  }  \lbrack s \rbrack  \lbrack  \Sigma  \rbrack ( \gamma  ^{  \Sigma  \leftrightarrow m } (f))( \varsigma ) \\
&\hspace{0em} =  \gamma  ^{  \Sigma  \leftrightarrow m } ( \lambda (x,s)  \rightarrow  S ^t  \lbrack s \rbrack ( \alpha  ^{  \Sigma  \leftrightarrow m } ( \gamma  ^{  \Sigma  \leftrightarrow m } (f)))(x)(s))( \varsigma )  \\
&\hspace{2em}  \Lbag\text{\footnotesize{}  definition of  \(  \Pi  ^{ S ^t  }  \lbrack s \rbrack  \lbrack  \Sigma  \rbrack  \)   }\Rbag  \\
&\hspace{0em}  \sqsubseteq   \gamma  ^{  \Sigma  \leftrightarrow m } ( \lambda (x,s)  \rightarrow  S ^t  \lbrack s \rbrack (f)(x)(s))( \varsigma )  \\
&\hspace{2em}  \Lbag\text{\footnotesize{}   \(  \alpha  ^{  \Sigma  \leftrightarrow m }   \circ   \gamma  ^{  \Sigma  \leftrightarrow m }  \)  reductive  }\Rbag  \\
&\hspace{0em} =  \gamma (S ^t  \lbrack s \rbrack (f))( \varsigma )  \hspace{1em}   \Lbag\text{\footnotesize{}  definition of  \(  \gamma  \)   }\Rbag   \hspace{1em}  \hspace{1em}   \blacksquare 
\end{align*}

\par

\paragraph{Nondeterminism}

\emph{$ \mathcal{P}  ^t $ is a Galois transformer.\quad} Recall the
definition of $ \mathcal{P}  ^t $ and $ \Pi  ^{  \mathcal{P}  ^t  } $:

\begin{align*}
&\hspace{0em}  \mathcal{P}  ^t (m)(A)  \coloneqq  m( \mathcal{P} (A)) \hspace{1em}  \hspace{1em}  \Pi  ^{  \mathcal{P}  ^t  } ( \Sigma )(A)  \coloneqq   \Sigma ( \mathcal{P} (A))
\end{align*}

\par

\noindent
\emph{Nondeterminism Property (1):\quad} The action $ \mathcal{P}  ^t $
on functions:

\begin{align*}
&\hspace{0em}  \mathcal{P}  ^t  : (A  \rightarrow  m(B))  \rightarrow  A  \rightarrow   \mathcal{P}  ^t (m)(B) \\
&\hspace{0em}  \mathcal{P}  ^t (f)(x)  \coloneqq  y  \leftarrow  ^m  f(x)  \;;\;   \ensuremath{\itop{return}} ^m  ( \{ y \} )
\end{align*}

\par

\noindent
To transport Galois connections, we assume a Galois connection
$A  \rightarrow  m _1 (B)  \galois{\alpha^{m}}{\gamma^{m}}  A  \rightarrow  m _2 (B)$
define $ \alpha $ and $ \gamma $:

\begin{align*}
&\hspace{0em}  \alpha  : (A  \rightarrow   \mathcal{P} (m _1 )(B))  \rightarrow  A  \rightarrow   \mathcal{P} (m _2 )(B) \\
&\hspace{0em}  \alpha (f)(x) =  \alpha  ^m ( \lambda ( \{ x _1 ..x _n  \} ). f(x _1 )  \sqcup  ^{ m _1  }  ..  \sqcup  ^{ m _1  }  f(x _n ))( \{ x \} ) \\
&\hspace{0em}  \gamma  : (A  \rightarrow   \mathcal{P} (m _2 )(B))  \rightarrow  A  \rightarrow   \mathcal{P} (m _1 )(B) \\
&\hspace{0em}  \gamma (f)(x) =  \gamma  ^m ( \lambda ( \{ x _1 ..x _n  \} ). f(x _1 )  \sqcup  ^{ m _2  }  ..  \sqcup  ^{ m _2  }  f(x _n ))( \{ x \} )
\end{align*}

\par

\noindent
$ \alpha $ and $ \gamma $ are monotonic by inspection, and extensive and
reductive:

\begin{align*}
&\hspace{0em} extensive :  \forall  f x, f(x)  \sqsubseteq   \gamma ( \alpha (f))(x) \\
&\hspace{0em}  \gamma ( \alpha (f))(x) \\
&\hspace{0em} =  \gamma  ^m ( \lambda ( \{ x _1 ..x _n  \} ).  \\
&\hspace{1em}  \alpha  ^m ( \lambda ( \{ x _1 ..x _n  \} ). f(x _1 )  \sqcup  ^{ m _1  }  ..  \sqcup  ^{ m _1  }  f(x _n ))( \{ x _1  \} ) \\
&\hspace{1em}  \sqcup  ^{ m _2  }  ..  \sqcup  ^{ m _2  }  \\
&\hspace{1em}  \alpha  ^m ( \lambda ( \{ x _1 ..x _n  \} ). f(x _1 )  \sqcup  ^{ m _1  }  ..  \sqcup  ^{ m _1  }  f(x _n ))( \{ x _n  \} ))( \{ x \} ) \\
&\hspace{2em}  \Lbag\text{\footnotesize{}  definition of  \(  \alpha  \)  and  \(  \gamma  \)   }\Rbag  \\
&\hspace{0em} =  \gamma  ^m ( \lambda ( \{ x _1 ..x _n  \} ).  \\
&\hspace{1em} ( \{ x _1 ..x _n  \}   \leftarrow  ^{ m _2  }   \ensuremath{\itop{return}} ^{  m _2  } ( \{ x _1  \} )  \;;\;   \alpha  ^m ( \lambda ( \{ x _1 ..x _n  \} ).  \\
&\hspace{2em} f(x _1 )  \sqcup  ^{ m _1  }  ..  \sqcup  ^{ m _1  }  f(x _n ))( \{ x _1 ..x _n  \} ))  \\
&\hspace{1em}  \sqcup  ^{ m _2  }  ..  \sqcup  ^{ m _2  }   \\
&\hspace{1em} ( \{ x _1 ..x _n  \}   \leftarrow  ^{ m _2  }   \ensuremath{\itop{return}} ^{  m _2  } ( \{ x _n  \} )  \;;\;   \alpha  ^m ( \lambda ( \{ x _1 ..x _n  \} ).  \\
&\hspace{2em} f(x _1 )  \sqcup  ^{ m _1  }  ..  \sqcup  ^{ m _1  }  f(x _n ))( \{ x _1 ..x _n  \} )))( \{ x \} ) \\
&\hspace{2em}  \Lbag\text{\footnotesize{}   \ensuremath{\itop{left-unit}}  of  \( m _2  \)   }\Rbag  \\
&\hspace{0em}  \sqsupseteq   \gamma  ^m ( \lambda ( \{ x _1 ..x _n  \} ). \\
&\hspace{1em} ( \{ x _1 ..x _n  \}   \leftarrow  ^{ m _2  }   \alpha  ^m ( \gamma  ^m  (\ensuremath{\itop{return}} ^{  m _2  } ( \{ x _1  \} )))  \;;\;   \\
&\hspace{2em}  \alpha  ^m ( \lambda ( \{ x _1 ..x _n  \} ). f(x _1 )  \sqcup  ^{ m _1  }  ..  \sqcup  ^{ m _1  }  f(x _n ))( \{ x _1 ..x _n  \} )) \\
&\hspace{1em}  \sqcup  ^{ m _2  }  ..  \sqcup  ^{ m _2  }  \\
&\hspace{1em} ( \{ x _1 ..x _n  \}   \leftarrow  ^{ m _2  }   \alpha  ^m ( \gamma  ^m  (\ensuremath{\itop{return}} ^{  m _2  } ( \{ x _n  \} )))  \;;\;   \\
&\hspace{2em}  \alpha  ^m ( \lambda ( \{ x _1 ..x _n  \} ). f(x _1 )  \sqcup  ^{ m _1  }  ..  \sqcup  ^{ m _1  }  f(x _n ))( \{ x _1 ..x _n  \} ))) \\
&\hspace{1em} ( \{ x \} )  \hspace{1em}   \Lbag\text{\footnotesize{}   \(  \alpha  ^m   \circ   \gamma  ^m  \)  reductive  }\Rbag 
\end{align*}

\begin{align*}
&\hspace{0em} =  \gamma  ^m ( \lambda ( \{ x _1 ..x _n  \} ).  \\
&\hspace{1em} ( \alpha  ^m ( \{ x _1 ..x _n  \}   \leftarrow  ^{ m _1  }   \ensuremath{\itop{return}} ^{  m _1  } ( \{ x _1  \} )  \;;\;   \\
&\hspace{2em} f(x _1 )  \sqcup  ^{ m _1  }  ..  \sqcup  ^{ m _1  }  f(x _n )))  \sqcup  ^{ m _2  }  ..  \sqcup  ^{ m _2  }  \\
&\hspace{1em} ( \alpha  ^m ( \{ x _1 ..x _n  \}   \leftarrow  ^{ m _1  }   \ensuremath{\itop{return}} ^{  m _1  } ( \{ x _n  \} )  \;;\;   \\
&\hspace{2em} f(x _1 )  \sqcup  ^{ m _1  }  ..  \sqcup  ^{ m _1  }  f(x _n ))))( \{ x \} ) \\
&\hspace{2em}  \Lbag\text{\footnotesize{}   \(  \alpha  ^m  \)  and  \(  \gamma  ^m  \)  homomorphic on  \(  \ensuremath{\itop{bind}} ^{  m _2  }  \)  and  \(  \ensuremath{\itop{return}} ^{  m _2  }  \)   }\Rbag  \\
&\hspace{0em} =  \gamma  ^m ( \alpha  ^m ( \lambda ( \{ x _1 ..x _n  \} ).  \{ x _1 ..x _n  \}   \leftarrow   \ensuremath{\itop{return}} ^{  m _1  } ( \{ x _1 ..x _n  \} )  \;;\;   \\
&\hspace{1em} f(x _1 )  \sqcup  ^{ m _1  }  ..  \sqcup  ^{ m _1  }  f(x _n )))( \{ x \} ) \\
&\hspace{2em}  \Lbag\text{\footnotesize{}  join-semilattice functorality of  \( m \)   }\Rbag  \\
&\hspace{0em}  \sqsupseteq   \{ x _1 ..x _n  \}   \leftarrow   \ensuremath{\itop{return}} ^{  m _1  } ( \{ x \} )  \;;\;  f(x _1 )  \sqcup  ^{ m _1  }  ..  \sqcup  ^{ m _1  }  f(x _n ) \\
&\hspace{2em}  \Lbag\text{\footnotesize{}   \(  \gamma  ^m   \circ   \alpha  ^m  \)  extensive  }\Rbag  \\
&\hspace{0em} = f(x)  \hspace{1em}   \Lbag\text{\footnotesize{}   \ensuremath{\itop{left-unit}}  of  \( m \)   }\Rbag   \hspace{1em}  \hspace{1em}   \blacksquare  \\
&\hspace{0em} reductive :  \forall  f x,  \alpha ( \gamma (f))(x)  \sqsubseteq  f(x) \\
&\hspace{0em}  \alpha ( \gamma (f))(x) \\
&\hspace{0em} =  \alpha  ^m ( \lambda ( \{ x _1 ..x _n  \} ).  \\
&\hspace{1em}  \gamma  ^m ( \lambda ( \{ x _1 ..x _n  \} ). f(x _1 )  \sqcup  ^{ m _2  }  ..  \sqcup  ^{ m _2  }  f(x _n ))( \{ x _1  \} ) \\
&\hspace{1em}  \sqcup  ^{ m _1  }  ..  \sqcup  ^{ m _1  }  \\
&\hspace{1em}  \gamma  ^m ( \lambda ( \{ x _1 ..x _n  \} ). f(x _1 )  \sqcup  ^{ m _2  }  ..  \sqcup  ^{ m _2  }  f(x _n ))( \{ x _n  \} ))( \{ x \} ) \\
&\hspace{2em}  \Lbag\text{\footnotesize{}  definition of  \(  \alpha  \)  and  \(  \gamma  \)   }\Rbag  \\
&\hspace{0em} =  \alpha  ^m ( \lambda ( \{ x _1 ..x _n  \} ). \\
&\hspace{1em} ( \{ x _1 ..x _n  \}   \leftarrow  ^{ m _1  }   \ensuremath{\itop{return}} ^{  m _1  } ( \{ x _1  \} )  \;;\;   \gamma  ^m ( \lambda ( \{ x _1 ..x _n  \} ).  \\
&\hspace{2em} f(x _1 )  \sqcup  ^{ m _2  }  ..  \sqcup  ^{ m _2  }  f(x _n ))( \{ x _1 ..x _n  \} )) \\
&\hspace{1em}  \sqcup  ^{ m _1  }  ..  \sqcup  ^{ m _1  }  \\
&\hspace{1em} ( \{ x _1 ..x _n  \}   \leftarrow  ^{ m _1  }   \ensuremath{\itop{return}} ^{  m _1  } ( \{ x _2  \} )  \;;\;   \gamma  ^m ( \lambda ( \{ x _1 ..x _n  \} ).  \\
&\hspace{2em} f(x _1 )  \sqcup  ^{ m _2  }  ..  \sqcup  ^{ m _2  }  f(x _n ))( \{ x _1 ..x _n  \} )))( \{ x \} ) \\
&\hspace{2em}  \Lbag\text{\footnotesize{}   \ensuremath{\itop{left-unit}}  of  \( m _1  \)   }\Rbag  \\
&\hspace{0em}  \sqsubseteq   \alpha  ^m ( \lambda ( \{ x _1 ..x _n  \} ). \\
&\hspace{1em} ( \{ x _1 ..x _n  \}   \leftarrow  ^{ m _1  }   \gamma  ^m ( \alpha  ^m  (\ensuremath{\itop{return}} ^{  m _1  } ( \{ x _1  \} )))  \;;\;   \\
&\hspace{2em}  \gamma  ^m ( \lambda ( \{ x _1 ..x _n  \} ). f(x _1 )  \sqcup  ^{ m _2  }  ..  \sqcup  ^{ m _2  }  f(x _n ))( \{ x _1 ..x _n  \} )) \\
&\hspace{1em}  \sqcup  ^{ m _1  }  ..  \sqcup  ^{ m _1  }  \\
&\hspace{1em} ( \{ x _1 ..x _n  \}   \leftarrow  ^{ m _1  }   \gamma  ^m ( \alpha  ^m  (\ensuremath{\itop{return}} ^{  m _1  } ( \{ x _n  \} )))  \;;\;   \\
&\hspace{2em}  \gamma  ^m ( \lambda ( \{ x _1 ..x _n  \} ). f(x _1 )  \sqcup  ^{ m _2  }  ..  \sqcup  ^{ m _2  }  f(x _n ))( \{ x _1 ..x _n  \} ))) \\
&\hspace{1em} ( \{ x \} )  \hspace{1em}   \Lbag\text{\footnotesize{}   \(  \gamma  ^m   \circ   \alpha  ^m  \)  extensive  }\Rbag  \\
&\hspace{0em} =  \alpha  ^m ( \lambda ( \{ x _1 ..x _n  \} ). \\
&\hspace{1em}  \gamma  ^m ( \{ x _1 ..x _n  \}   \leftarrow  ^{ m _2  }   \ensuremath{\itop{return}} ^{  m _2  } ( \{ x _1  \} )  \;;\;  f(x _1 )  \sqcup  ^{ m _2  }  ..  \sqcup  ^{ m _2  }  f(x _n )) \\
&\hspace{1em}  \sqcup  ^{ m _1  }  ..  \sqcup  ^{ m _1  }  \\
&\hspace{1em}  \gamma  ^m ( \{ x _1 ..x _n  \}   \leftarrow  ^{ m _2  }   \ensuremath{\itop{return}} ^{  m _2  } ( \{ x _1  \} )  \;;\;  f(x _1 )  \sqcup  ^{ m _2  }  ..  \sqcup  ^{ m _2  }  f(x _n ))) \\
&\hspace{1em} ( \{ x \} )  \hspace{1em}   \Lbag\text{\footnotesize{}   \(  \alpha  ^m  and  \gamma  ^m  \)  homomorphic on  \(  \ensuremath{\itop{bind}} ^{  m _1  }  \)  and  \(  \ensuremath{\itop{return}} ^{  m _1  }  \)   }\Rbag  \\
&\hspace{0em} =  \alpha  ^m ( \gamma  ^m ( \lambda  \{ x _1 ..x _n  \} ).  \{ x _1 ..x _n  \}   \leftarrow  ^{ m _2  }   \ensuremath{\itop{return}} ^{  m _2  } ( \{ x _1 ..x _n  \} )  \;;\;   \\
&\hspace{1em} f(x _1 )  \sqcup  ^{ m _2  }  ..  \sqcup  ^{ m _2  }  f(x _n ))( \{ x \} ) \\
&\hspace{2em}  \Lbag\text{\footnotesize{}  join-semilattice functorailty of  \( m \)   }\Rbag  \\
&\hspace{0em}  \sqsubseteq   \{ x _1 ..x _n  \}   \leftarrow  ^{ m _2  }   \ensuremath{\itop{return}} ^{  m _2  } ( \{ x \} )  \;;\;  f(x _1 )  \sqcup  ^{ m _2  }  ..  \sqcup  ^{ m _2  }  f(x _n ) \\
&\hspace{2em}  \Lbag\text{\footnotesize{}   \(  \alpha  ^m   \circ   \gamma  ^m  \)  reductive  }\Rbag  \\
&\hspace{0em} = f(x)  \hspace{1em}   \Lbag\text{\footnotesize{}   \ensuremath{\itop{left-unit}}  of  \( m \)   }\Rbag   \hspace{1em}  \hspace{1em}   \blacksquare 
\end{align*}

\par

\noindent
Finally, Property (1) commutes, assuming that the Galois connection
$A  \rightarrow  m _1 (B)  \galois{\alpha^{m}}{\gamma^{m}}  A  \rightarrow  m _2 (B)$
is homomorphic:

\begin{align*}
&\hspace{0em} goal:  \forall  f s,  \mathcal{P}  ^t  \lbrack m _2  \rbrack ( \alpha  ^m (f))(x) =  \alpha ( \mathcal{P}  ^t  \lbrack m _1  \rbrack (f))(x) \\
&\hspace{0em}  \alpha ( \mathcal{P}  ^t  \lbrack m _1  \rbrack (f))(x) \\
&\hspace{0em} =  \alpha  ^m ( \lambda ( \{ x _1 ..x _n  \} ).  \\
&\hspace{1em} (y  \leftarrow  ^{ m _1  }  f(x _1 )  \;;\;   \ensuremath{\itop{return}} ^{  m _1  } ( \{ y \} ))  \sqcup  ^{ m _1  }  ..  \sqcup  ^{ m _1  }   \\
&\hspace{1em} (y  \leftarrow  ^{ m _1  }  f(x _n )  \;;\;   \ensuremath{\itop{return}} ^{  m _1  } ( \{ y \} )))( \{ x \} ) \\
&\hspace{2em}  \Lbag\text{\footnotesize{}  definition of  \(  \alpha  \)  and  \(  \mathcal{P}  ^t  \lbrack m _1  \rbrack (f) \)   }\Rbag  \\
&\hspace{0em} = y  \leftarrow  ^{ m _2  }   \alpha  ^m (f)(x)  \;;\;   \ensuremath{\itop{return}} ^{  m _2  } ( \{ y \} ) \\
&\hspace{2em}  \Lbag\text{\footnotesize{}  homomorphic on  \(  \ensuremath{\itop{bind}} ^{  m _1  }  \)  and  \(  \ensuremath{\itop{return}} ^{  m _1  }  \)   }\Rbag  \\
&\hspace{0em} =  \mathcal{P}  ^t  \lbrack m _2  \rbrack ( \alpha  ^m (f))(x)  \hspace{1em}   \Lbag\text{\footnotesize{}  definition of  \(  \mathcal{P}  ^t  \lbrack m _2  \rbrack  \)   }\Rbag   \hspace{1em}  \hspace{1em}   \blacksquare 
\end{align*}

\par

\noindent
\emph{Nondeterminism Property (2):\quad} The action
$ \Pi  ^{  \mathcal{P}  ^t  } $ on functions uses the mapping to monadic
functions defined in Property (3):

\begin{align*}
&\hspace{0em}  \Pi  ^{  \mathcal{P}  ^t  }  : ( \Sigma (A)  \rightarrow   \Sigma (B))  \rightarrow   \Pi  ^{  \mathcal{P}  ^t  } ( \Sigma )(A)  \rightarrow   \Pi  ^{  \mathcal{P}  ^t  } ( \Sigma )(B) \\
&\hspace{0em}  \Pi  ^{  \mathcal{P}  ^t  } (f)( \varsigma )  \coloneqq   \gamma  ^{  \Sigma  \leftrightarrow  \gamma  } ( \mathcal{P}  ^t ( \alpha  ^{  \Sigma  \leftrightarrow  \gamma  } (f)))
\end{align*}

\par

\noindent
To transport Galois connections, we assume a Galois connection
$ \Sigma  _1 (A)  \rightarrow   \Sigma  _1 (B)  \galois{\alpha^{\Sigma}}{\gamma^{\Sigma}}   \Sigma  _2 (A)  \rightarrow   \Sigma  _2 (B)$
and define $ \alpha $ and $ \gamma $ as instantiations of
$ \alpha  ^{  \Sigma  } $ and $ \gamma  ^{  \Sigma  } $:

\begin{align*}
&\hspace{0em}  \alpha  : ( \Pi  ^{  \mathcal{P}  ^t  } ( \Sigma  _1 )(A)  \rightarrow   \Pi  ^{  \mathcal{P}  ^t  } ( \Sigma  _1 )(B))  \rightarrow   \Pi  ^{  \mathcal{P}  ^t  } ( \Sigma  _2 )(A)  \rightarrow   \Pi  ^{  \mathcal{P}  ^t  } ( \Sigma  _2 )(B) \\
&\hspace{0em}  \gamma  : ( \Pi  ^{  \mathcal{P}  ^t  } ( \Sigma  _2 )(A)  \rightarrow   \Pi  ^{  \mathcal{P}  ^t  } ( \Sigma  _2 )(B))  \rightarrow   \Pi  ^{  \mathcal{P}  ^t  } ( \Sigma  _1 )(A)  \rightarrow   \Pi  ^{  \mathcal{P}  ^t  } ( \Sigma  _1 )(B) \\
&\hspace{0em}  \hspace{1em}  \hspace{1em}   \alpha (f)( \varsigma )  \coloneqq   \alpha  ^{  \Sigma  } (f)( \varsigma )  \hspace{1em}  \hspace{1em}   \gamma (f)( \varsigma )  \coloneqq   \gamma  ^{  \Sigma  } (f)( \varsigma )
\end{align*}

\par

\noindent
Monotonicity, reductive and extensive properties carry over by
definition. Finally, Property (2) commutes, assuming that
$ \alpha  ^{  \Sigma  } $ and $ \alpha  ^m $ commute with both
$ \gamma  ^{  \Sigma  \leftrightarrow m } $ and
$ \alpha  ^{  \Sigma  \leftrightarrow m } $:

\begin{align*}
&\hspace{0em} goal:  \Pi  ^{  \mathcal{P}  ^t  }  \lbrack  \Sigma  _2  \rbrack ( \alpha  ^{  \Sigma  } (f))( \varsigma ) =  \alpha  ^{  \Sigma  } ( \Pi  ^{  \mathcal{P}  ^t  }  \lbrack  \Sigma  _1  \rbrack (f))( \varsigma ) \\
&\hspace{0em}  \alpha  ^{  \Sigma  } ( \Pi  ^{  \mathcal{P}  ^t  }  \lbrack  \Sigma  _1  \rbrack (f))( \varsigma ) \\
&\hspace{0em} =  \alpha  ^{  \Sigma  } ( \gamma  ^{  \Sigma  \leftrightarrow  \gamma  } ( \mathcal{P}  ^t ( \alpha  ^{  \Sigma  \leftrightarrow  \gamma  } (f))))( \varsigma )  \hspace{1em}   \Lbag\text{\footnotesize{}  definition of  \(  \Pi  ^{  \mathcal{P}  ^t  }  \)   }\Rbag  \\
&\hspace{0em} =  \alpha  ^{  \Sigma  } ( \gamma  ^{  \Sigma  \leftrightarrow  \gamma  } ( \lambda (x). y  \leftarrow  ^{ m _1  }   \alpha  ^{  \Sigma  \leftrightarrow  \gamma  } (f)(x)  \;;\;   \ensuremath{\itop{return}} ^{  m _1  } ( \{ y \} )))( \varsigma ) \\
&\hspace{2em}  \Lbag\text{\footnotesize{}  definition of  \(  \mathcal{P}  ^t  \)   }\Rbag  \\
&\hspace{0em} =  \gamma  ^{  \Sigma  \leftrightarrow  \gamma  } ( \alpha  ^m ( \lambda (x). y  \leftarrow  ^{ m _1  }   \alpha  ^{  \Sigma  \leftrightarrow  \gamma  } (f)(x)  \;;\;   \ensuremath{\itop{return}} ^{  m _1  } ( \{ y \} )))( \varsigma ) \\
&\hspace{2em}  \Lbag\text{\footnotesize{}   \(  \alpha  ^{  \Sigma  }  \)  and  \(  \gamma  ^{  \Sigma  \leftrightarrow  \gamma  }  \)  commute  }\Rbag  \\
&\hspace{0em} =  \gamma  ^{  \Sigma  \leftrightarrow  \gamma  } ( \lambda (x). y  \leftarrow  ^{ m _2  }   \alpha  ^m ( \alpha  ^{  \Sigma  \leftrightarrow  \gamma  } (f))(x)  \;;\;   \ensuremath{\itop{return}} ^{  m _2  } ( \{ y \} ))( \varsigma ) \\
&\hspace{2em}  \Lbag\text{\footnotesize{}   \(  \alpha  ^m  \)  homomorphic on  \(  \ensuremath{\itop{bind}} ^{  m _1  }  \)  and  \(  \ensuremath{\itop{return}} ^{  m _2  }  \)   }\Rbag  \\
&\hspace{0em} =  \gamma  ^{  \Sigma  \leftrightarrow  \gamma  } ( \lambda (x). y  \leftarrow  ^{ m _2  }   \alpha  ^{  \Sigma  \leftrightarrow  \gamma  } ( \alpha  ^{  \Sigma  } (f))(x)  \;;\;   \ensuremath{\itop{return}} ^{  m _2  } ( \{ y \} ))( \varsigma ) \\
&\hspace{2em}  \Lbag\text{\footnotesize{}   \(  \alpha  ^{ m }  \)  and  \(  \alpha  ^{  \Sigma  \leftrightarrow  \gamma  }  \)  commute  }\Rbag  \\
&\hspace{0em} =  \Pi  ^{  \mathcal{P}  ^t  }  \lbrack  \Sigma  _2  \rbrack ( \alpha  ^{  \Sigma  } (f))( \varsigma )  \hspace{1em}   \Lbag\text{\footnotesize{}  definition of  \(  \Pi  ^{  \mathcal{P}  ^t  }  \lbrack  \Sigma  _2  \rbrack  \)  and  \(  \alpha  ^{  \Sigma  }  \)   }\Rbag   \hspace{1em}  \hspace{1em}   \blacksquare 
\end{align*}

\par

\noindent
\emph{Nondeterminism Property (3):\quad} Assume a Galois connection
$ \Sigma (A)  \rightarrow   \Sigma (B)  \galois{\alpha^{\Sigma\leftrightarrow m}}{\gamma^{\Sigma\leftrightarrow m}}  A  \rightarrow  m(B)$.
The Galois connection between $ \mathcal{P}  ^t (m)$ and
$ \Pi  ^{  \mathcal{P}  ^t  } ( \Sigma )$ is:

\begin{align*}
&\hspace{0em}  \alpha  : ( \Pi  ^{  \mathcal{P}  ^t  } ( \Sigma )(A)  \rightarrow   \Pi  ^{  \mathcal{P}  ^t  } ( \Sigma )(B))  \rightarrow  A  \rightarrow   \mathcal{P}  ^t (m)(B) \\
&\hspace{0em}  \alpha (f)(x)  \coloneqq   \alpha  ^{  \Sigma  \leftrightarrow m } (f)( \{ x \} )
\end{align*}

\begin{align*}
&\hspace{0em}  \gamma  : (A  \rightarrow   \mathcal{P}  ^t (m)(B))  \rightarrow   \Pi  ^{  \mathcal{P}  ^t  } ( \Sigma )(A)  \rightarrow   \Pi  ^{  \mathcal{P}  ^t  } ( \Sigma )(B) \\
&\hspace{0em}  \gamma (f)( \varsigma )  \coloneqq   \gamma  ^{  \Sigma  \leftrightarrow m } ( \lambda ( \{ x _1 ..x _n  \} ). f(x _1 )  \sqcup  ^m  ..  \sqcup  ^m  f(x _n ))( \varsigma )
\end{align*}

\par

\noindent
$ \alpha $ and $ \gamma $ are monotonic by inspection, and extensive and
reductive:

\begin{align*}
&\hspace{0em} extensive :  \forall  f  \varsigma , f( \varsigma )  \sqsubseteq   \gamma ( \alpha (f))( \varsigma ) \\
&\hspace{0em}  \gamma ( \alpha (f))( \varsigma ) \\
&\hspace{0em} =  \gamma  ^{  \Sigma  \leftrightarrow m } ( \lambda ( \{ x _1 ..x _n  \} ).  \\
&\hspace{1em}  \alpha  ^{  \Sigma  \leftrightarrow m } (f)( \{ x _1  \} )  \sqcup  ^m  ..  \sqcup  ^m   \alpha  ^{  \Sigma  \leftrightarrow m } (f)( \{ x _n  \} ))( \varsigma ) \\
&\hspace{2em}  \Lbag\text{\footnotesize{}  definition of  \(  \alpha  \)  and  \(  \gamma  \)   }\Rbag  \\
&\hspace{0em} =  \gamma  ^{  \Sigma  \leftrightarrow m } ( \lambda ( \{ x _1 ..x _n  \} ).  \alpha  ^{  \Sigma  \leftrightarrow m } (f)( \{ x _1 ..x _n  \} ))( \varsigma ) \\
&\hspace{2em}  \Lbag\text{\footnotesize{}  join-semilattice functorality of  \( m \)   }\Rbag  \\
&\hspace{0em}  \sqsupseteq  f( \varsigma )  \hspace{1em}   \Lbag\text{\footnotesize{}   \(  \gamma  ^{  \Sigma  \leftrightarrow m }   \circ   \alpha  ^{  \Sigma  \leftrightarrow m }  \)  extensive and  \(  \eta  \) -reduction  }\Rbag   \hspace{1em}  \hspace{1em}   \blacksquare  \\
&\hspace{0em} reductive :  \forall  f x,  \alpha ( \gamma (f))(x)  \sqsubseteq  f(x) \\
&\hspace{0em}  \alpha ( \gamma (f))(x) \\
&\hspace{0em} =  \alpha  ^{  \Sigma  \leftrightarrow m } ( \gamma  ^{  \Sigma  \leftrightarrow m } ( \lambda ( \{ x _1 ..x _n  \} ). f(x _1 )  \sqcup  ^m  ..  \sqcup  ^m  f(x _n )))( \{ x \} )  \\
&\hspace{2em}  \Lbag\text{\footnotesize{}  definition of  \(  \alpha  \)  and  \(  \gamma  \)   }\Rbag  \\
&\hspace{0em}  \sqsubseteq  ( \lambda ( \{ x _1 ..x _n  \} ). f(x _1 )  \sqcup  ^m  ..  \sqcup  ^m  f(x _n ))( \{ x \} ) \\
&\hspace{2em}  \Lbag\text{\footnotesize{}   \(  \alpha  ^{  \Sigma  \leftrightarrow m }   \circ   \gamma  ^{  \Sigma  \leftrightarrow m }  \)  reductive  }\Rbag  \\
&\hspace{0em} = f(x)  \hspace{1em}   \Lbag\text{\footnotesize{}   \(  \beta  \) -reduction  }\Rbag   \hspace{1em}  \hspace{1em}   \blacksquare 
\end{align*}

\par

\noindent
Finally, Property (3) commutes:

\begin{align*}
&\hspace{0em} goal:  \Pi  ^{  \mathcal{P}  ^t  } ( \gamma  ^{  \Sigma  \leftrightarrow m } (f))( \varsigma )  \sqsubseteq   \gamma ( \mathcal{P}  ^t (f))( \varsigma ) \\
&\hspace{0em}  \Pi  ^{  \mathcal{P}  ^t  } ( \gamma  ^{  \Sigma  \leftrightarrow m } (f))( \varsigma ) \\
&\hspace{0em} =  \gamma  ^{  \Sigma  \leftrightarrow m } ( \mathcal{P}  ^t ( \alpha  ^{  \Sigma  \leftrightarrow m } ( \gamma  ^{  \Sigma  \leftrightarrow m } (f))))( \varsigma )  \hspace{1em}   \Lbag\text{\footnotesize{}  definition of  \(  \Pi  ^{  \mathcal{P}  ^t  }  \)   }\Rbag  \\
&\hspace{0em}  \sqsubseteq   \gamma  ^{  \Sigma  \leftrightarrow m } ( \mathcal{P}  ^t (f))( \varsigma )  \hspace{1em}   \Lbag\text{\footnotesize{}   \(  \alpha  ^{  \Sigma  \leftrightarrow m }   \circ   \gamma  ^{  \Sigma  \leftrightarrow m }  \)  reductive  }\Rbag  \\
&\hspace{0em} =  \gamma ( \mathcal{P}  ^t (f))( \varsigma )  \hspace{1em}   \Lbag\text{\footnotesize{}  definition of  \(  \gamma  \)   }\Rbag   \hspace{1em}  \hspace{1em}   \blacksquare 
\end{align*}

\par

\paragraph{Flow Sensitivity}

\emph{$F ^t  \lbrack s \rbrack $ is a Galois transformer.\quad} Recall
the definition of $F ^t  \lbrack s \rbrack $ and
$ \Pi  ^{ F ^t  }  \lbrack s \rbrack $:

\begin{align*}
&\hspace{0em} F ^t  \lbrack s \rbrack (m)(A)   \coloneqq  s  \rightarrow  m( \lbrack A  \mapsto  s \rbrack )  \hspace{1em}  \hspace{1em}   \Pi  ^{ F ^t  }  \lbrack s \rbrack ( \Sigma )(A)   \coloneqq   \Sigma ( \lbrack A  \mapsto  s \rbrack )
\end{align*}

\par

\noindent
\emph{Flow Sensitivity Property (1):\quad} The action
$F ^t  \lbrack s \rbrack $ on functions:

\begin{align*}
&\hspace{0em} F ^t  \lbrack s \rbrack  : (A  \rightarrow  m(B))  \rightarrow  A  \rightarrow  F ^t  \lbrack s \rbrack (m)(B) \\
&\hspace{0em} F ^t  \lbrack s \rbrack (f)(x)(s)  \coloneqq  y  \leftarrow  ^m  f(x)  \;;\;   \ensuremath{\itop{return}} ^m  ( \{ y  \mapsto  s \} )
\end{align*}

\par

\noindent
To transport Galois connections we assume a Galois connection
$A  \rightarrow  m _1 (B)  \galois{\alpha^{m}}{\gamma^{m}}  A  \rightarrow  m _2 (B)$
and define $ \alpha $ and $ \gamma $:

\begin{align*}
&\hspace{0em}  \alpha  : (A  \rightarrow  F ^t  \lbrack s \rbrack (m _1 )(B))  \rightarrow  A  \rightarrow  F ^t  \lbrack s \rbrack (m _2 )(B) \\
&\hspace{0em}  \alpha (f)(x)(s)  \coloneqq   \alpha  ^m ( \lambda ( \{ x _1  \mapsto s _1 ..x _n  \mapsto s _n  \} ).  \\
&\hspace{1em} f(x _1 )(s _1 )  \sqcup  ^m  ..  \sqcup  ^m  f(x _n )(s _n ))( \{ x \mapsto s \} ) \\
&\hspace{0em}  \gamma  : (A  \rightarrow  F _t  \lbrack s \rbrack (m _2 )(B))  \rightarrow  A  \rightarrow  F ^t  \lbrack s \rbrack (m _1 )(B) \\
&\hspace{0em}  \gamma (f)(x)(s)  \coloneqq   \gamma  ^m ( \lambda ( \{ x _1  \mapsto s _1 ..x _n  \mapsto s _n  \} ).  \\
&\hspace{1em} f(x _1 )(s _1 )  \sqcup  ^m  ..  \sqcup  ^m  f(x _n )(s _n ))( \{ x \mapsto s \} )
\end{align*}

\par

\noindent
$ \alpha $ and $ \gamma $ are monotonic by inspection. $ \alpha $ and
$ \gamma $ are extensive and reductive:

\begin{align*}
&\hspace{0em} extensive :  \forall  f x s, f(x)(s)  \sqsubseteq   \gamma ( \alpha (f))(x)(s) \\
&\hspace{0em}  \gamma ( \alpha (f))(x)(s) \\
&\hspace{0em} =  \gamma  ^m ( \lambda ( \{ x _1  \mapsto s _1 ..x _n  \mapsto s _n  \} ).  \\
&\hspace{1em}  \alpha  ^m ( \lambda ( \{ x _1  \mapsto s _1 ..x _n  \mapsto s _n  \} ).  \\
&\hspace{2em} f(x _1 )(s _1 )  \sqcup  ^{ m _1  }  ..  \sqcup  ^{ m _1  }  f(x _n )(s _n ))( \{ x _1  \mapsto s _1  \} ) \\
&\hspace{1em}  \sqcup  ^{ m _2  }  ..  \sqcup  ^{ m _2  }  \\
&\hspace{1em}  \alpha  ^m ( \lambda ( \{ x _1  \mapsto s _1 ..x _n  \mapsto s _n  \} ).  \\
&\hspace{2em} f(x _1 )(s _1 )  \sqcup  ^{ m _1  }  ..  \sqcup  ^{ m _1  }  f(x _n )(s _n ))( \{ x _n  \mapsto s _n  \} )) \\
&\hspace{1em} ( \{ x \mapsto s \} )  \hspace{1em}   \Lbag\text{\footnotesize{}  definition of  \(  \alpha  \)  and  \(  \gamma  \)   }\Rbag  \\
&\hspace{0em}  \sqsupseteq   \gamma  ^m ( \lambda ( \{ x _1  \mapsto s _1 ..x _n  \mapsto s _n  \} ). \\
&\hspace{1em} ( \{ x _1  \mapsto s _1 ..x _n  \mapsto s _n  \}   \leftarrow  ^{ m _2  }   \\
&\hspace{2em}  \alpha  ^m ( \gamma  ^m  (\ensuremath{\itop{return}} ^{  m _2  } ( \{ x _1  \mapsto s _1  \} )))  \;;\;   \\
&\hspace{2em}  \alpha  ^m (f(x _1 )(s _1 )  \sqcup  ^{ m _1  }  ..  \sqcup  ^{ m _1  }  f(x _n )(s _n ))) \\
&\hspace{1em}  \sqcup  ^{ m _2  }  ..  \sqcup  ^{ m _2  }  \\
&\hspace{1em} ( \{ x _1  \mapsto s _1 ..x _n  \mapsto s _n  \}   \leftarrow  ^{ m _2  }   \\
&\hspace{2em}  \alpha  ^m ( \gamma  ^m  (\ensuremath{\itop{return}} ^{  m _2  } ( \{ x _n  \mapsto s _n  \} )))  \;;\;   \\
&\hspace{2em}  \alpha  ^m (f(x _1 )(s _1 )  \sqcup  ^{ m _1  }  ..  \sqcup  ^{ m _1  }  f(x _n )(s _n ))))( \{ x \mapsto s \} ) \\
&\hspace{2em}  \Lbag\text{\footnotesize{}   \ensuremath{\itop{left-unit}}  of  \( m \)  and  \(  \alpha  ^m   \circ   \gamma  ^m  \)  reductive  }\Rbag  \\
&\hspace{0em} =  \gamma  ^m ( \alpha  ^m ( \lambda ( \{ x _1  \mapsto s _1 ..x _n  \mapsto s _n  \} ). \\
&\hspace{1em}  \{ x _1  \mapsto s _1 ..x _n  \mapsto s _n  \}   \leftarrow  ^{ m _1  }   \\
&\hspace{2em}  \ensuremath{\itop{return}} ^{  m _1  } ( \{ x _1  \mapsto s _1 ..x _n  \mapsto s _n  \} )  \;;\;   \\
&\hspace{2em} f(x _1 )(s _1 )  \sqcup  ^{ m _1  }  ..  \sqcup  ^{ m _1  }  f(x _n )(s _n )))( \{ x \mapsto s \} ) \\
&\hspace{2em}  \Lbag\text{\footnotesize{}   \(  \alpha  ^m  \)  and  \(  \gamma  ^m  \)  homomorphic and join functorality  }\Rbag  \\
&\hspace{0em}  \sqsupseteq  f(x)(s)  \hspace{1em}   \Lbag\text{\footnotesize{}   \(  \gamma  ^m   \circ   \alpha  ^m  \)  extensive and  \ensuremath{\itop{left-unit}}  of  \( m \)   }\Rbag   \hspace{1em}  \hspace{1em}   \blacksquare  \\
&\hspace{0em} reductive :  \forall  f x s,  \alpha ( \gamma (f))(x)(s)  \sqsubseteq  f(x)(s) \\
&\hspace{0em}  \alpha ( \gamma (f))(x)(s) \\
&\hspace{0em} =  \alpha  ^m ( \lambda ( \{ x _1  \mapsto s _1 ..x _n  \mapsto s _n  \} ).  \\
&\hspace{1em}  \gamma  ^m ( \lambda ( \{ x _1  \mapsto s _1 ..x _n  \mapsto s _n  \} ).  \\
&\hspace{2em} f(x _1 )(s _1 )  \sqcup  ^{ m _2  }  ..  \sqcup  ^{ m _2  }  f(x _n )(s _n ))( \{ x _1  \mapsto s _1  \} ) \\
&\hspace{1em}  \sqcup  ^{ m _1  }  ..  \sqcup  ^{ m _1  }  \\
&\hspace{1em}  \gamma  ^m ( \lambda ( \{ x _1  \mapsto s _1 ..x _n  \mapsto s _n  \} ).  \\
&\hspace{2em} f(x _1 )(s _1 )  \sqcup  ^{ m _2  }  ..  \sqcup  ^{ m _2  }  f(x _n )(s _n ))( \{ x _n  \mapsto s _n  \} )) \\
&\hspace{1em} ( \{ x \mapsto s \} )  \hspace{1em}   \Lbag\text{\footnotesize{}  definition of  \(  \alpha  \)  and  \(  \gamma  \)   }\Rbag  \\
&\hspace{0em}  \sqsubseteq   \alpha  ^m ( \lambda ( \{ x _1  \mapsto s _1 ..x _n  \mapsto s _n  \} ). \\
&\hspace{1em} ( \{ x _1  \mapsto s _1 ..x _n  \mapsto s _n  \}   \leftarrow  ^{ m _1  }   \gamma  ^m ( \alpha  ^m  (\ensuremath{\itop{return}} ^{  m _1  } ( \{ x _1  \mapsto s _1  \} )))  \;;\;   \\
&\hspace{2em}  \gamma  ^m (f(x _1 )(s _1 )  \sqcup  ^{ m _2  }  ..  \sqcup  ^{ m _2  }  f(x _n )(s _n ))) \\
&\hspace{1em}  \sqcup  ^{ m _1  }  ..  \sqcup  ^{ m _1  }  \\
&\hspace{1em} ( \{ x _1  \mapsto s _1 ..x _n  \mapsto s _n  \}   \leftarrow  ^{ m _1  }   \gamma  ^m ( \alpha  ^m  (\ensuremath{\itop{return}} ^{  m _1  } ( \{ x _n  \mapsto s _n  \} )))  \;;\;   \\
&\hspace{2em}  \gamma  ^m (f(x _1 )(s _1 )  \sqcup  ^{ m _2  }  ..  \sqcup  ^{ m _2  }  f(x _n )(s _n ))))( \{ x \mapsto s \} ) \\
&\hspace{2em}  \Lbag\text{\footnotesize{}   \ensuremath{\itop{left-unit}}  of  \( m \)  and  \(  \gamma  ^m   \circ   \alpha  ^m  \)  extensive  }\Rbag 
\end{align*}

\begin{align*}
&\hspace{0em} =  \alpha  ^m ( \gamma  ^m ( \lambda ( \{ x _1  \mapsto s _1 ..x _n  \mapsto s _n  \} ). \\
&\hspace{1em}  \{ x _1  \mapsto s _1 ..x _n  \mapsto s _n  \}   \leftarrow  ^{ m _2  }   \ensuremath{\itop{return}} ^{  m _2  } ( \{ x _1  \mapsto s _1 ..x _n  \mapsto s _n  \} )  \;;\;   \\
&\hspace{1em} f(x _1 )(s _1 )  \sqcup  ^{ m _2  }  ..  \sqcup  ^{ m _2  }  f(x _n )(s _n )))( \{ x \mapsto s \} ) \\
&\hspace{2em}  \Lbag\text{\footnotesize{}   \(  \alpha  ^m  \)  and  \(  \gamma  ^m  \)  homomorphic and join functorality  }\Rbag  \\
&\hspace{0em}  \sqsubseteq  f(x)(s)  \hspace{1em}   \Lbag\text{\footnotesize{}   \(  \alpha  ^m   \circ   \gamma  ^m  \)  extensive and  \ensuremath{\itop{left-unit}}  of  \( m \)   }\Rbag   \hspace{1em}  \hspace{1em}   \blacksquare 
\end{align*}

\par

\noindent
Finally, Property (1) commutes, assuming that
$A  \rightarrow  m _1 (B)  \galois{\alpha^{m}}{\gamma^{m}}  A  \rightarrow  m _2 (B)$
is homomorphic:

\begin{align*}
&\hspace{0em} goal:  \forall  f s, F ^t  \lbrack s \rbrack  \lbrack m _2  \rbrack ( \alpha  ^m (f))(x)(s) =  \alpha (F ^t  \lbrack s \rbrack  \lbrack m _1  \rbrack (f))(x)(s) \\
&\hspace{0em}  \alpha (F ^t  \lbrack s \rbrack  \lbrack m _1  \rbrack (f))(x)(s) \\
&\hspace{0em} =  \alpha  ^m ( \lambda ( \{ x _1  \mapsto s _1 ..x _n  \mapsto s _n  \} ).  \\
&\hspace{1em} (y  \leftarrow  ^{ m _1  }  f(x)  \;;\;   \ensuremath{\itop{return}} ^{  m _1  } (y _1 )(s _1 ))  \sqcup  ^{ m _1  }  ..  \sqcup  ^{ m _1  }   \\
&\hspace{1em} (y  \leftarrow  ^{ m _1  }  f(x)  \;;\;   \ensuremath{\itop{return}} ^{  m _1  } (y _n )(s _n )))( \{ x \mapsto s \} ) \\
&\hspace{2em}  \Lbag\text{\footnotesize{}  definition of  \(  \alpha  \)  and  \( F ^t  \lbrack s \rbrack  \lbrack m _1  \rbrack  \)   }\Rbag  \\
&\hspace{0em} = y  \leftarrow  ^{ m _2  }   \alpha  ^m (f)(x)  \;;\;   \ensuremath{\itop{return}} ^{  m _2  } (y)(s) \\
&\hspace{2em}  \Lbag\text{\footnotesize{}  homomorphic on  \(  \ensuremath{\itop{bind}} ^{  m _1  }  \)  and  \(  \ensuremath{\itop{return}} ^{  m _1  }  \)   }\Rbag  \\
&\hspace{0em} = F ^t  \lbrack s \rbrack  \lbrack m _2  \rbrack ( \alpha  ^m (f))(x)  \hspace{1em}   \Lbag\text{\footnotesize{}  definition of  \( F ^t  \lbrack s \rbrack  \lbrack m _2  \rbrack  \)   }\Rbag   \hspace{1em}  \hspace{1em}   \blacksquare 
\end{align*}

\par

\noindent
\emph{Flow Sensitivity Property (2):\quad} The action
$ \Pi  ^{ F ^t  \lbrack s \rbrack  } $ on functions uses the mapping to
monadic functions defined in Property (3):

\begin{align*}
&\hspace{0em}  \Pi  ^{ F ^t  }  \lbrack s \rbrack  : ( \Sigma (A)  \rightarrow   \Sigma (B))  \rightarrow   \Pi  ^{ F ^t  }  \lbrack s \rbrack ( \Sigma )(A)  \rightarrow   \Pi  ^{ F ^t  }  \lbrack s \rbrack ( \Sigma )(B) \\
&\hspace{0em}  \Pi  ^{ F ^t  }  \lbrack s \rbrack (f)( \varsigma )  \coloneqq   \gamma  ^{  \Sigma  \leftrightarrow  \gamma  } (F ^t  \lbrack s \rbrack ( \alpha  ^{  \Sigma  \leftrightarrow  \gamma  } (f)))
\end{align*}

\par

\noindent
To transport Galois connections, we assume a Galois connection
$ \Sigma  _1 (A)  \rightarrow   \Sigma  _1 (B)  \galois{\alpha^{\Sigma}}{\gamma^{\Sigma}}   \Sigma  _2 (A)  \rightarrow   \Sigma  _2 (B)$
and define $ \alpha $ and $ \gamma $ as instantiations of
$ \alpha  ^{  \Sigma  } $ and $ \gamma  ^{  \Sigma  } $:

\begin{align*}
&\hspace{0em}  \alpha  : ( \Pi  ^{ F ^t  }  \lbrack s \rbrack ( \Sigma  _1 )(A)  \rightarrow   \Pi  ^{ F ^t  }  \lbrack s \rbrack ( \Sigma  _1 )(B)) \\
&\hspace{1em}  \rightarrow   \Pi  ^{ F ^t  }  \lbrack s \rbrack ( \Sigma  _2 )(A)  \rightarrow   \Pi  ^{ F ^t  }  \lbrack s \rbrack ( \Sigma  _2 )(B) \\
&\hspace{0em}  \gamma  : ( \Pi  ^{ F ^t  }  \lbrack s \rbrack ( \Sigma  _2 )(A)  \rightarrow   \Pi  ^{ F ^t  }  \lbrack s \rbrack ( \Sigma  _2 )(B)) \\
&\hspace{1em}  \rightarrow   \Pi  ^{ F ^t  }  \lbrack s \rbrack ( \Sigma  _1 )(A)  \rightarrow   \Pi  ^{ F ^t  }  \lbrack s \rbrack ( \Sigma  _1 )(B) \\
&\hspace{0em}  \alpha (f)( \varsigma ) =  \alpha  ^{  \Sigma  } (f)( \varsigma )  \hspace{1em}  \hspace{1em}   \gamma (f)( \varsigma ) =  \gamma  ^{  \Sigma  } (f)( \varsigma )
\end{align*}

\par

\noindent
Monotonicity, reductive and extensive properties carry over by
definition. Finally, Property (2) commutes, assuming that
$ \alpha  ^{  \Sigma  } $ and $ \alpha  ^m $ commute with both
$ \gamma  ^{  \Sigma  \leftrightarrow m } $ and
$ \alpha  ^{  \Sigma  \leftrightarrow m } $:

\begin{align*}
&\hspace{0em} goal:  \Pi  ^{ F ^t  }  \lbrack s \rbrack  \lbrack  \Sigma  _2  \rbrack ( \alpha  ^{  \Sigma  } (f))( \varsigma ) =  \alpha  ^{  \Sigma  } ( \Pi  ^{ F ^t  }  \lbrack s \rbrack  \lbrack  \Sigma  _1  \rbrack (f))( \varsigma ) \\
&\hspace{0em}  \alpha  ^{  \Sigma  } ( \Pi  ^{ F ^t  }  \lbrack s \rbrack  \lbrack  \Sigma  _1  \rbrack (f))( \varsigma ) \\
&\hspace{0em} =  \alpha  ^{  \Sigma  } ( \gamma  ^{  \Sigma  \leftrightarrow  \gamma  } (F ^t  \lbrack s \rbrack ( \alpha  ^{  \Sigma  \leftrightarrow  \gamma  } (f))))( \varsigma )  \hspace{1em}   \Lbag\text{\footnotesize{}  definition of  \(  \Pi  ^{ F ^t  }  \lbrack s \rbrack  \)   }\Rbag  \\
&\hspace{0em} =  \alpha  ^{  \Sigma  } ( \gamma  ^{  \Sigma  \leftrightarrow  \gamma  } ( \lambda (x)(s). y  \leftarrow  ^{ m _1  }   \alpha  ^{  \Sigma  \leftrightarrow  \gamma  } (f)(x)  \;;\;   \\
&\hspace{1em}  \ensuremath{\itop{return}} ^{  m _1  } ( \{ y \mapsto s \} )))( \varsigma )  \hspace{1em}   \Lbag\text{\footnotesize{}  definition of  \( F ^t  \lbrack s \rbrack  \)   }\Rbag  \\
&\hspace{0em} =  \gamma  ^{  \Sigma  \leftrightarrow  \gamma  } ( \alpha  ^m ( \lambda (x)(s). y  \leftarrow  ^{ m _1  }   \alpha  ^{  \Sigma  \leftrightarrow  \gamma  } (f)(x)  \;;\;   \\
&\hspace{1em}  \ensuremath{\itop{return}} ^{  m _1  } ( \{ y \mapsto s \} )))( \varsigma )  \hspace{1em}   \Lbag\text{\footnotesize{}   \(  \alpha  ^{  \Sigma  }  \)  and  \(  \gamma  ^{  \Sigma  \leftrightarrow  \gamma  }  \)  commute  }\Rbag 
\end{align*}

\begin{align*}
&\hspace{0em} =  \gamma  ^{  \Sigma  \leftrightarrow  \gamma  } ( \lambda (x)(s). y  \leftarrow  ^{ m _2  }   \alpha  ^m ( \alpha  ^{  \Sigma  \leftrightarrow  \gamma  } (f))(x)  \;;\;   \\
&\hspace{1em}  \ensuremath{\itop{return}} ^{  m _2  } ( \{ y \mapsto s \} ))( \varsigma )  \hspace{1em}   \Lbag\text{\footnotesize{}   \(  \alpha  ^m  \)  homomorphic  }\Rbag  \\
&\hspace{0em} =  \gamma  ^{  \Sigma  \leftrightarrow  \gamma  } ( \lambda (x)(s). y  \leftarrow  ^{ m _2  }   \alpha  ^{  \Sigma  \leftrightarrow  \gamma  } ( \alpha  ^{  \Sigma  } (f))(x)  \;;\;   \\
&\hspace{1em}  \ensuremath{\itop{return}} ^{  m _2  } ( \{ y \mapsto s \} ))( \varsigma )  \hspace{1em}   \Lbag\text{\footnotesize{}   \(  \alpha  ^{ m }  \)  and  \(  \alpha  ^{  \Sigma  \leftrightarrow  \gamma  }  \)  commute  }\Rbag  \\
&\hspace{0em} =  \Pi  ^{  \mathcal{P}  ^t  }  \lbrack  \Sigma  _2  \rbrack ( \alpha  ^{  \Sigma  } (f))( \varsigma )  \hspace{1em}   \Lbag\text{\footnotesize{}  definition of  \(  \Pi  ^{  \mathcal{P}  ^t  }  \lbrack  \Sigma  _2  \rbrack  \)  and  \(  \alpha  ^{  \Sigma  }  \)   }\Rbag   \hspace{1em}  \hspace{1em}   \blacksquare 
\end{align*}

\par

\noindent
\emph{Flow Sensitivity Property (3):\quad} Assume a Galois connection:

\begin{align*}
&\hspace{0em}  \Sigma (A)  \rightarrow   \Sigma (B)  \galois{\alpha^{\Sigma\leftrightarrow m}}{\gamma^{\Sigma\leftrightarrow m}}  A  \rightarrow  m(B)
\end{align*}

\par

\noindent
The Galois connection between $F ^t  \lbrack s \rbrack (m)$ and
$ \Pi  ^{ F ^t  }  \lbrack s \rbrack ( \Sigma )$ is:

\begin{align*}
&\hspace{0em}  \alpha  : ( \Pi  ^{ F ^t  }  \lbrack s \rbrack ( \Sigma )(A)  \rightarrow   \Pi  ^{ F ^t  }  \lbrack s \rbrack ( \Sigma )(B))  \rightarrow  A  \rightarrow  F ^t  \lbrack s \rbrack (m)(B) \\
&\hspace{0em}  \alpha (f)(x)(s)  \coloneqq   \alpha  ^{  \Sigma  \leftrightarrow m } (f)( \{ x \mapsto s \} ) \\
&\hspace{0em}  \gamma  : (A  \rightarrow  F ^t  \lbrack s \rbrack (m)(B))  \rightarrow   \Pi  ^{ F ^t  }  \lbrack s \rbrack ( \Sigma )(A)  \rightarrow   \Pi  ^{ F ^t  }  \lbrack s \rbrack ( \Sigma )(B) \\
&\hspace{0em}  \gamma (f)( \varsigma )  \coloneqq   \gamma  ^{  \Sigma  \leftrightarrow m } ( \lambda ( \{ x _1  \mapsto s _1 ..x _n  \mapsto s _n  \} ).  \\
&\hspace{2em} f(x _1 )(s _1 )  \sqcup  ^m  ..  \sqcup  ^m  f(x _n )(s _n ))( \varsigma )
\end{align*}

\par

\noindent
$ \alpha $ and $ \gamma $ are monotonic by inspection. $ \alpha $ and
$ \gamma $ are extensive and reductive:

\begin{align*}
&\hspace{0em} extensive :  \forall  f  \varsigma , f( \varsigma )  \sqsubseteq   \gamma ( \alpha (f))( \varsigma ) \\
&\hspace{0em}  \gamma ( \alpha (f))( \varsigma ) \\
&\hspace{0em} =  \gamma  ^{  \Sigma  \leftrightarrow m } ( \lambda ( \{ x _1  \mapsto s _1 ..x _n  \mapsto s _n  \} ).  \\
&\hspace{1em}  \alpha  ^{  \Sigma  \leftrightarrow m } (f)( \{ x _1  \mapsto s _1  \} )  \sqcup  ^m  ..  \sqcup  ^m   \alpha  ^{  \Sigma  \leftrightarrow m } (f)( \{ x _n  \mapsto s _n  \} ))( \varsigma ) \\
&\hspace{2em}  \Lbag\text{\footnotesize{}  definition of  \(  \alpha  \)  and  \(  \gamma  \)   }\Rbag  \\
&\hspace{0em} =  \gamma  ^{  \Sigma  \leftrightarrow m } ( \alpha  ^{  \Sigma  \leftrightarrow m } (f))( \varsigma )  \hspace{1em}   \Lbag\text{\footnotesize{}  join-semilattice functorality of  \( m \)   }\Rbag  \\
&\hspace{0em}  \sqsupseteq  f( \varsigma )  \hspace{1em}   \Lbag\text{\footnotesize{}   \(  \gamma  ^{  \Sigma  \leftrightarrow m }   \circ   \alpha  ^{  \Sigma  \leftrightarrow m }  \)  extensive  }\Rbag   \hspace{1em}  \hspace{1em}   \blacksquare  \\
&\hspace{0em} reductive :  \forall  f x,  \alpha ( \gamma (f))(x)(s)  \sqsubseteq  f(x)(s) \\
&\hspace{0em}  \alpha ( \gamma (f))(x)(s) \\
&\hspace{0em} =  \alpha  ^{  \Sigma  \leftrightarrow m } ( \gamma  ^{  \Sigma  \leftrightarrow m } ( \lambda ( \{ x _1  \mapsto s _1 ..x _n  \mapsto s _n  \} ).  \\
&\hspace{1em} f(x _1 )(s _1 )  \sqcup  ^m  ..  \sqcup  ^m  f(x _n )(s _n )))( \{ x \mapsto s \} ) \\
&\hspace{2em}  \Lbag\text{\footnotesize{}  definition of  \(  \alpha  \)  and  \(  \gamma  \)   }\Rbag  \\
&\hspace{0em}  \sqsubseteq  ( \lambda ( \{ x _1  \mapsto s _1 ..x _n  \mapsto s _n  \} ).  \\
&\hspace{1em} f(x _1 )(s _1 )  \sqcup  ^m  ..  \sqcup  ^m  f(x _n )(s _n ))( \{ x \mapsto s \} ) \\
&\hspace{2em}  \Lbag\text{\footnotesize{}   \(  \alpha  ^{  \Sigma  \leftrightarrow m }   \circ   \gamma  ^{  \Sigma  \leftrightarrow m }  \)  reductive  }\Rbag  \\
&\hspace{0em} = f(x)(s)  \hspace{1em}   \Lbag\text{\footnotesize{}   \(  \beta  \) -reduction  }\Rbag   \hspace{1em}  \hspace{1em}   \blacksquare 
\end{align*}

\par

\noindent
Finally, Property (3) commutes:

\begin{align*}
&\hspace{0em} goal:  \Pi  ^{ F ^t  }  \lbrack s \rbrack ( \gamma  ^{  \Sigma  \leftrightarrow m } (f))( \varsigma )  \sqsubseteq   \gamma (F ^t  \lbrack s \rbrack (f))( \varsigma ) \\
&\hspace{0em}  \Pi  ^{ F ^t  }  \lbrack s \rbrack ( \gamma  ^{  \Sigma  \leftrightarrow m } (f))( \varsigma ) \\
&\hspace{0em} =  \gamma  ^{  \Sigma  \leftrightarrow m } (F ^t  \lbrack s \rbrack ( \alpha  ^{  \Sigma  \leftrightarrow m } ( \gamma  ^{  \Sigma  \leftrightarrow m } (f))))( \varsigma )  \hspace{1em}   \Lbag\text{\footnotesize{}  definition of  \(  \Pi  ^{ F ^t  }  \lbrack s \rbrack  \)   }\Rbag  \\
&\hspace{0em}  \sqsubseteq   \gamma  ^{  \Sigma  \leftrightarrow m } (F ^t  \lbrack s \rbrack (f))( \varsigma )  \hspace{1em}   \Lbag\text{\footnotesize{}   \(  \alpha  ^{  \Sigma  \leftrightarrow m }   \circ   \gamma  ^{  \Sigma  \leftrightarrow m }  \)  reductive  }\Rbag  \\
&\hspace{0em} =  \gamma (F ^t  \lbrack s \rbrack (f))( \varsigma )  \hspace{1em}   \Lbag\text{\footnotesize{}  definition of  \(  \gamma  \)   }\Rbag   \hspace{1em}  \hspace{1em}   \blacksquare 
\end{align*}

\end{document}